\RequirePackage{ifpdf}
\documentclass[12pt,letterpaper]{article}
\usepackage{jheppub}
\usepackage{epsfig}
\usepackage[latin1]{inputenc}
\usepackage{bbm,amsfonts}
\usepackage{graphicx}
\usepackage{amssymb,amsmath,amsfonts,mathtools}
\usepackage{fancybox}
\usepackage{enumerate}
\usepackage{dsfont}
\usepackage{verbatim}
\usepackage{wrapfig}
\usepackage{slashed}

\author[a]{Marco S. Bianchi,}
\author[b]{ Luca Griguolo,}
\author[c,d]{Andrea Mauri,}
\author[c,d]{\\Silvia Penati,}
\author[e]{Michelangelo Preti}
\author[f]{and Domenico Seminara}
 
\affiliation[a]{Center for Research in String Theory - School of Physics and Astronomy Queen Mary University of London, Mile End Road, London E1 4NS, UK}
\affiliation[b]{Dipartimento di Fisica e Scienze della Terra, Universit\`a di Parma and INFN Gruppo Collegato di Parma, Viale G.P. Usberti 7/A, 43100 Parma, Italy}
\affiliation[c]{ Dipartimento di Fisica, Universit\`a degli studi di Milano--Bicocca, Piazza della Scienza 3, I-20126 Milano, Italy }
\affiliation[d]{INFN, Sezione di Milano--Bicocca, Piazza della Scienza 3, I-20126 Milano, Italy }
\affiliation[e]{DESY Hamburg, Theory Group, Notkestra\ss e 85, 22607 Hamburg, Germany} 
\affiliation[f]{Dipartimento di Fisica, Universit\`a di Firenze and INFN Sezione di Firenze, via G. Sansone 1, 50019 Sesto Fiorentino, Italy\\}

\emailAdd{m.s.bianchi@qmul.ac.uk}  
\emailAdd{luca.griguolo@pr.infn.it} 
\emailAdd{andrea.mauri@mi.infn.it} 
\emailAdd{silvia.penati@mib.infn.it} 
\emailAdd{michelangelo.preti@desy.de} 
\emailAdd{seminara@fi.infn.it}
   \preprint{QMUL-PH-17-09}
  
\abstract{We present the three-loop calculation of the  Bremsstrahlung function associated to the 1/2--BPS cusp in ABJM theory, including color subleading corrections.
Using the BPS condition we reduce the computation to that of a cusp with vanishing angle. We work within the framework of heavy quark effective theory (HQET) that further simplifies the analytic evaluation of the relevant cusp anomalous dimension in the near--BPS limit. The result passes nontrivial tests, such as exponentiation, and is in agreement with the conjecture made in \cite{Bianchi:2014laa} for the exact expression of the Bremsstrahlung function, based on the relation with fermionic latitude Wilson loops.  
}

\title{Towards the exact Bremsstrahlung  function of ABJM theory}


\newcommand{\be}{\begin{equation}}
\newcommand{\ee}{\end{equation}}
\newcommand{\beq}{\begin{equation}}
\newcommand{\eeq}{\end{equation}}
\newcommand{\bea}{\begin{eqnarray}}
\newcommand{\eea}{\end{eqnarray}}
\newcommand{\ena}{\end{eqnarray}}
\newcommand {\non}{\nonumber}

\renewcommand{\a}{\alpha}
\renewcommand{\b}{\beta}

\renewcommand{\d}{\delta}
\renewcommand{\th}{\theta}
\renewcommand{\TH}{\Theta}
\newcommand{\pa}{\partial}
\newcommand{\g}{\gamma}
\newcommand{\G}{\Gamma}

\newcommand{\e}{\epsilon}

\renewcommand{\l}{\lambda}

\newcommand{\m}{\mu}

\newcommand{\n}{\nu}

\newcommand{\p}{\pi}

\newcommand{\s}{\sigma}
\renewcommand{\S}{\Sigma}
\renewcommand{\t}{\tau}

\def\Tr{\textrm{Tr}}

\numberwithin{equation}{section}

\def\clock{{\count0=\time
           \divide\count0 60
           \ifnum\count0<10 0\fi\the\count0
           \multiply\count0 -60 \advance\count0 \time
           :\ifnum\count0<10 0\fi \the\count0
         }}
\newcommand{\timestamp}{{\small\vbox{\hbox{\tt\jobname.tex}
\hbox{\the\day/\the\month/\the\year, \clock}}}}
\begin{document}

\maketitle
\allowdisplaybreaks

\section{Introduction}
Wilson loops (WLs) play an ubiquitous role in gauge theories, both at perturbative and non--perturbative level. Depending on the particular contour, they encode the potential between heavy colored sources and control important aspects of scattering between charged particles. At the same time they provide the basis for the lattice formulation of any gauge theory and represent a general class of observables with deep mathematical meaning. The AdS/CFT correspondence \cite{Maldacena:1997re,Witten:1998qj,Gubser:1998bc} has triggered further interest for WLs in supersymmetric gauge theories. In fact, they are directly related to  fundamental string or brane configurations of the dual theory and are natural observables capable of testing the correspondence itself \cite{Maldacena:1998im,Rey:1998ik}. For example, the vacuum expectation values of some BPS WLs are non--trivial functions of the gauge coupling, potentially providing an interpolation between weak and strong coupling regimes. The exact computation of these quantities is sometimes possible, for instance using localization techniques \cite{Pestun:2007rz,Kapustin:2009kz}, offering non--trivial tests of the AdS/CFT predictions \cite{Erickson:2000af,Drukker:2000rr}. 

When the theory is conformal, WLs govern the computation of the energy radiated by a moving quark in the low energy limit, the so--called Bremsstrahlung function $B(\lambda)$ \cite{Correa:2012at}. This function also enters the small angle expansion of the cusp anomalous dimension $\G_{cusp}$ that, in turn, controls the short distance divergences of a WL in the proximity of a cusp, according to the universal behaviour $\langle WL \rangle \sim \exp{(-\G_{cusp} \log{\frac{\Lambda}{\mu}})}$ (where $\Lambda$ and $\mu$ are IR and UV cutoffs, respectively). Precisely, for supersymmetric WLs given as the holonomy of generalized connections that include also couplings to matter, we can consider a cusped WL depending on two parameters, $\varphi$ representing the geometric angle between the two Wilson lines defining the cusp, and the latitude angle $\theta$ describing the change in the orientation of the couplings to matter between the two rays \cite{Drukker:1999zq,Drukker:2011za}. At $\theta^2=\varphi^2$ the cusped WL is BPS, therefore finite, and its anomalous dimension vanishes. For small angles, $\th, \varphi \ll 1$, the expansion of $\G_{cusp}$ around the BPS point reads
\beq\label{eq:BPS}
\Gamma_{cusp} (\l, \th, \varphi) \sim B(\lambda) (\th^2 - \varphi^2)   
\eeq
where $B(\lambda)$ is the Bremsstrahlung function, given as a function of the coupling constant of the theory.  

In ${\cal N} = 4$ SYM theory an exact prescription to extract this non--BPS observable from BPS loops has been given in the seminal paper \cite{Correa:2012at}, where results from localization where explicitly used. The very same result can be obtained by starting from an exact set of TBA equations \cite{Bombardelli:2009ns,Gromov:2009bc,Arutyunov:2009ur} describing the generalized cusp \cite{Correa:2012hh,Drukker:2012de}, solving the system in the near--BPS limit \cite{Gromov:2012eu}. Actually, using integrability, one can further obtain a Bremsstrahlung function $B_L(\lambda)$, by considering the insertion of some chiral operator of R--charge L on the tip of the cusp \cite{Gromov:2013qga}. This last result still calls for a localization explanation (see some progress in this direction in \cite{Bonini:2015fng}).  Moreover, the use of the quantum spectral curve techniques \cite{Gromov:2013pga,Gromov:2014caa} has allowed to obtain results away from the BPS point \cite{Gromov:2015dfa}. A proposal for the Bremsstrahlung function in four dimensional ${\cal N}=2$ SYM appeared more recently in \cite{Fiol:2015spa}. 

It is quite natural to extend the above investigations to the three--dimensional case, exploring the Bremsstrahlung function in ${\cal N}=6$ superconformal  ABJ(M) theories \cite{Aharony:2008ug,Aharony:2008gk}. One immediately faces a first difference with the four-dimensional case: In ABJ(M) models not only bosonic but also fermionic matter can be used to build up generalized (super)connections whose holonomy gives rise to supersymmetric loop operators. Precisely, there are two prototypes of supersymmetric Wilson lines, one associated to a generalized gauge connection that includes couplings to bosonic matter only, so preserving 1/6 of the original supersymmetries \cite{Berenstein:2008dc,Drukker:2008zx,Chen:2008bp,Rey:2008bh}, and another one in which the addition of local couplings to the fermions enhances the operator to be 1/2--BPS \cite{Drukker:2009hy}. The latter should be dual to the fundamental string on $AdS_4\times CP^3$. While the 1/2 BPS operator is cohomologically equivalent (and therefore basically indistinguishable) to a linear combination of 1/6 BPS ones, the fact that they preserve different portions of supersymmetry allows for constructing different non--BPS observables starting from them. Generalized cusps formed with 1/6-BPS rays or 1/2-BPS rays are actually different \cite{Griguolo:2012iq,Lewkowycz:2013laa} and, consequently, different Bremsstrahlung functions can be defined and potentially evaluated exactly. In particular, in \cite{Lewkowycz:2013laa} a proposal for the exact Bremsstrahlung function of the 1/6-BPS cusp was put forward, based on the localization result for the 1/6-BPS circular WL, and an extension to the 1/2-BPS case was conjectured.

In \cite{Bianchi:2014laa} an exact formula  appeared 
that expresses the Bremsstrahlung function $B_{1/2}$ for 1/2--BPS quark configurations as the derivative of a fermionic latitude WL with respect to the latitude angle. This proposal was suggested by the analogy with the ${\cal N}=4$ SYM case \cite{Correa:2012at} and supported by an explicit two--loop computation that agrees with the result obtained directly from the cusp with 1/2--BPS rays \cite{Griguolo:2012iq}. Given that the fermionic latitude WL is cohomologically equivalent to a bosonic latitude WL \cite{Bianchi:2014laa}, $B_{1/2}$ can be eventually expressed as
\begin{equation}\label{eq:conjecture0}
B_{1/2} = -\frac{i}{8\pi}\, \frac{\langle W_{1/6}\rangle-\langle \hat W_{1/6}\rangle}{\langle W_{1/6}\rangle+\langle \hat W_{1/6}\rangle}
\end{equation}
in terms of the well--known vev of 1/6-BPS WLs associated to the two $U(N)$ gauge groups and computed exactly at framing one using localization \cite{Kapustin:2009kz,Marino:2009jd,Drukker:2010nc}. Remarkably, this formula coincides with the proposal of \cite{Lewkowycz:2013laa}, despite being derived from different arguments. Such an expression is also in agreement with a one-loop string computation at strong coupling \cite{Forini:2012bb,Aguilera-Damia:2014bqa,Correa:2014aga}.

Expanding the matrix model result for $\langle W_{1/6}\rangle$ and $\langle \hat W_{1/6}\rangle$ at weak coupling one can infer that in the planar limit this expression contains odd powers of the 't Hooft coupling $\lambda=N/k$  only. In principle, there should be a second contribution to (\ref{eq:conjecture0}) proportional to the derivative of a multiply $n$--wound 1/6--BPS circular WL with respect to $n$. This term should provide the even power expansion in $\lambda$. However, using the exact expression for the $n$--wound WL known from localization, we can verify that both at weak and strong coupling this term vanishes (at least) in the planar limit \cite{Lewkowycz:2013laa,Bianchi:2014laa}. In particular, up to two loops it vanishes also at finite $N$.

Based on this observation and using the weak coupling expansion of $\langle W_{1/6}\rangle$ and $\langle \hat W_{1/6}\rangle$ as derived from the matrix model (including the color subleading contributions), eq. (\ref{eq:conjecture0}) gives the following prediction for the $B_{1/2}$ function, at any finite $N$
\beq \label{eq:prediction}
B_{1/2}(k,N) = \frac{N}{8\, k}-\frac{\pi ^2\, N \left(N^2-3\right)}{48\, k^3}+{\cal O}\left(k^{-4}\right)
\eeq
The first term has been already checked in \cite{Bianchi:2014laa}. In order to put this conjecture on more solid bases, further checks at higher orders are desirable. 
In this paper we perform a three--loop computation of $B_{1/2}$ directly from its definition in term of the fermionic cusped WL. As a result we obtain 
exactly the predicted $1/k^3$ coefficient in (\ref{eq:prediction}), so providing a highly non--trivial test of the proposal \cite{Bianchi:2014laa,Lewkowycz:2013laa}. In particular, at least up to three loops,  
we have checked that (\ref{eq:conjecture0}) is valid not only in the planar limit, but also at finite $N$. 
We remark that since we are checking the Bremsstrahlung function through an odd perturbative order, our test is independent of any assumption on the even--$\lambda$ part that was discarded in \eqref{eq:conjecture0}, according to the discussion above.

To simplify the three--loop calculation, we rely on the BPS condition satisfied by the fermionic cusped WL \cite{Griguolo:2012iq} that implies the small angle expansion (\ref{eq:BPS}). From this equation we can read the Bremsstrahlung function equivalently from the $\theta$ or the $\varphi$ expansions of $\G_{1/2}(\l, \th,\varphi)$, setting the other angle to zero. It turns out to be  particularly convenient to set $\varphi = 0$ (flat cusp limit), working with the internal angle $\theta$ only. As will be discussed in details, this choice leads to crucial simplifications at several stages of the calculation. 

As usual, when computing  the cusped WL in the near--BPS limit UV and IR divergences arise. Here we employ dimensional regularization with dimensional reduction (DRED) to regulate the UV divergences and introduce an IR suppression factor to tame the IR divergences coming from the integration region at large distances. The cusp anomalous dimension is then obtained through the usual renormalization group equation for the renormalized cusped WL  \cite{Korchemsky:1987wg}. 
The calculation is carried out in momentum space where the Wilson lines are effectively described by non--relativistic, eikonal propagators.   This procedure has a physical interpretation in the context of heavy quark effective theory (HQET) \cite{Gervais:1979fv,Dotsenko:1979wb,Arefeva:1980zd}. 


 
The plan of the paper is the following. In section 2 we review the construction of 1/2--BPS WLs in ABJM theories and present the proposal for the exact 1/2--BPS Bremsstrahlung function, as derived in \cite{Bianchi:2014laa}. In section 3 we outline the strategy of the three--loop computation, describing in particular the HQET formalism. 
Section 4 contains a review of the one-- and two--loop results that are important in order to check the correct exponentiation. The actual three--loop calculation appears in section 5, while the Bremsstrahlung function is extracted in section 6. Section 7 contains our conclusions. Many technical details, together with the results for the relevant two-- and three--loop diagrams are deferred to eight appendices.

\section{Exact 1/2--BPS Bremsstrahlung in ABJM theory}

\subsection{The 1/2--BPS cusp in ABJM}

\begin{figure}[ht]
\centering
	\includegraphics[width=.6 \textwidth]{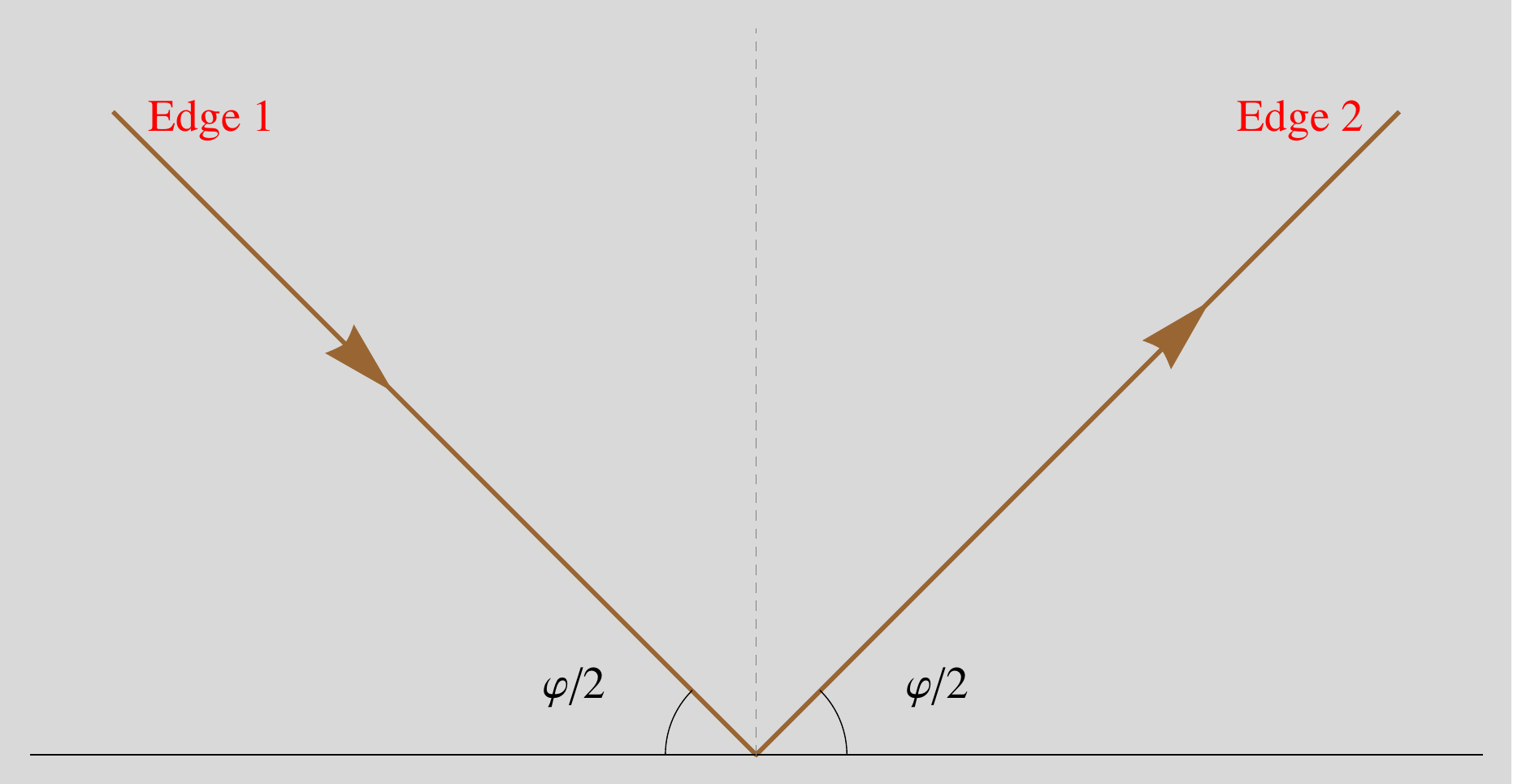}
\caption{\label{cusp1} Eq. \eqref{cusp} represents a planar cusp, whose angular extension is given by $\pi-\varphi$.}
\end{figure}

We start reviewing the definition of the locally 1/2--BPS cusp in ABJM theory \footnote{Generalities on the ABJM theory in our conventions are given in appendices \ref{app:ABJM}, \ref{app:FeynmanRules}.}.
This is constructed by considering  the fermionic WL \footnote{
Here the path--exponential is defined by
 \[
P\, \exp{ \left( -i \int_\G d\tau {\cal L}(\tau)\right) } \equiv 1 - i \int_\G d\tau \, {\cal L}(\tau) - \int_\G  d\tau_{1>2}  \, {\cal L}(\tau_1)  {\cal L}(\tau_2)  + \cdots
\]}
\beq
\label{WL2}
W[\G] = \frac{1}{2N}\, \Tr \left[ \,\textrm{P}\exp{ \left( -i \int_\G d\tau~ {\cal L}(\tau)\right) } \right]
\eeq 
evaluated along the contour $\Gamma$ formed by two rays lying in the $(1,2)$ plane and intersecting at the origin, as illustrated in figure \ref{cusp1}.
The angle between the rays is $\pi-\varphi$, such that for $\varphi= 0$ they form a continuous straight line.
Explicitly, the contour drawn in figure \ref{cusp1} is parametrized as
\be
\label{cusp}
 x^{0}=0 \ \ \ \ \ x^{1}=s\cos\frac{\varphi}{2} \ \ \ \ \  x^{2}=|s|\sin\frac{\varphi}{2}\ \ \ \ \  -\infty\le s\le \infty
\ee
In \eqref{WL2}   $\mathcal{L}$ is a superconnection belonging to the Lie-superalgebra $U(N|N)$ and given by 
\begin{align}
\label{supermatrix}
\mathcal{L} = \begin{pmatrix}
\mathcal{A}
&i \sqrt{\frac{2\pi}{k}}  |\dot x | \eta_{I}^\a\bar\psi^{I}_\a \\
-i \sqrt{\frac{2\pi}{k}}   |\dot x | \psi_{I}^\a \bar{\eta}^{I}_\a &
\hat{\mathcal{A}}
\end{pmatrix} \ \  \  \mathrm{with}\ \ \  \left\{\begin{matrix} \mathcal{A}\equiv A_{\mu} \dot x^{\mu}-\frac{2 \pi i}{k} |\dot x| {\cal M}_{J}^{\ \ I} C_{I}\bar C^{J}\\
 \\
\hat{\mathcal{A}}\equiv\hat  A_{\mu} \dot x^{\mu}-\frac{2 \pi i}{k} |\dot x| {\cal M}_{J}^{\ \ I} \bar C^{J} C_{I}
\end{matrix}\ \right.
\end{align}
Here $\mathrm{Tr}$ denotes the standard matrix trace (and not the super--trace, see \cite{Griguolo:2012iq} for a detailed discussion).  The fermionic couplings on each straight--line  possess the  factorized structure \cite{Drukker:2009hy,Griguolo:2012iq}
\be
\label{cuspcoupling}
\eta_{i M}^{\alpha}=n_{i M} \eta_i^{\alpha}\ \ \ \ \ \mathrm{and}\ \ \ \ \  \bar\eta^{M}_{i\alpha}=\bar n_{i}^{M} \bar\eta_{i\alpha}
\ee
The lowercase index $i$ distinguishes the two different rays forming the cusp.
On the first edge, the spinor and R--symmetry factors  are respectively given  by 
\be
n_{1M}=\mbox{\small$ \left(\cos\frac{\theta}{4}\ \ \sin\frac{\theta}{4}\ \ 0\ \ 0\right)$}\ \ \ \ \
\eta_{1}^{\alpha}= ( e^{-i\frac{\varphi}{4}}\ \ \  e^{i\frac{\varphi}{4}})\ \ \ \ \ 
\bar n_{1}^{M}=\mbox{\footnotesize $
\begin{pmatrix}\cos\frac{\theta}{4}\\ \sin\frac{\theta}{4}\\ 0\\ 0\end{pmatrix}$} \ \ \ \ \
\bar\eta_{1\alpha} =i\begin{pmatrix} e^{i\frac{\varphi}{4}}\\ e^{-i\frac{\varphi}{4}}\end{pmatrix}
\ee
while on the second one we have
\be  
n_{2M}=\mbox{\small$ \left(\cos\frac{\theta}{4}\ \ -\sin\frac{\theta}{4}\ \ 0\ \ 0\right)$}\ \ \ \ 
\eta_{2}^{\alpha}= ( e^{i\frac{\varphi}{4}}\ \ \  e^{-i\frac{\varphi}{4}})\ \ \ \ 
\bar n_{2}^{M}=\mbox{\footnotesize $
\begin{pmatrix}\cos\frac{\theta}{4}\\ -\sin\frac{\theta}{4}\\ 0\\ 0\end{pmatrix}$}\ \ \ \  
\bar\eta_{2\alpha} =i \begin{pmatrix} e^{-i\frac{\varphi}{4}}\\ e^{i\frac{\varphi}{4}}\end{pmatrix}
\ee
The $\theta$ angle is the counterpart of $\varphi$ in the  R--symmetry space. It denotes  the angular separation of 
the two edges in the internal space ({\it i.e.}~$CP^3$).
The two matrices which couple the scalars are determined in terms of the vectors $n_{iM}$ and $\bar n^N_j$  through the relations $M_{iJ}^{\ \ I}=\delta^I_J-2 (n^I_i \bar n_{iJ})$. We have respectively
\be
M_{1J}^{\ \ I}=
\widehat M_{1J}^{\ \ I}=\mbox{\small $\left(
\begin{array}{cccc}
 -\cos \frac{\theta }{2}& -\sin \frac{\theta }{2} & 0 & 0 \\
 -\sin \frac{\theta }{2}& \cos\frac{\theta }{2} & 0 & 0 \\
 0 & 0 & 1 & 0 \\
 0 & 0 & 0 & 1
\end{array}
\right)$}\ \ \ \ \mathrm{and}\ \ \ \  M_{2J}^{\ \ I}=
\widehat M_{2J}^{\ \ I}=\mbox{\small $\left(
\begin{array}{cccc}
 -\cos \frac{\theta }{2} & \sin \frac{\theta }{2} & 0 & 0 \\
 \sin \frac{\theta }{2} & \cos\frac{\theta }{2} & 0 & 0 \\
 0 & 0 & 1 & 0 \\
 0 & 0 & 0 & 1
\end{array}
\right)$}
\ee
As discussed in details in \cite{Griguolo:2012iq}  the presence of two angles allows us to find BPS configurations. 
Explicitly, for $\theta^2=\varphi^2$  the two straight lines forming the cusp share two Poincar\`e supercharges and two conformal supercharges. In other words we deal with a globally $1/6-$BPS WL.
As a consequence, the divergent contributions arising from the cusp geometry are expected to vanish if the BPS condition is satisfied, a fact which was explicitly checked up to second order in perturbation theory \cite{Griguolo:2012iq}.

\subsection{The exact 1/2--BPS Bremsstrahlung function}

Away from the BPS point $\theta^2=\varphi^2$, the cusped WL defined above suffers in general from UV divergences.  
At generic angles it has to be renormalized and we expect that the operator possesses an anomalous dimension $\Gamma_{cusp}$, according to the universal behaviour
\begin{equation}
\label{espo}
\langle W_{cusp} \rangle = e^{-\Gamma_{cusp}(k,N,\varphi,\theta)\, \log \frac{\Lambda}{\mu}} + finite
\end{equation}
where $\Lambda$ is an IR cutoff and $\mu$ stems for the renormalization scale.
For the familiar bosonic WL,  the  divergences associated to a cusp singularity along the contour  can be cast   in the exponential form \eqref{espo} \cite{Korchemsky:1987wg}. This  property follows from the presence of general exponentiation theorems for  these operators.  More precisely, their expectation value  can be written as the  exponential of the  sum of all two--particle irreducible diagrams \cite{Dotsenko:1979wb,Gatheral:1983cz}.  Extending these results to our case is not straightforward. In fact, the additional couplings between the contour and the fermions appearing in the superconnection may affect the  standard analysis presented there.  However, despite of these potential technical issues,  we expect the trace of the Wilson line in ABJM to respect the usual exponentiation process, so that we can define an anomalous dimension for the cusp  according to the  standard text--book procedure. Our three--loop result will explicitly confirm the consistency of this picture (see discussion in section \ref{result}).

In the limit where the cusp angle is small, the cusp anomalous dimension has a Taylor expansion in even powers of $\varphi$, whose first coefficient is defined as the Bremsstrahlung function \cite{Correa:2012at}.
The 1/2--BPS cusp $\Gamma_{1/2}$, associated to the 1/2--BPS Wilson lines \eqref{WL2} of the ABJM theory, vanishes for $\varphi=\pm \theta$ and consequently the small angle expansion reads
\begin{equation}\label{eq:Brems}
\Gamma_{1/2}(k,N,\varphi,\theta) = -B_{1/2}(k,N)\, \left( \varphi^2-\theta^2 \right) + {\cal O}\left((\varphi^2-\theta^2)^2\right)
\end{equation}
where $B_{1/2}$ is a function of the coupling and the number of colors only.

In \cite{Bianchi:2014laa} a conjecture was put forward for the exact expression of $B_{1/2}$. 
The argument was articulated in a few steps, which we briefly review here.
\begin{itemize}
\item
First, the Bremsstrahlung function was related to the expectation value of a supersymmetric fermionic circular WL $W_F^{\circ}$ evaluated on latitude contours in the $S^2$ sphere, that is displaced by an angle $\th_0$ from the maximal circle, and potentially by an additional internal angle $\a$ in the R--symmetry space. Remarkably, the expectation value of such WL seems to depend only on a particular combination of the angles, $\nu= \sin{2\a} \cos{\th_0}$ \cite{Bianchi:2014laa}. The undeformed contour yields the 1/2--BPS circular WL of \cite{Drukker:2009hy} and corresponds to $\nu=1$.

Paralleling an analogous derivation for ${\cal N}=4$ SYM \cite{Correa:2012at}, the conjecture of \cite{Bianchi:2014laa} states that the Bremsstrahlung function is retrieved by taking derivatives of the WL expectation value with respect to the characteristic parameter $\nu$ measuring the deformation 
\beq
\label{conjecture}
B_{1/2} (k,N) = \frac{1}{4\pi^2}\, \partial_\nu\,\log\, \langle W_F^{\circ}(\nu,k,N)\rangle_0 ~ \Big|_{\nu=1}
\eeq
Unfortunately, in contrast with the ${\cal N}=4$ SYM case, the latitude WLs have not been yet evaluated via localization, so more work is necessary to derive an exact formula for the Bremsstrahlung function.
\item As a second step, the expectation value of the latitude WL $W_{F}^{\circ}$ was argued to be expressible, via a cohomological equivalence, in terms of bosonic WL operators $W_{B}^{\circ}$ and $\hat W_{B}^{\circ}$ preserving lower supersymmetry. Using this equivalence, eq. (\ref{conjecture}) can then be rewritten as
\bea
\label{Bfin}
B_{1/2}(k,N)
&=& \frac{1}{4\pi^2}\left[ \partial_\nu\,\log\left(\langle W_B^{\circ}(\nu)\rangle + \langle \hat W_B^{\circ}(\nu)\rangle \right) +\frac{\pi}{2} \, {\rm tg}\Phi_B   \right]\Big|_{\nu=1}
\eea
where $\Phi_B$ stands for the phase of the WL $W_B^{\circ}$.
Again, in the $\nu=1$ limit, such WL operators land on the 1/6--BPS circular WL of \cite{Chen:2008bp,Rey:2008bh,Drukker:2008zx} and the cohomological equivalence parallels that for 1/2--BPS operators, described in \cite{Drukker:2009hy}.

\item The expectation value of $W_B^{\circ}$ operators is not known exactly, either.
Nevertheless, it was conjectured that at least the relevant derivative in \eqref{Bfin} could be traded with the derivative of the expectation value of a multiply wound 1/6--BPS circular WL $W_n^{1/6}$ with respect to the winding number $n$, upon proper identification of the latitude and winding parameters 
\beq\label{crazy}
\left. \partial_\nu\, \log\left(\langle W_B^{\circ}(\nu)\rangle_\nu+\langle \hat W_B^{\circ}(\nu)\rangle_\nu\right)\right|_{\nu=1}=\left. \partial_n\, \log\left(\langle W_n^{1/6}\rangle + \langle \hat W_n^{1/6}\rangle\right) \, \frac{\partial n(\nu)}{\partial \nu}\right|_{\nu=1}
\eeq
Such a trick was also proposed in \cite{Lewkowycz:2013laa} in the case of the Bremsstrahlung function associated to a cusp constructed with two locally 1/6--BPS rays in ABJM theory. Multiply wound supersymmetric WLs  were explored in details in \cite{Klemm:2012ii}, where their expectation value was given using a matrix model average description. 

\item The expansion of the results for $\langle W_n^{1/6}\rangle $ and $\langle \hat W_n^{1/6}\rangle$ provides evidence that \eqref{crazy} vanishes at all orders, as advocated in \cite{Lewkowycz:2013laa,Bianchi:2014laa}, at least in the planar limit. 
Under these circumstances, we then arrive at the following simple proposal for the 1/2--BPS Bremsstrahlung function at all orders \footnote{The subscript 1 indicates that the computation is performed at framing 1.}
\begin{equation}\label{eq:conjecture}
B_{1/2} = -\frac{i}{8\pi}\, \frac{\langle W_{1/6}\rangle_1-\langle \hat W_{1/6}\rangle_1}{\langle W_{1/6}\rangle_1+\langle \hat W_{1/6}\rangle_1}
\end{equation}
The Bremsstrahlung function associated to a 1/2--BPS cusped WL is given in terms of the expectation value of the 1/6--BPS circular WL, well--known from localization  \cite{Kapustin:2009kz,Drukker:2010nc}. 
In \cite{Griguolo:2012iq} it has been checked that at two loops (\ref{crazy}) vanishes also for finite $N$. Therefore, up to three loops expression (\ref{eq:prediction})  provides the expected result for $B_{1/2}$, without requiring the planar limit. This is the prediction that we are going to check with an explicit three--loop calculation of $\G_{1/2}$.  
\end{itemize}
Somewhat unusually in ABJM theory, such an object seems to possess an odd dependence on the coupling $k$.
Nevertheless, the proposal is in agreement with previously derived results for the cusp anomalous dimension at weak coupling \cite{Griguolo:2012iq}, that is up to two--loop order and up to subleading order at strong coupling \cite{Forini:2012bb,Aguilera-Damia:2014bqa,Correa:2014aga}. 
In this paper we put the conjecture on much firmer grounds, by providing an explicit check of it up to three loop order in perturbation theory, including the sub-leading corrections in $N$, which first appear at this order.

\section{Overview of the computation}\label{sec:computation}

\begin{figure}[ht]
\begin{center}
\includegraphics[scale=0.6]{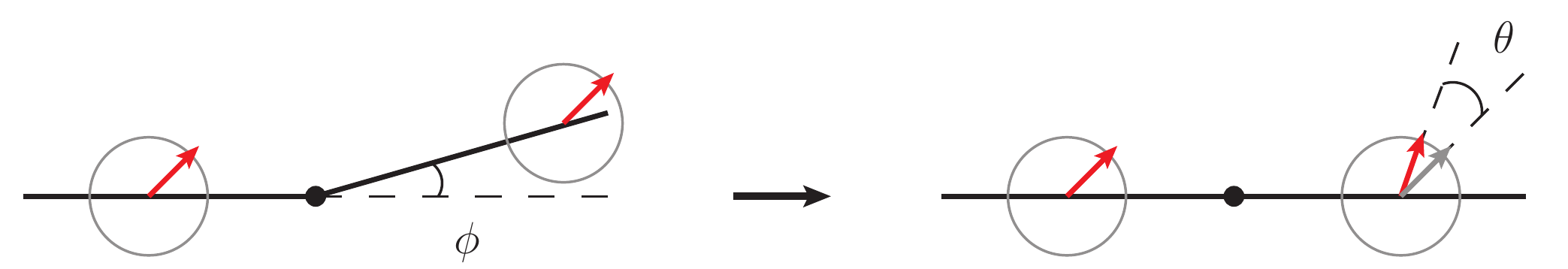}
\caption{For the 1/2--BPS cusp the small $\varphi$ limit at null internal angle (the red arrow in the cartoon stems for a direction in the internal space) is equivalent to the small $\theta$ limit at zero $\varphi$.}
\label{fig:cusplimit}
\end{center}
\end{figure}

 Referring to eq. (\ref{eq:Brems}),  the Bremsstrahlung function can be read equivalently from the $\theta$ or the $\varphi$ expansions, setting the other angle to 0
\begin{equation}\label{prescription2}
B_{1/2} = \frac12\, \frac{\partial^2}{\partial \theta^2}\, \Gamma_{1/2} \Big|_{\varphi=\theta=0} = -\frac12\, \frac{\partial^2}{\partial \varphi^2}\, \Gamma_{1/2} \Big|_{\varphi=\theta=0}
\end{equation}
We take advantage of this fact for simplifying the evaluation of the Bremsstrahlung function, circumventing the computation of the complete cusp anomalous dimension. In particular, it proves convenient to set $\varphi=0$ and work with the internal angle $\theta$  only, as shown in the cartoon of figure \ref{fig:cusplimit}. 
This strategy brings dramatic simplifications at several stages of the computation, both in the number of diagrams to be taken into account and in their evaluation.

\subsection{The diagrams}\label{sec:ovdiag}

We now discuss how to suitably choose the computational setting in order to pin down the calculation to a relatively small number of Feynman diagrams to be considered at three loops. 

First of all, we recall that in Chern--Simons--matter theories remarkable simplifications arise when the WL contour lies in a two--dimensional plane. In fact, this choice triggers the vanishing of several tensor contractions, by virtue of the antisymmetric nature of Chern--Simons gauge propagators  \eqref{treevector} and cubic vertex (\ref{gaugecubic}) in their index structure.
In particular, whenever an odd number of antisymmetric tensors $\varepsilon_{\mu \nu \rho}$ arises in the algebra of a diagram, their product eventually vanishes. In fact, one can always reduce them to a single Levi--Civita tensor whose indices have to be contracted with three external vectors. The latter all come from the WL contour and hence, lying on a plane, are not linearly independent and always give vanishing expressions when contracted with the $\varepsilon$ tensor.

This feature of perturbation theory becomes particularly powerful at odd loop orders. Namely, from ABJM Feynman rules (see appendix \ref{app:FeynmanRules}), one can infer that purely bosonic diagrams with gauge interactions vanish identically, by virtue of the aforementioned mechanism. For supersymmetric WLs  this singles out contributions containing matter exchanges only.  Nonetheless, there are still a number of simplifications. 

For example, bosonic 1/6--BPS WL diagrams with scalar exchanges also evaluate to zero, thanks to the vanishing of the trace of their coupling matrices $M$.
Therefore, in ABJM theory bosonic WLs with planar contours, when computed at framing zero, automatically have vanishing expectation values at odd loops \cite{Rey:2008bh} \footnote{ A three--loop calculation at non--zero framing leads instead to a non--vanishing result \cite{Bianchi:2016yzj} that matches the matrix model prediction.}. 

As a consequence, only fermionic WLs  of the Drukker--Trancanelli type \cite{Drukker:2009hy} allow for nontrivial contributions at odd orders,  as shown in \cite{Bianchi:2016vvm}. This is precisely the class of WLs  we are computing in this paper.

The arguments provided above imply that in our computation we have to consider only graphs containing at least one matter propagator. This condition substantially narrows down the number of diagrams to be considered. 

\begin{figure}[ht]
\begin{center}
\includegraphics[scale=0.5]{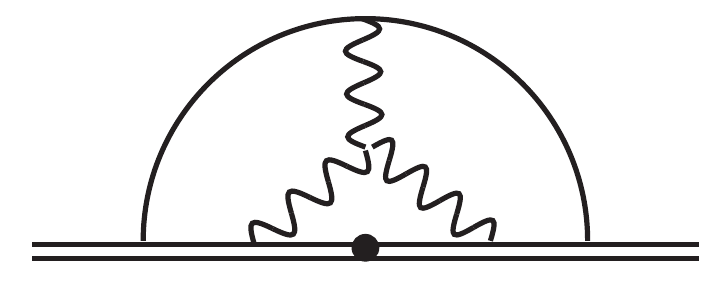}
\caption{An example of a diagram identically vanishing in the $\varphi=0$ limit.}
\label{fig:vanishing}
\end{center}
\end{figure}

A further simplification arises from performing the calculation at $\varphi=0$, as previously discussed. In fact, taking the contour to lie on a straight line triggers the additional vanishing of diagrams such as the one in figure \ref{fig:vanishing}, again by antisymmetry in the contraction of $\varepsilon$ tensors with vectors on the line.

Finally, aimed at computing the Bremsstrahlung function, one could in principle restrict to diagrams producing factors of $\cos \theta/2$ only (see eq. (\ref{prescription2})). According to the rules in appendix \ref{App:cusp}, these occur only if a fermion line or a scalar bubble stretch between opposite sides of the cusp.
Nevertheless, it turns out that we can easily compute the whole cusp at $\varphi=0$ including also $\theta$ independent terms.

Technically, evaluating the various diagrams entails a few steps. After using the Feynman rules reviewed in appendix \ref{app:FeynmanRules}, the integral associated to a generic diagram with fermion exchanges  has the following structure 
\bea
\label{genint}
&& \int d\t_{1>2> \cdots} \; \underbrace{\dot{x}_1^{\mu_1} \dot{x}_2^{\mu_2}\cdots }_\text{velocities}\; \underbrace{\int d^n y \; d^n w \cdots}_\text{internal integrations} \; \underbrace{(\eta_{p_1} \g^{\nu_1} \cdots  \bar{\eta}_{p_2}) (\eta_{p_3} \g^{\rho_1} \cdots \bar{\eta}_{p_4}) \cdots}_\text{spinor structure}
\non \\
&& \underbrace{\pa_{\s_1} \cdots \pa_{\s_q}}_\text{derivatives acting on propagators} \underbrace{\frac{1}{(\t_{i_1 i_2})^{2p} (x_{i_3 w})^{2p} (x_{i_4 y})^{2p} (x_{yw})^{2p} \cdots}}_\text{propagators} 
\eea
where $\dot{x}_i \equiv \dot{x}(\tau_i)$ and we have used the shortening notation $\t_{ij} \equiv \t_i - \t_j, \, x_{jw}^2 \equiv (x(\t_j) - w)^2, \, x_{yw}^2 \equiv (y-w)^2 $.  The number of internal and contour integrations, spinor structures, derivatives and propagators depend on the particular diagram. Lorentz indices carried by velocities, gamma matrices and derivatives are contracted among themselves or with external $\varepsilon$ tensors. If couplings with scalars are also present, ${\cal M}$--matrix structures will appear.

We use identities in appendix \ref{App:cusp} to reduce the $\eta$ bilinears and the ${\cal M}$--matrix structures. We perform the relevant tensor algebra strictly in three dimensions and in an automated manner with a computer program. In this process we drop all the terms containing an odd number of Levi--Civita tensors and reduce all the products of an even number of them to products of metric tensors. We finally arrive at expressions containing scalar products of velocities and derivatives only, where the ultimate step consists in integrating over internal vertices and WL parameters.

\subsection{The integrals and the HQET formalism}\label{sec:HQET}

The most striking simplification of working at $\varphi=0$ is that the integrals arising from Feynman graphs reduce to propagator--type contributions which evaluate to numbers rather than functions of the angle, considerably reducing the effort needed for their determination.  

The integrals could in principle be evaluated directly by solving internal integrations in configuration space and then integrating over the Wilson line parameters. In this case one could make use of the Gegenbauer polynomial $x$--space technique GPXT \cite{Chetyrkin:1980pr} to solve the three--dimensional internal integrals  along the lines of \cite{Mauri:2013vd}. However this approach becomes cumbersome and impractical for contributions with more than one internal integration, such as those appearing at three loops.

A more powerful strategy consists in Fourier transforming the integrals to momentum space and perform the contour integrations first.  
This step casts the various contributions into the form of the non--relativistic Feynman integrals appearing in the effective theory of heavy quarks (HQET) (see \cite{Grozin:2014hna,Grozin:2015kna}, for a recent application in four dimensions), or rather an Euclidean version thereof, as we perform the whole computation with $(+,+,+)$ signature.
\begin{figure}[ht]
\begin{center}
\hspace{0.4cm}\includegraphics[width=0.99\textwidth]{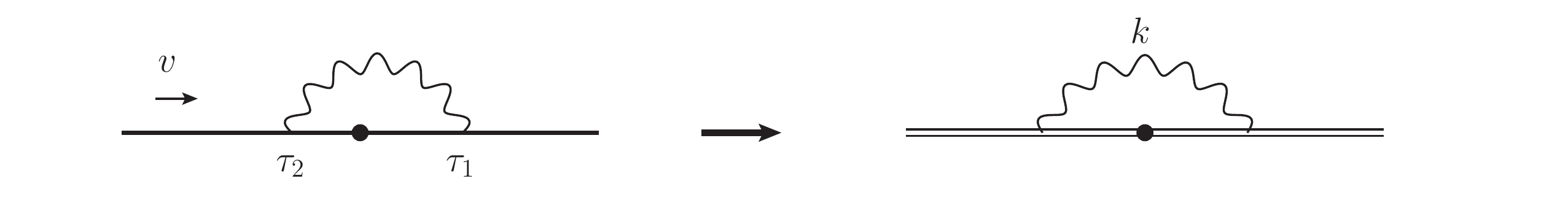}
\begin{equation}
\hspace{-1.5cm} \left[\frac{\G(\frac{1}{2}-\e)}{4 \pi^{3/2-\e}} \right]^2 \!\! \int_{0}^{+\infty}\!\!\!\! d\tau_1\, \int_{-\infty}^{0}\!\! d\tau_2\, \frac{1}{\left[(x_1-x_2)^2\right]^{1/2-\epsilon}} 
\quad \longrightarrow \quad 
\int\, \frac{d^{3-2\epsilon}\,k}{(2\pi)^{3-2\epsilon}}\, \frac{1}{k^2\, (-i\, k\cdot v)^2} \nonumber
\end{equation}
\caption{A cartoon of the WL Fourier transform to a HQET integral.}
\label{fig:HQET}
\end{center}
\end{figure}
This procedure, applied to a single propagator, is sketched pictorially in figure \ref{fig:HQET}. A more detailed application is described in sections \ref{se:onetwoloop} and \ref{sec:3loop}, when dealing with the example of the one--loop and a particular graph of the three--loop computation. The velocity of the heavy quark is the vector tangent to the WL contour, $v=\dot x(\tau)$.

In general, the resulting integrals suffer from both IR and UV divergences. 
UV divergences are treated within the framework of dimensional regularization, shifting spacetime integrations to $d=3-2\epsilon$ dimensions. As a consequence, the integrals evaluate to Laurent series in the regularization parameter $\epsilon$. In order to preserve supersymmetry we work in the dimensional reduction scheme (DRED) \cite{Siegel:1979wq}. In practice this translates in performing the tensor algebra in the numerators strictly in three dimensions. This prescription can introduce evanescent terms if some tensor integration is attained (in $d=3-2\epsilon$) and then contractions with three dimensional objects are carried out. Such evanescent terms can be taken into account by suitable regulator dependent factors (see e.g. 
\cite{Bianchi:2013rma,Griguolo:2013sma,Bianchi:2013pva} for a discussion in the context of WLs in three dimensions).
However, in the present case this subtlety is already built--in in our method and does not cause any practical worry. In fact, the index algebra is always handled (in integer dimensions) before any integration, as recalled in the previous section. After this reduction, the only potentially dangerous integrals occur whenever a tensor numerator contains contractions with an $\varepsilon$ tensor, but such contributions were argued to vanish and promptly discarded.
 
In order to tame IR divergences at long distances along the WL contour, following \cite{Grozin:2015kna} we introduce an exponential damping factor $e^{\delta\, \tau}$ ($Re(\delta)<0$) for each contour integral, enforcing the finiteness of the integrals at large radius. This introduces a residual energy in the heavy quark, shifting the HQET propagators by
\begin{equation}\label{eq:IRreg}
\frac{1}{-i\, k\cdot v} \longrightarrow \frac{1}{-i\, k\cdot v - \delta}
\end{equation}
The parameter $\delta$ is customarily set to $\delta = -1/2$, a choice which simplifies the result of the relevant integrals.
The final result for the cusp anomalous dimension is scheme independent and is not affected by the particular value of $\delta$, albeit all intermediate steps might be.
 
Up to three loops, HQET propagator integrals were studied in detail in \cite{Grozin:2000jv,Chetyrkin:2003vi} for QCD in four dimensions. The analysis performed there reveals that all topologies can be reduced to a set of 1, 2 and 8 master integrals (seven planar and one non--planar) at one, two and three loop order, respectively, by means of integration by parts identities \cite{Tkachov:1981wb,Chetyrkin:1981qh}. Such a statement is true, independently of the space--time dimensions and applies to our case as well.
This reduction step can be made automatic thanks to the Laporta algorithm \cite{Laporta:1996mq,Laporta:2001dd} and carried out with one of its available implementations \cite{Smirnov:2008iw,Smirnov:2013dia,Smirnov:2014hma,Lee:2012cn,Lee:2013mka}.
In practice we have used FIRE5 \cite{Smirnov:2014hma}, whose C++ version is able to perform the reduction of all the required integrals in a few minutes on a PC.

The relevant master integrals have to be evaluated in $d=3-2\epsilon$ dimensions. This task can be carried out straightforwardly relying on previously derived results for such topologies with general integration dimension $d$ and powers of the propagators \cite{Grozin:2000jv} (some of them are non--integer because arising from the integration of bubble sub--topologies).
The final evaluation of the master integrals and expansion in series of $\epsilon$ up to the desired order for the cusp anomalous dimension is spelled out in appendix \ref{app:masters}.

\vskip 10pt
The procedure described above that reduces the cusp loop integrals to propagator--type integrals for non--relativistic heavy particles, has a very suggestive physical interpretation. 
The anomalous dimension of the cusped WL captures the renormalization of the current associated to a massive quark, undergoing a transition from a velocity $v_1$ to a velocity $v_2$ at an angle $\varphi$.
In fact Wilson lines have been thoroughly employed as a convenient tool for describing the dynamics of such objects.
According to this physical picture, the 1/2--BPS Wilson line in ABJM theory is associated to heavy quarks, which are interpreted as ${\cal N}=3$ supermultiplet $W$ bosons, transforming in the fundamental representations of the two gauge groups.
These were shown to arise when moving away from the origin of the moduli space \cite{Lee:2010hk,Lietti:2017gtc}. 

The Bremsstrahlung function controls the energy loss by radiation of these massive particles undergoing a deviation by an infinitesimal angle.
In particular, the Higgsing procedure mentioned above reveals that this superparticle is able to radiate off pairs of bifundamental fields, both scalars and fermions, quite exotically.
The coupling to such matter fields enters the superconnection \eqref{supermatrix} of the 1/2--BPS Wilson line via a parameter $\theta$, which can be interpreted as an angle in the internal $SU(4)_R$ R--symmetry space.
According to the discussion above, we determine the Bremsstrahlung function by considering the equivalent (but computationally simpler) picture of a heavy quark subject to a small kick in R--symmetry space at fixed vanishing geometrical angle.

In particular, thanks to the $\varphi=0$ condition, the integrals involved in the computation of the Bremsstrahlung function are precisely those contributing to the self--energy corrections of a heavy quark, i.e. propagator--type. The presence of the cusp point on the line has the simple effect of increasing the power of a HQET propagator in the diagram. Integration by parts identities can  then be employed to reduce the power of the doubled propagator to unity.

\subsection{The cusp anomalous dimension at zero angle}\label{sec:detailscusp}

In the previous subsection we explained how the computation of the expectation value of the cusped WL is performed up to three loops, in the flat cusp limit. 
In this section we provide the details on how the cusp anomalous dimension and the Bremsstrahlung function can be practically extracted from this result.

As recalled above, the expectation value of the cusped WL suffers from UV and IR divergences. The former determine the renormalization properties of the WL and are the crucial object of our investigation. The latter can be regulated in different ways, either by assigning the heavy quarks a residual energy offsetting them from the mass shell as described above or by computing the loop on a finite interval $(-L,L)$ as done in \cite{Griguolo:2012iq}. Although these divergences are uninteresting for our purposes, different kinds of regularizations can be more or less convenient according to the approach we use to compute the integrals. Therefore, while discussing how to extract the UV divergent behavior of the cusped WL, we will also compare the effectiveness of different IR regularization strategies.  
 
For a cusped WL in a gauge theory UV divergences may have different origins.  
The first source consists in the divergences of the Lagrangian. In particular, this can cause a running coupling and the presence of a nontrivial $\beta$ function. The ABJM model is superconformal and such a phenomenon does not occur. 
Then, divergences associated to the short distance behavior of the Wilson line can arise. In the language of HQET these are the (potentially divergent) radiative corrections to the self--energy of the heavy quarks. Finally, the cusp geometry itself introduces further divergent contributions, which are the ones we are interested in.

\begin{figure}[ht]
\begin{center}
\includegraphics[width=0.75\textwidth]{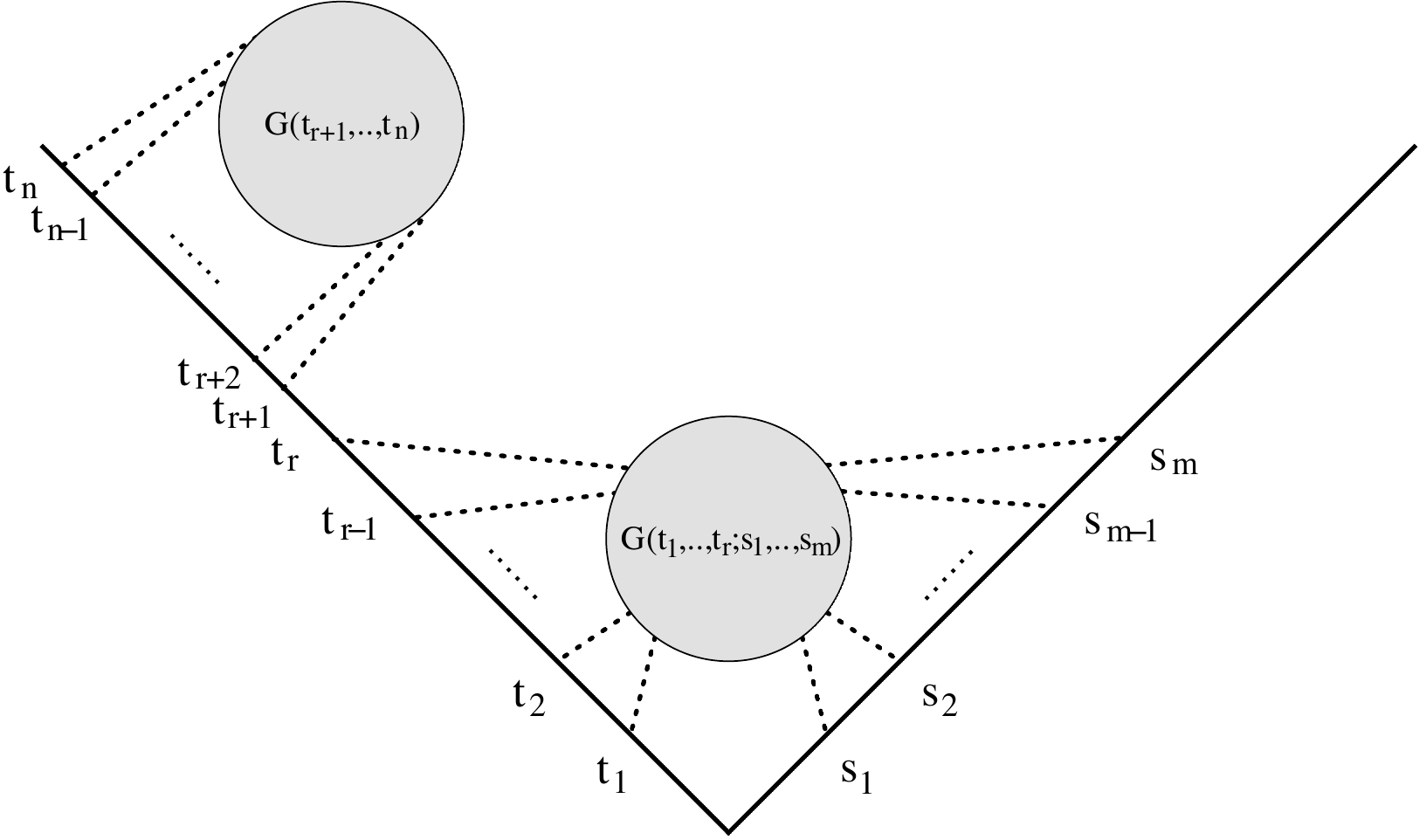}
\caption{Factorization of quantum corrections to the cusped WL.}
\label{fig:factor}
\end{center}
\end{figure}
The systematics of the renormalization properties of WLs with cusps were discussed in \cite{Dorn:1986dt,Korchemsky:1987wg}. Here  we recall some basic and general features.   

First of all we can focus our attention only on contributions which are 1PI vertex diagrams in  HQET language. For instance  if we deal with a configuration of the type   represented in figure \ref{fig:factor} the subsector governed by the Green function $G(t_{r+1}, \cdot\cdot, t_{n})$ completely  decouples from the rest. In fact the corresponding contribution is factorized as
\be
\mbox{\small $\displaystyle
\int_{-\infty}^0\!\!\!\!\!\! dt_1\cdot\cdot\! \!\int_{-\infty}^{t_{r-1}}\!\!\!\!\!\!\!\!d t_{r}\! \int^{t_{r}}_{-\infty}\!\!\!\!\!dt_{r+1}\cdot\cdot\!\! \int_{-\infty}^{t_{n-1}}\!\!\!\!\!\!\!\!dt_{n}\int_0^\infty\!\!\!\!\! ds_m\!\!\cdots\!\!\int_0^{s_2}\!\!\!\!\! ds_1 e^{\delta (t_n-s_m)}
G(t_{r+1}, \cdot\cdot, t_{n}) G(t_1,\cdot\cdot, t_r; s_1,\cdot\cdot, s_m)$}
\ee
If we perform the change of variables $t_i\mapsto t_i+t_r~  (\mbox{with}~i=r+1,\cdots,n)$  and we use that $G(t_{r+1}, \cdot\cdot, t_{n})$
is a translationally invariant function we find 
\be
\label{factorization}
\mbox{\small $\displaystyle
 \underset{(A)}{\int^{0}_{-\infty}\!\!\!\!\!dt_{r+1}\cdot\cdot\!\! \int_{-\infty}^{t_{n-1}}\!\!\!\!\!\!\!\!dt_{n}e^{\delta t_n} G(t_{r+1}, \cdot\cdot, t_{n})}\underset{(B)}{
\int_{-\infty}^0\!\!\!\!\!\! dt_1\cdot\cdot\! \!\int_{-\infty}^{t_{r-1}}\!\!\!\!\!\!\!\!d t_{r}\!\int_0^\infty\!\!\!\!\! ds_m\!\!\cdots\!\!\int_0^{s_2}\!\!\!\!\! ds_1 e^{\delta (t_r-s_m)}
 G(t_1,\cdot\cdot, t_r; s_1,\cdot\cdot, s_m)}$}
\ee
The factor $(B)$ in \eqref{factorization} provides the value of the  diagram obtained from the one in figure \ref{fig:factor} removing the subsector  governed by $G(t_{r+1}, \cdot\cdot, t_{n})$, while the factor $(A)$  is the contribution  due to $G(t_{r+1}, \cdot\cdot, t_{n})$  to the vacuum expectation value of a  straight-line running from $-\infty$ to $0$. In HQET language this factorization is a trivial manifestation of momentum conservation. On the other hand, we have to stress that eq. \eqref{factorization}  also relies on the judicious choice of the IR regulator which breaks translation invariance along the two rays of the cusp in a controlled way. For instance, if we were to tame the IR divergences by considering a cusp with edges of finite length $L$, it is easy to see that the above argument would fail and the two contributions would remain in general intertwined.

Because of \eqref{factorization}  the cusped WL is given by the sum of all the 1PI diagrams, which we shall denote  by $V(\theta,\varphi)$, times the vacuum expectation value of the Wilson lines $[S(-\infty,0),S(0,\infty)]$, running from $-\infty$ to $0$ and   from $0$ to $\infty$
\be
\label{factorization2}
\langle W(\theta,\varphi)\rangle=S(-\infty,0)V(\theta,\varphi) S(0,\infty)
\ee
The two Wilson line factors in HQET language are identified with the two point functions of the heavy quark and eq. \eqref{factorization2} is the usual and well--known  decomposition of the vacuum expectation value in terms of its  1PI  sector and two-point functions. 

To single out the cusp anomalous dimension from $\langle W(\theta,\varphi)\rangle$, we have first to eliminate the spurious divergences  which  are due to the fact that the presence of the IR regulator $e^{\pm \delta t}$ mildly breaks gauge invariance \footnote{The integral of a total derivative along of the whole contour is no longer zero because of the presence of this factor.}. A similar phenomenon also  occurs when we  consider edges of finite length $L$ \cite{Craigie:1980qs,Aoyama:1981ev,Knauss:1984rx,Dorn:1986dt}: There the gauge invariance is lost because the loop is open.  In both cases the gauge--dependent divergent terms can be eliminated by introducing a multiplicative renormalization $Z_{\rm open}$ \cite{Craigie:1980qs,Aoyama:1981ev,Knauss:1984rx,Dorn:1986dt}, which, in practice,   is equivalent to the subtraction  of the straight line or  the cusp at $\theta=\varphi=0$
\be
\label{eq:KorchRad}
\log(\widetilde{W} (\theta,\varphi))\equiv
\log(Z^{-1}_{\rm open} W(\theta,\varphi))=\log\frac{W(\theta,\varphi)}{W(0,0)}=\log \frac{V(\theta,\varphi)}{V(0,0)}
\ee
with $V(\theta,\varphi)$ defined in (\ref{factorization2}). 
In the formalism of \cite{Gervais:1979fv} $Z_{open}$ corresponds to the wave-function renormalization of the fermions in the one-dimensional effective description of the Wilson line operator. This factor needs to be eliminated in order to single out the proper renormalization of the relevant operator describing the cusp \cite{Korchemsky:1987wg}.

Finally we can  extract $Z_{cusp}$  by defining  the renormalized WL as follows
\begin{equation}
\langle W_R(\theta,\varphi) \rangle = Z_{cusp}^{-1}\, \langle \widetilde W(\theta,\varphi) \rangle
\end{equation}
The cusp anomalous dimension is then evaluated through the  relation
\begin{equation}
\Gamma_{cusp}(k,N) = \frac{d \log Z_{cusp}}{d \log \mu}
\end{equation}
where $\mu$ stems for the renormalization scale and we have suppressed the dependence on the angles.

The perturbative computation yields $V(\theta,\varphi)$ as an expansion in $\frac{1}{k}$
\beq\label{eq:Vexpansion}
V(\theta,\varphi)=  \left(\frac{2\pi}{k}\right)V^{(1)}(\theta,\varphi)+ \left(\frac{2\pi}{k}\right)^2 V^{(2)}(\theta,\varphi)+  \left(\frac{2\pi}{k}\right)^3 V^{(3)}(\theta,\varphi) + {\cal O}\left(k^{-4}\right)
\eeq
In dimensional regularization every $V^{(i)}$ coefficient is given by a Laurent expansion in $\epsilon$.  According to the standard textbook dictionary, the cusp anomalous dimension can be read from the residues of the simple poles of $Z_{cusp}$ in the $\epsilon$ plane
\begin{equation}\label{eq:cusp}
\log Z_{cusp} = \left. \log\left(\frac{V(\theta,\varphi)}{V(0,0)}\right)\right|_{\frac{1}{\epsilon}\rm terms}\!\!\!\!= -\frac{1}{2\epsilon\, k}\, \Gamma^{(1)} - \frac{1}{4\epsilon\, k^2}\, \Gamma^{(2)} - \frac{1}{6\epsilon\, k^3}\, \Gamma^{(3)} + {\cal O}\left(k^{-4}\right)
\end{equation}

This completes the evaluation of the cusp anomalous dimension, which in our case is restricted to the $\varphi=0$ limit $\Gamma_{1/2}\Big|_{\varphi=0}$.
Finally the Bremsstrahlung function can be evaluated by taking the double derivative
\begin{equation}\label{eq:brems}
B_{1/2}(k,N) = \frac12\, \partial^2_\theta\, \Gamma_{1/2}(k,N,\varphi=0,\theta)\, \Big|_{\theta=0}
\end{equation}

\section{Review of one-- and two--loop results}\label{se:onetwoloop}

The cusp anomalous dimension is extracted from the divergent part of the logarithm of the cusped WL expectation value.
In the absence of a well-established exponentiation theorem for WLs in ABJM, we do not compute this logarithm directly, rather we consider the whole WL and take the logarithm perturbatively.
In particular at three loops, this entails subtracting contributions from the one-- and two--loop order corrections. At the order in the $\epsilon$ expansion to which we compute the three--loop expectation value and its logarithm, namely up to $1/\epsilon$ terms, we need to consider the one--loop correction up to order ${\cal O}(\epsilon)$ and the two--loop contribution up to finite terms in the regulator.
In this section we perform such a computation. 

The relevant 1PI diagrams were already determined in \cite{Griguolo:2012iq}, but here we expand them to higher orders in $\epsilon$ and treat them within the HQET formalism, for consistency with the three--loop results.
This produces a mismatch with respect to the results in \cite{Griguolo:2012iq} by scheme dependent terms. However, the leading poles in the $\epsilon$ regulator are expected to coincide, diagram by diagram. This is indeed the case and provides a handy consistency check.

\begin{figure}[ht]
\begin{center}
\includegraphics[scale=0.33]{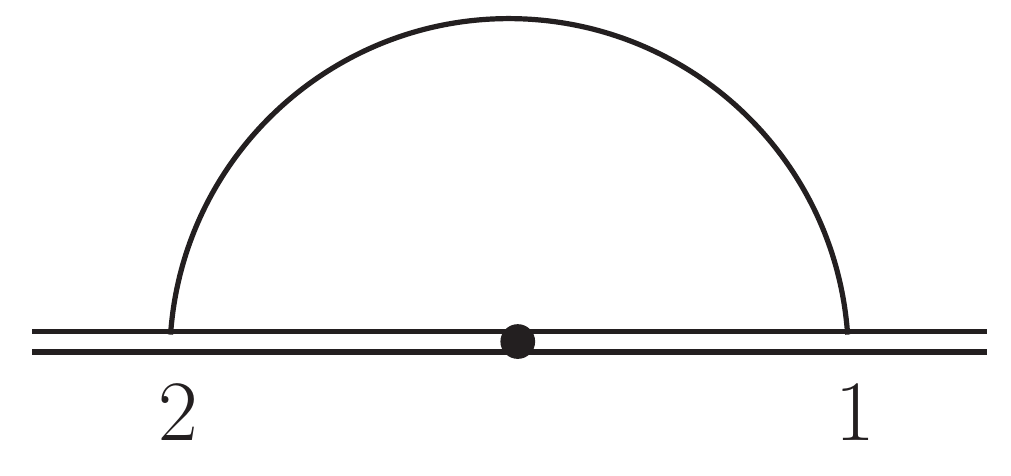}
\caption{One--loop fermion exchange diagram $( 1)_a$. The double line represents the WL contour, the single line is the fermionic propagator and the black dot is the cusp point.}
\label{fig:1loop}
\end{center}
\end{figure}

At one loop, only one non--vanishing diagram arises, which is a fermion exchange across the cusp point, as depicted in figure \ref{fig:1loop}.
Throughout this computation, the contributions from the two diagonal blocks of the supermatrix \eqref{supermatrix} are the same and their sum simply cancels the factor 2 in the overall normalization \eqref{WL2}.
Parametrizing the points on the cusp line as $x^{\mu}_i(s)= v^{\mu} \t_i$, the algebra of the diagram gives
\begin{equation}
( 1)_a = \, i\, N\,  \left( \frac{2\pi }{k}\right)  \left[\frac{\G(\frac{1}{2}-\e)}{4 \pi^{3/2-\e}} \right]   \int^{\infty}_0 d \t_1 \int_{-\infty}^0 d \t_2  \, (\eta_1 \gamma_{\mu} \bar{\eta}_2)  \, \partial^{\mu}_{1}  \frac{1}{(x_{12}^2)^{1/2-\e}} 
\end{equation}
with $x_{12}^2 = (x_1(\t_1)-x_2(\t_2))^2$.
According to the method outlined above, we Fourier transform this to momentum space. Introducing the exponential IR regulator $\delta$, we obtain
\begin{align}
( 1)_a &=  -  N\, \left( \frac{2\pi }{k}\right)  \int^{\infty}_0 d \t_1 \int_{-\infty}^0 d \t_2 \int \frac{d^{3-2\e}k}{(2\pi)^{3-2\e}}   \, e^{i (\t_1-\t_2) v \cdot k + \delta (\t_1-\t_2)}   \, (\eta_1 \gamma_{\mu} \bar{\eta}_2)  \,  \frac{k^{\mu}}{k^2} \non \\ 
&=  - N\,  \left( \frac{2\pi }{k}\right)  \int \frac{d^{3-2\e}k}{(2\pi)^{3-2\e}}     \, (\eta_1 \gamma_{\mu} \bar{\eta}_2)  \,  \frac{k^{\mu}}{k^2 (i\, k\cdot v + \delta)^2}
\end{align}
In the last line, the squared propagator arises from the two sides of the cusp which degenerate to the same HQET propagator in the $\varphi=0$ limit. 

We set $\delta=-1/2$ and absorb the imaginary unit in the HQET propagator into the velocity $v=i\, \tilde{v}$. The resulting imaginary vector with $\tilde{v}^2=-1$ is suitable for the evaluation of the master integrals in euclidean space, making them manifestly real.

Then we make use of identity \eqref{eq:etagamma} to express the $\eta$--bilinear in terms of the external velocity $v$. This gives rise to the $\theta$ dependence of the diagram, according to \eqref{eq:etagamma2}, and expresses the numerator of the integrand as $k\cdot v$.
This in turn can be rewritten as an inverse HQET propagator, so that at this stage the diagram can be expressed as a linear combination of scalar HQET one-loop integrals \eqref{eq:masterintegrals}
\begin{equation}
( 1)_a =   \left( \frac{2\pi }{k}\right) 4 N\, C_{\theta}\, \left( G_{1,1}-G_{2,1} \right)
\end{equation}
where $C_{\theta }=\cos \frac{\theta}{2}$.  
An integration--by--parts identity allows to express everything only in terms of the master integral $G_{1,1}$  
\begin{equation}
( 1)_a =   \left( \frac{2\pi }{k}\right) 4 N\, C_{\theta}\, (d-2)\, G_{1,1}
\end{equation}
Finally, using formula \eqref{1lmast} we expand this integral in $\epsilon$ up to order ${\cal O}(\epsilon)$ as required by the three--loop computation and obtain the one--loop 1PI HQET vertex correction at zero cusp angle
\begin{equation}\label{eq:V1}
V^{(1)}(\theta) = N C_{\theta } \bigg(\frac{1}{ 2  \epsilon }-1+\frac{7\pi^2}{24} \, \epsilon \bigg )+{\cal O}\left(\epsilon ^2\right)
\end{equation}
up to an overall factor $\frac{(e^{-\gamma_E}16\pi)^{\epsilon}}{k}$, omitted to keep the expression compact.\\

\begin{figure}[ht]
\begin{center}
\includegraphics[scale=0.42]{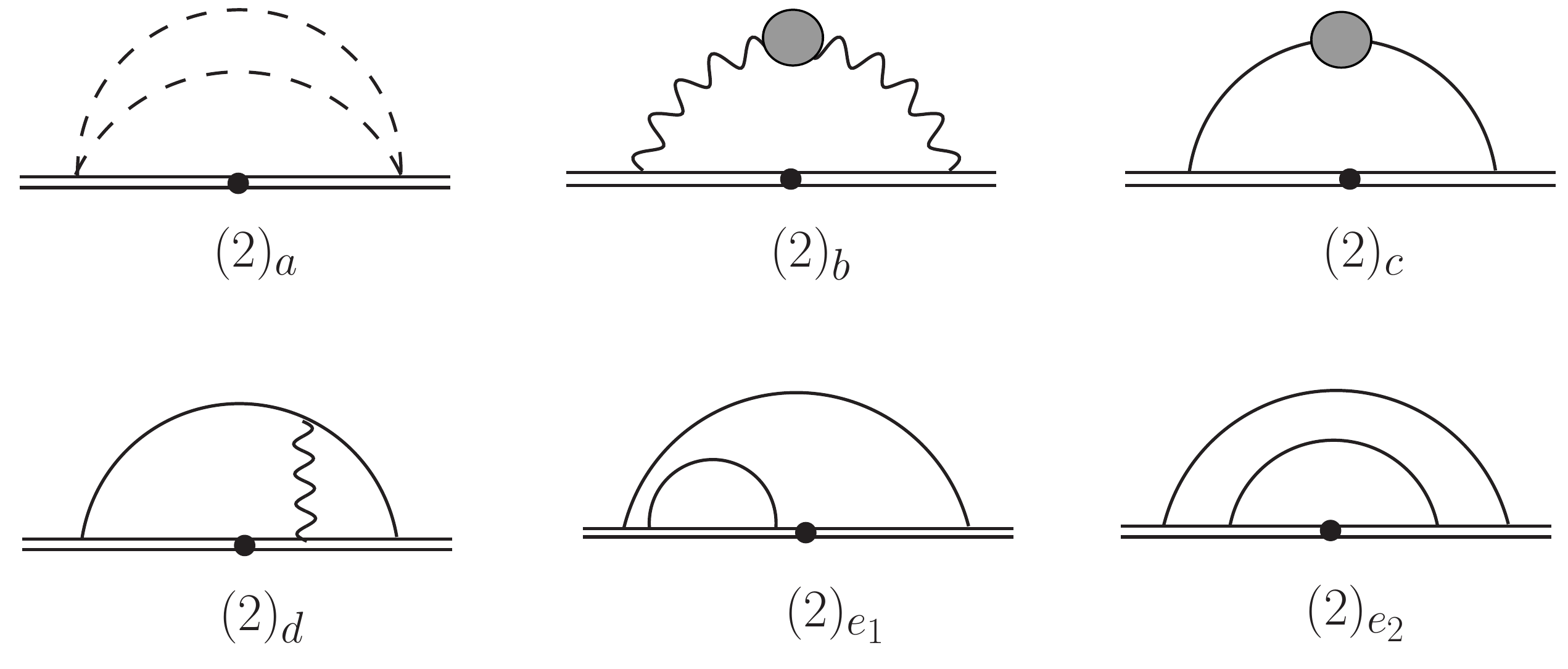}
\caption{Two-loop diagrams. Solid lines represent fermions, while dashed lines correspond to scalars. Wave lines are gauge vectors, as usual.}
\label{2ldiag}
\end{center}
\end{figure}

At two loops the relevant diagrams are shown in figure \ref{2ldiag}.
In general, each topology of diagrams is associated to a number of independent 1PI configurations, according to the position of the cusp point.  Different configurations give rise to different powers of  $\cos \theta/2$. As an example, in figure \ref{2ldiag} we depict the two possible configurations of diagrams with double fermion exchanges,  $(2)_{e_1}$ and $(2)_{e_2}$, which give contributions proportional to $\cos \theta/2$ and $\cos^2 \theta/2$, respectively. 

All the diagrams are computed in the HQET framework, following what have been described above for the one--loop case. The integration--by--parts reduction lands on to two master integrals only, which are defined and evaluated in appendix \ref{app:masters}.
The list of results for each individual diagram is given in appendix \ref{app:diagrams2loops}.
Combining these expressions and expanding them up to order ${\cal O}(\epsilon^0)$ leads to the following (unrenormalized) expectation value of the 1PI part for the cusped WL at two loops
\beq\label{eq:V2}
V^{(2)}(\theta)  = N^2  \left( \frac{ C_{\theta } ^2}{8 \epsilon^2}-\frac{ 1+2 C_{\theta } ^2 }{4 \epsilon }
 + \frac{\pi^2(-4 -4 C_{\theta }+11 C_{\theta }^2 ) + 24 \left(1-2C_{\theta }+4 C_{\theta }^2 \right)}{48 } \right)+{\cal O}\left(\epsilon \right) 
\eeq
once again, omitting a factor  $\frac{(e^{-\gamma_E}16\pi)^{2\epsilon}}{k^2}$. We stress that no sub--leading corrections in $N$ arise up to two loops.

\section{Three loop calculation}

\subsection{Diagrams}\label{sec:diag3l}

At three loops it is convenient to classify the diagrams according to the number of fermion and scalar insertions. We provide only a sketch of the possible topologies of diagrams arising in the computation, but each of them can contribute with several different relative positions of the insertion and the cusp points, all giving rise to 1PI configurations, and/or different choices of the gauge vectors, $A_\mu$ or $\hat{A}_\mu$.

The planar diagrams can be organized according to  
\begin{itemize}
\item Diagrams containing scalar insertions, figure \ref{fig:scalar}.
\begin{figure}[ht]
\begin{center}
\includegraphics[scale=0.49]{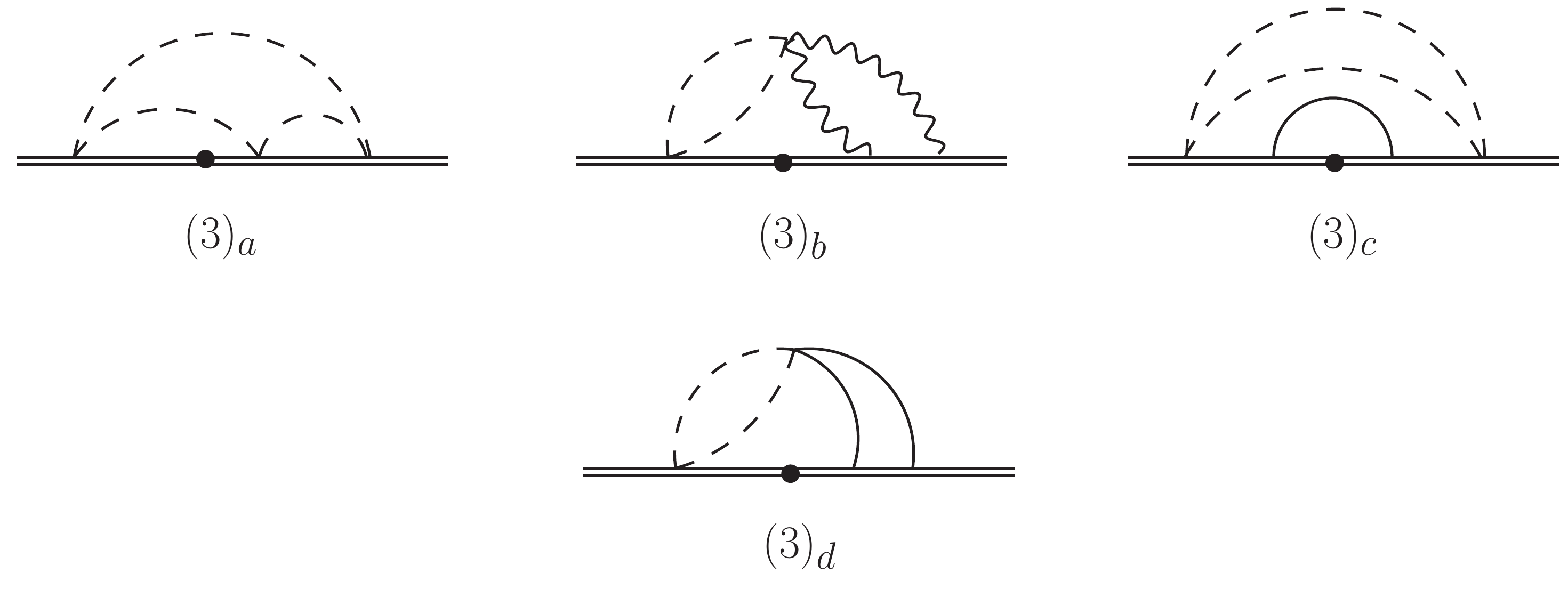}
\caption{Diagrams with scalar insertions.}
\label{fig:scalar}
\end{center}
\end{figure}

$(3)_a$ is the only diagram obtained by contracting three bi--scalar insertions, with no fermion lines.
This, along with diagram $(3)_b$, does not actually contribute to the Bremsstrahlung function. In fact they possess a trivial dependence on the internal angle $\theta$, as is immediate to infer using identities \eqref{Iden2}. Nonetheless, we take them into account in order to provide the full result for the cusp at vanishing $\varphi$ angle.

Diagram $(3)_c$ comes in different configurations, each with a specific $\theta$ dependence, according to the relative positions between the insertions and the cusp point.
 
Apart from the diagrams shown in figure \ref{fig:scalar}, other topologies with scalar insertions arise, which are not directly vanishing according to the criteria of section \ref{sec:ovdiag}. Nevertheless, these evaluate to zero after performing the algebra and integrating. Therefore we have not displayed them here.

\item Diagrams with no scalar insertions and a fermion line attached to the WL, as depicted in figure \ref{fig:2fermions}.
\begin{figure}[ht]
\begin{center}
\includegraphics[width=0.87\textwidth]{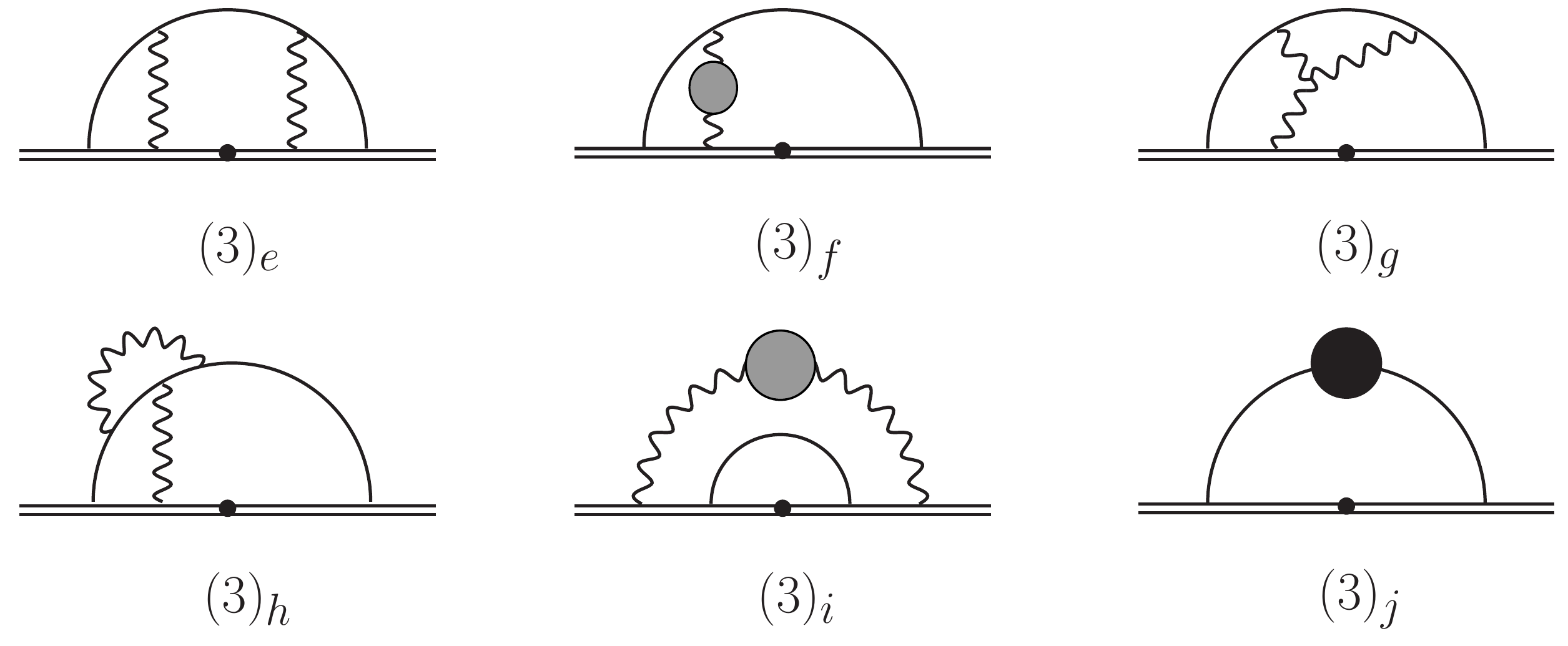}
\caption{Diagrams with two fermion insertions. Grey and black bullets represent one and two loops self--energy corrections, respectively.}
\label{fig:2fermions}
\end{center}
\end{figure}

Such diagrams arise in a variety of fashions, with up to three internal integration vertices.
Several graphs feature lower order corrections to the field propagators.
We recall that the one--loop correction to the scalar two--point function vanishes identically and the one for the fermion line \eqref{1fermion}  is proportional to the difference between the ranks of the two gauge groups, hence it vanishes in the ABJM theory.
Diagrams $(3)_f$  and $(3)_i$ involve the one--loop gluon self--energy correction, whose result is reviewed in appendix \ref{app:FeynmanRules}. In our computation we employed an effective propagator in momentum space with an unintegrated scalar bubble, in order to make the corresponding cusp contribution fit into the integral classification \eqref{eq:masterintegrals}.
Diagram $(3)_j$ contains the two--loop self--energy correction of the fermion propagator.
Its computation in configuration space is spelled out in appendix \ref{sec:2loopferm}. 

\item Diagrams with four fermion insertions along the line, as drawn in figure \ref{fig:4fermions}.
\begin{figure}[h]
\begin{center}
\includegraphics[width=0.67\textwidth]{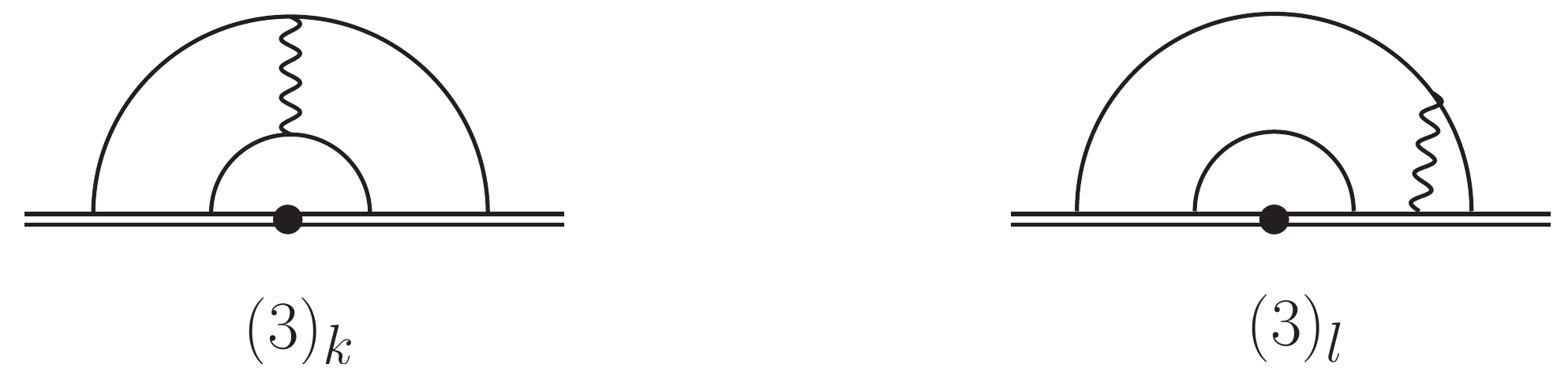}
\caption{Diagrams with four fermion insertions.}
\label{fig:4fermions}
\end{center}
\end{figure}

The topologies $(3)_k$ and $(3)_l$ possess 4 and 5 different inequivalent 1PI vertex configurations, respectively, among which we displayed explicitly two representatives.

\item Diagrams with three fermion lines departing and landing on the Wilson line, as pictured in figure \ref{fig:ladder}.
\begin{figure}[ht]
\begin{center}
\includegraphics[width=0.73\textwidth]{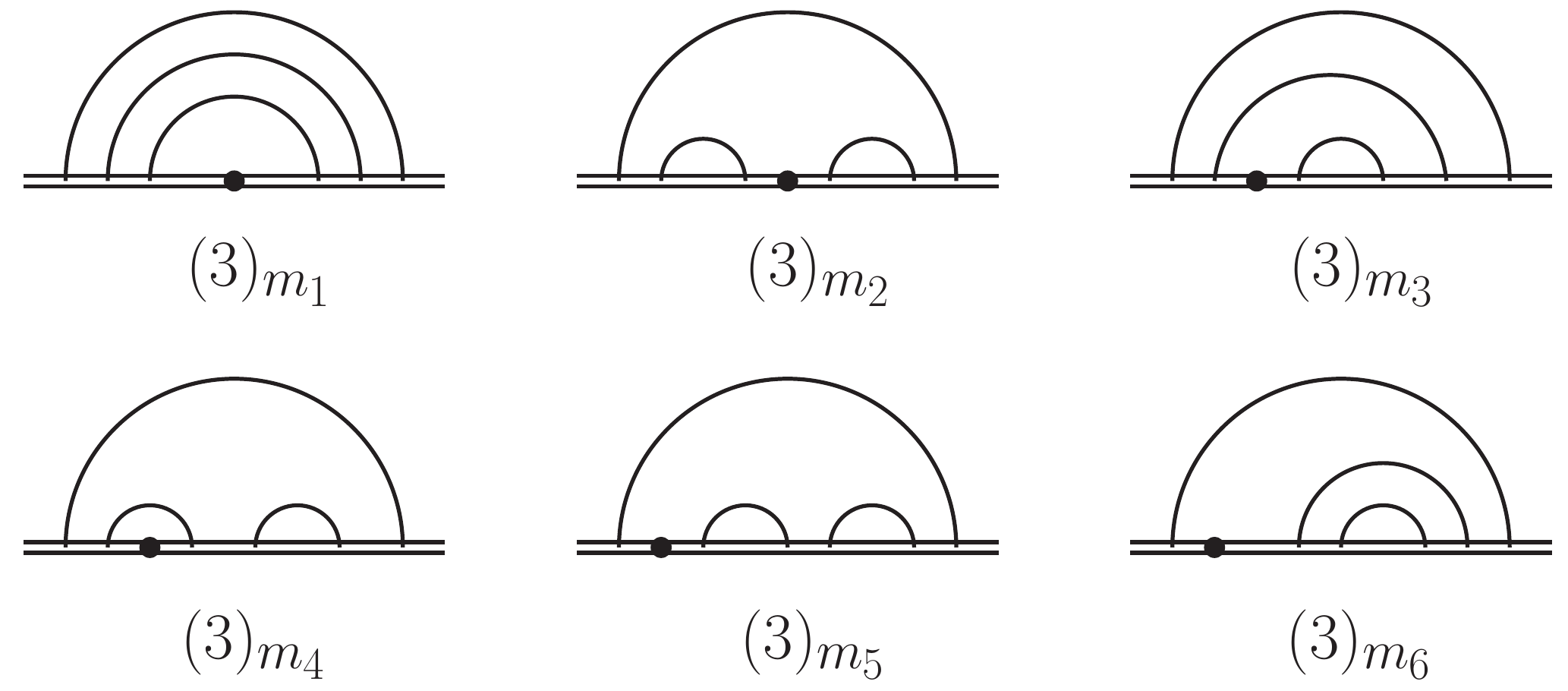}
\caption{Fermion ladder diagrams.}
\label{fig:ladder}
\end{center}
\end{figure}

These are all ladder contributions with no internal integrations, making their evaluation easier than the previous contributions. In the cartoon we have only represented all possible 1PI vertex contractions, which are relevant for our computation.

\end{itemize}

\begin{figure}[ht]
\begin{center}
\includegraphics[width=0.88\textwidth]{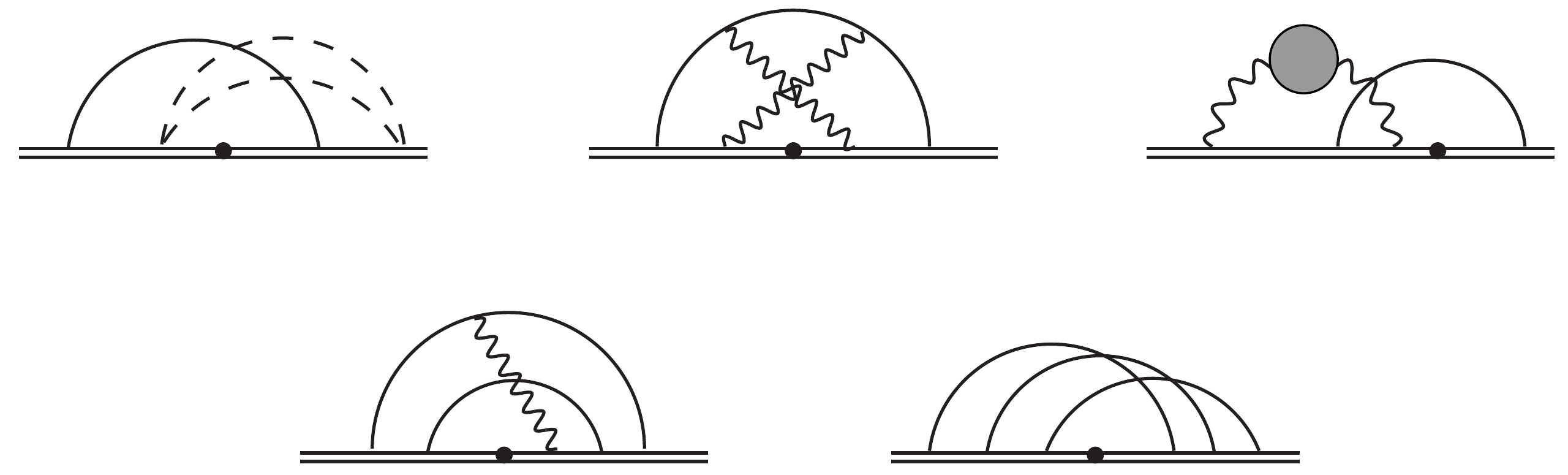}
\caption{Nonplanar topologies.}
\label{fig:nonplanar}
\end{center}
\end{figure}

Finally, nonplanar diagrams can be constructed starting from the same field insertions as in graphs $(3)_c$, $(3)_e$, $(3)_i$, $(3)_l$, $(3)_m$. These are sketched in figure \ref{fig:nonplanar}.

\subsection{The computation}\label{sec:3loop}

We tackle the evaluation of three--loop Feynman diagrams by applying the HQET method described in section \ref{sec:computation}.  
The integrands related to the diagrams listed in the previous section can be computed using the Feynman rules in appendix \ref{app:FeynmanRules} or using the \textit{Mathematica}\textsuperscript{\textregistered} package \texttt{WiLE} \cite{Preti:2017}.
Since the complete calculation is quite long and cumbersome, here we spell out the explicit details of one particular diagram. 

As an example, we consider diagram $(3)_h$, selecting in particular the configuration drawn in figure \ref{fig:2fermions} plus the one where the gluon insertion on the contour and the cusp point are interchanged. The two configurations, being symmetric and thus equivalent, yield an overall factor 2. To be definite, we call the sum of these two configurations $(3)_h^1$. Further possible configurations from the $(3)_h$ topology arise, which correspond to other orderings of insertion points on the contour.  

Considering configurations $(3)_h^1$,  we may obtain leading and sub--leading contributions, depending whether the gauge lines correspond to $A_\mu$ or $\hat A_\mu$. 

We start computing the planar contributions. 
The starting string in configuration space is given by
\begin{align}
(3)_h^1 & = K \int d^{3-2\e}w  \int d^{3-2\e}y \int d^{3-2\e}t  \int^0_{-L} d\t_2 \int^L_{0} d\t_1 \int^{\t_2}_{-L} d\t_3\,   \partial^{\mu_{10}}_{1}\frac{1 }{(x_{1t}^2)^{1/2-\e}}   \non \\  &\partial^{\mu_3}_{2}\frac{1}{(x_{2y}^2)^{1/2-\e}}\, \partial^{\mu_7}_{3}    \frac{1}{(x_{3w}^2)^{1/2-\e}}\,\partial^{\mu_9}_{t}\frac{1}{(x_{yt}^2)^{1/2-\e}}\,  \partial^{\mu_6}_{w}\frac{1}{(x_{wt}^2)^{1/2-\e}}\,\partial^{\mu_8}_{y }\frac{1}{(x_{wy}^2)^{1/2-\e}} \non \\ &  v^{\mu_1}_1\, \e^{\mu_1\mu_2 \mu_3}\,\e^{\mu_4\mu_5 \mu_6}\,  (\eta_1\, \gamma^{\mu_7}\,\gamma^{\mu_5}\,\gamma^{\mu_8}\,\gamma^{\mu_2}\,\gamma^{\mu_9}\,\gamma^{\mu_4}\,\gamma^{\mu_{10}}\,\bar{\eta}_2)
\end{align}
where the constant factor $K$ reads
\begin{equation}
K= 2i\, N^3   \left( \frac{2\pi }{k}\right)^3 \left[\frac{\G(\frac{1}{2}-\e)}{4 \pi^{3/2-\e}} \right]^6
\end{equation}
Next, we Fourier transform this expression using \eqref{eq:Fourier} (which eats up part of $K$) and arrive to an equivalent HQET--like contribution
\begin{align}
(3)_h^1 & = 2^4\, i\, N^3 \left( \frac{2\pi }{k}\right)^3  \int\frac{d^{3-2\e}k_{1/2/3}}{\left((2\pi)^{3-2\e}\right)^3}\, v^{\mu_1}_1\, \e^{\mu_1\mu_2 \mu_3}\,\e^{\mu_4\mu_5 \mu_6}\,  (\eta_1\, \gamma^{\mu_7}\,\gamma^{\mu_5}\,\gamma^{\mu_8}\,\gamma^{\mu_2}\,\gamma^{\mu_9}\,\gamma^{\mu_4}\,\gamma^{\mu_{10}}\,\bar{\eta}_2)   \non \\  & \times
\frac{k_1^{\mu_{10}}\, k_2^{\mu_7}\, k_3^{\mu_6}\, (k_1-k_2)^{\mu_3}\,(k_1-k_3)^{\mu_9}\, (k_2-k_3)^{\mu_8}}{\left( 2k_1\cdot \tilde{v} + 1 \right)^2\,\left( 2k_2\cdot \tilde{v} + 1 \right)\, k_1^2\, k_2^2\, k_3^2\, (k_1-k_2)^2\, (k_1-k_3)^2\, (k_2-k_3)^2}
\end{align}
where we have already introduced the IR regulator $\delta = -1/2$. 

Next we use \eqref{eq:etagamma} to reduce the $\eta$--bilinear and extract the corresponding $\theta$ dependence.
We perform the tensor algebra within the DRED scheme, using in particular a repeated (automatized) application of \eqref{prod} to simplify the product of $\gamma$ matrices to products of metrics and Levi--Civita tensors.
In this process we discard all the contributions proportional to one final $\varepsilon$ tensor, as it is doomed to vanish after integration, as explained in section \ref{sec:ovdiag}.
The result of this procedure leads to a numerator which can be expressed in terms of scalar products between integrated momenta and external velocities only. It looks like

{\scriptsize 
\begin{align}
& 4 \left(k_2.k_3\right){}^2 \left(k_1.v\right){}^2
-4 k_1.k_3 k_2.k_2 \left(k_1.v\right){}^2
+8 k_1.k_2 k_2.k_3 \left(k_1.v\right){}^2
-4 k_1.k_3 k_2.k_3 \left(k_1.v\right){}^2-8 k_2.k_2 k_2.k_3 \left(k_1.v\right){}^2
\nonumber\\&
+8 k_2.k_2 k_3.k_3 \left(k_1.v\right){}^2
-4 k_2.k_3 k_3.k_3 \left(k_1.v\right){}^2-4 k_2.v \left(k_2.k_3\right){}^2 k_1.v-4 \left(k_1.k_3\right){}^2 k_2.v k_1.v+8 k_1.k_2 k_1.k_3 k_2.v k_1.v
\nonumber\\&
-4 k_1.k_3 k_2.v k_2.k_2 k_1.v-4 k_1.k_1 k_2.v k_2.k_3 k_1.v
+8 k_1.k_2 k_2.v k_2.k_3 k_1.v+8 k_1.k_3 k_2.v k_2.k_3 k_1.v-8 \left(k_1.k_2\right){}^2 k_3.v k_1.v
\nonumber\\&
+4 k_1.k_2 k_1.k_3 k_3.v k_1.v+8 k_1.k_2 k_2.k_2 k_3.v k_1.v+4 k_1.k_1 k_2.k_3 k_3.v k_1.v
-12 k_1.k_2 k_2.k_3 k_3.v k_1.v+4 k_2.k_2 k_2.k_3 k_3.v k_1.v
\nonumber\\&
-16 k_1.k_2 k_2.v k_3.k_3 k_1.v+4 k_1.k_3 k_2.v k_3.k_3 k_1.v+4 k_2.v k_2.k_3 k_3.k_3 k_1.v+4 k_1.k_2 k_3.v k_3.k_3 k_1.v-4 k_2.k_2 k_3.v k_3.k_3 k_1.v
\nonumber\\&
+4 \left(k_1.k_3\right){}^2 \left(k_2.v\right){}^2-8 k_1.k_1 k_1.k_3 \left(k_2.v\right){}^2
+8 k_1.k_2 k_1.k_3 \left(k_2.v\right){}^2+8 \left(k_1.k_2\right){}^2 \left(k_3.v\right){}^2-4 k_1.k_1 k_1.k_2 \left(k_3.v\right){}^2
\nonumber\\&
-4 k_1.k_2 k_2.k_2 \left(k_3.v\right){}^2+4 k_1.k_1 k_1.k_3 k_2.k_2-4 k_1.k_2 k_1.k_3 k_2.k_2-4 k_1.k_1 \left(k_2.v\right){}^2 k_2.k_3-4 k_1.k_3 \left(k_2.v\right){}^2 k_2.k_3
\nonumber\\&
-4 k_1.k_1 k_1.k_2 k_2.k_3
+4 k_1.k_1 k_2.k_2 k_2.k_3-8 \left(k_1.k_2\right){}^2 k_2.v k_3.v+8 k_1.k_1 k_1.k_2 k_2.v k_3.v+4 k_1.k_1 k_1.k_3 k_2.v k_3.v
\nonumber\\&
-12 k_1.k_2 k_1.k_3 k_2.v k_3.v+4 k_1.k_3 k_2.v k_2.k_2 k_3.v
+4 k_1.k_2 k_2.v k_2.k_3 k_3.v+8 \left(k_1.k_2\right){}^2 k_3.k_3+8 k_1.k_1 \left(k_2.v\right){}^2 k_3.k_3
\nonumber\\&
-4 k_1.k_3 \left(k_2.v\right){}^2 k_3.k_3-4 k_1.k_2 k_1.k_3 k_3.k_3-8 k_1.k_1 k_2.k_2 k_3.k_3+4 k_1.k_3 k_2.k_2 k_3.k_3+4 k_1.k_1 k_2.k_3 k_3.k_3
+\nonumber\\&
-4 k_1.k_2 k_2.k_3 k_3.k_3
-4 k_1.k_1 k_2.v k_3.v k_3.k_3+4 k_1.k_2 k_2.v k_3.v k_3.k_3 \nonumber
\end{align}}
As a final step, we rewrite the above expression in terms of inverse propagators and arrive at a sum of scalar integrals with different powers of the propagators, in a form which FIRE can be fed with.
After the FIRE digestion, the diagram evaluates to a simple sum over the master integrals listed in appendix \ref{app:masters}
\begin{align}
& (3)_h^1 = \left( \frac{2\pi\, N }{k}\right)^3  C_{\theta} \bigg[-\frac{16(d-2)(5d-13)}{d-3} G_{0,1,1,1,1,0,0,0,1}
\nonumber\\&+\frac{8(3d-7)(134+d(19d-101))}{(d-3)^2 (2d-5)}
G_{0,0,1,0,1,0,1,0,1}
+  \frac{8(3d-7)(3d-8)}{d-3} G_{0,1,1,1,0,0,1,0,1}
\nonumber\\&
+ 8 (d-1) G_{0,1,1,0,1,1,1,0,1} 
+ (32-14d) G_{1,0,0,0,1,1,1,0,1} \bigg]
\end{align}
which can directly be expanded in $\epsilon$ to the desired order
\begin{equation}
(3)_h^1 = \left( \frac{ N }{k}\right)^3 \, \frac{\left(\pi ^2-60\right)\, C_{\theta}}{576 \epsilon }+{\cal O}\left(\epsilon ^0\right)
\end{equation}
Repeating the exercise for all the diagrams we obtain the results listed in appendix \ref{app:diagrams}.\\

We compute the nonplanar corrections as well, which are produced by two possible sources. At first they can come from planar diagrams, when for instance the choice of the vector boson gives rise to a sub--leading in color contraction of the color matrices. Second,  they can be produced by genuinely nonplanar diagrams.  We  remark that, since we consider color subleading contributions, we also have to take into account the mixed gauge propagator $\langle A\hat A \rangle$ contribution \eqref{1vector3},  arising at one--loop from the creation and annihilation of a matter pair.

The first class of contributions arises from diagrams $(3)_a$, $(3)_b$, $(3)_d$, $(3)_f$, $(3)_g$, $(3)_h$, $(3)_j$ and $(3)_k$, whereas genuine nonplanar Wick contractions arise in topologies $(3)_c$, $(3)_e$, $(3)_i$, $(3)_l$ and $(3)_m$. Examples of such graphs are shown in figure \ref{fig:nonplanar}, where we have shown only one particular nonplanar configuration for each topology.
Thanks to the linear nature of the HQET propagators as functions of loop momenta, such nonplanar integrals contain linearly dependent propagators. These can be manipulated by partial fractioning (using an automated routine) in such a way to reduce them all to either planar integrals or a single nonplanar topology depicted in figure \ref{fig:3lmasters}, with arbitrary powers of the propagators. These can be finally reduced by integration--by--parts identities to a single nonplanar master integral, which is defined and computed in appendix \ref{app:masters}. 
In particular, after decomposition by partial fractioning, the nonplanar diagrams arising from topologies $(3)_c$ and $(3)_i$ evaluate to planar integrals. Only diagrams $(3)_e$, $(3)_l$ and $(3)_m$ give rise to the nonplanar topology $J$ of figure \ref{fig:3lmasters}.

\section{The three--loop cusp and the Bremsstrahlung function}\label{result}

We are now ready to provide the full result for the 1/2--BPS cusp anomalous dimension up to three loops, at vanishing geometric angle $\varphi=0$.
Omitting an overall factor $\left(\frac{(e^{-\gamma_E}16\pi)^{\epsilon}}{k}\right)^3$, the sum of all the three--loop contributions to the 1PI vertex function $V$ reads
\begin{align}\label{eq:V3}
& V^{(3)}(\theta) = \frac{N^3\, C_{\theta}^3}{48\, \epsilon ^3}-\frac{N^3 \left(C_{\theta}^3+C_{\theta}\right)}{8\, \epsilon ^2}
+\frac{1}{576\, \epsilon }\big[288\, N^3\, C_{\theta} \left(C_{\theta} \left(2 C_{\theta}-1\right)+1\right)
\\& +
\pi ^2 N \left[ N^2  \left(C_{\theta}^2 \left(45 C_{\theta}-8\right)-72C_{\theta}+25\right)+ 16 C_{\theta} \left(\left(C_{\theta}-1\right) C_{\theta}+2\right)-17\right]\big]+{\cal O}\left(\epsilon ^0\right) \nonumber
\end{align}

As a first consistency check of the result, we probe exponentiation (\ref{espo}). This test was already successfully performed up to two loops in \cite{Griguolo:2012iq} and here we extend the analysis to the next perturbative order. To this end we insert the explicit expressions (\ref{eq:V1}, \ref{eq:V2}, \ref{eq:V3})  for $V^{(1)},V^{(2)},V^{(3)}$ in the general expansion for the vertex function (\ref{eq:Vexpansion}). Now, taking the logarithm and expand the resulting expression in powers of $1/k$ we do find 
\begin{align}\label{eq:log}
\log V(\theta) &= \frac{N\, C_{\theta}}{2\, k\, \epsilon } - \frac{N^2}{4\, k^2\, \epsilon } + \frac{\pi ^2\, N}{576\, k^3\, \epsilon } \Big[N^2 \left( 16 C_{\theta}^2 -48 C_{\theta}+25 \right) \nonumber\\& \qquad + (16C_{\theta}^3-16C_{\theta}^2+32C_{\theta}-17)\Big] +{\cal O}\left(k^{-4}\right) +{\cal O}(\epsilon^0)
\end{align}
Although the $V^{(2)}$ and $V^{(3)}$ coefficients originally contained higher order poles in $1/\e$, only single poles appear in (\ref{eq:log}). This proves that quadratic and cubic poles in $V^{(2)}(\theta)$ and $V^{(3)}(\theta)$, eqs. (\ref{eq:V2}, \ref{eq:V3}), are ascribable to powers of lower order divergent contributions to $V$. Therefore, as expected, the divergent part of the expectation value of the cusped WL exponentiates.
In particular, since no nonplanar corrections appear at one-- and two--loop orders, their contribution at three loops has only a simple pole in the regulator.

We note that the two--loop part in (\ref{eq:log}) is independent of $\theta$. Applying prescription (\ref{eq:KorchRad}) to obtain the whole expression for the cusped WL this contribution will then get removed. The final result for the (unrenormalized) cusped WL  at $\varphi=0$ reads
\begin{equation}\label{eq:logW}
\log \langle W(\theta)\rangle = \frac{N(C_{\theta}-1)}{2\, k\, \epsilon }+\frac{\pi ^2\, N \left(C_{\theta}-1\right) \left(N^2 \left(C_{\theta}-2\right)+ C_{\theta}^2+2\right)}{36\, k^3\, \epsilon } +{\cal O}\left(k^{-4}\right) +{\cal O}(\epsilon^0)
\end{equation}
This expression displays uniform transcendentality. Moreover, it satisfies the BPS condition, namely it vanishes for $\theta=0$, albeit this statement is quite trivial at $\varphi=0$, being it an obvious consequence of \eqref{eq:KorchRad}. 

As a further consistency check, we have also directly verified prescription \eqref{eq:KorchRad} by computing the non--1PI contributions as well.

We renormalize \eqref{eq:logW} by multiplying by a renormalization function $Z_{cusp}$. Following the steps outlined in section \ref{sec:detailscusp}, from $Z_{cusp}$ we obtain the final result for the anomalous dimension at three loops
\begin{equation}
\Gamma_{cusp}(k,N,\varphi=0) = \frac{N \left(1-C_{\theta}\right)}{k}-\frac{\pi ^2\, N \left(C_{\theta}-1\right) \left(C_{\theta}^2+N^2 \left(C_{\theta}-2\right)+2\right)}{6\, k^3}+{\cal O}\left(k^{-4}\right)
\end{equation}
The Bremsstrahlung function is finally evaluated according to \eqref{eq:brems} and reads
\begin{equation}\label{eq:bremsstrahlung}
B_{1/2}(k,N) = \frac{N}{8\, k}-\frac{\pi ^2\, N \left(N^2-3\right)}{48\, k^3}+{\cal O}\left(k^{-4}\right)
\end{equation}
This result is in perfect agreement with conjecture \eqref{eq:conjecture} that was formulated in \cite{Lewkowycz:2013laa, Bianchi:2014laa}. In fact, using the expectation value for the 1/6--BPS WLs at framing one, as derived from localization
\begin{align}
\langle W_{1/6}\rangle_1 &= 1 + i\, \pi\, \frac{N}{k} + \frac{1}{6} \left(1+2 N^2\right) \pi ^2\, \frac{1}{k^2} + \frac{1}{6}\, i\, N \left(4+N^2\right) \pi ^3\, \frac{1}{k^3} + O(k^{-4})
\nonumber\\
\langle \hat W_{1/6}\rangle_1 &= \langle W_{1/6}\rangle_1^*
\end{align}
and inserting them in the r.h.s. of \eqref{eq:conjecture} we find exactly expression \eqref{eq:bremsstrahlung}. Remarkably, the agreement applies also to the nonplanar part. This is consistent with the fact that up to three loops the arguments of \cite{Bianchi:2014laa} do not rely on restricting to the planar limit.

\section{Conclusions}

In this paper we have studied the Bremsstrahlung function for the locally 1/2--BPS cusp in the ABJM model at weak coupling with the aim of proving conjecture \eqref{eq:conjecture} for its exact determination.
We remind that this conjecture is based on linking the Bremsstrahlung function to the expectation value of supersymmetric 1/6--BPS WLs with multiple winding, which are computable exactly via localization \cite{Klemm:2012ii}.
To begin with, it relies on the expectation values of supersymmetric WLs on latitude contours in $S^2$ (see eq. (\ref{conjecture})) and hinges on a few steps, some of which lack a rigorous proof.
Therefore, in order to substantiate the conjecture, we have performed a precision perturbative check at three loops.

Technically, we have taken considerable advantage from the BPS condition for the cusp, in order to make the computation simpler. We have considered the expansion of the cusp anomalous dimension in the small internal angle in the R--symmetry space, at vanishing geometric angle. This setting entails several simplifications at both the level of the diagrams involved, and in their practical evaluation, especially in handling the integrals.
These reduce to propagator--type integrals, which we computed by switching to the HQET picture (via Fourier transform to momentum space) and employing integration by parts identities to reduce them to a restricted set of master topologies.

Result \eqref{eq:bremsstrahlung} for the three--loop Bremsstrahlung function agrees with the prediction based on conjecture \eqref{eq:conjecture}.
Remarkably, such an agreement extends to the nonplanar part as well.
This provides a compelling test in favour of proposal \eqref{eq:conjecture} and its validity beyond the planar limit.

This exact result is interesting per se, but acquires additional appeal in view of a potential integrability based computation of the same quantity.
Recent developments on the Quantum Spectral Curve approach \cite{Gromov:2013pga,Gromov:2014caa,Gromov:2015dfa,Gromov:2016rrp} in the ABJM model \cite{Cavaglia:2014exa,Bombardelli:2017vhk} seem to point to such a possibility. This would provide a crossed check of the localization based proposal and, by comparison with the integrable description, of the conjecture on the exact expression for the interpolating $h$ function of ABJM \cite{Nishioka:2008gz,Grignani:2008is,Gaiotto:2008cg,Gromov:2014eha}.

We conclude with comments on future perspectives and extensions of this paper. 

A first, sensible generalization of the three--loop computation of the cusped 1/2--BPS WL consists in lifting the ranks of the two gauge groups to unequal values, or in other words to compute the same quantity in the ABJ model.
From the technical standpoint, this requires modifying the color factors of the diagrams analysed in this article and supplying further graphs which were discarded, because they vanish in the ABJM limit.

Despite the fact that such a result is expected to be straightforward to derive, lower order contributions suggest that the expectation value of the locally 1/2--BPS WL on a cusped contour does not exponentiate \cite{Griguolo:2012iq,Bonini:2016fnc}, at least not in a standard fashion. This might hamper the interpretation of the three--loop correction and the extraction of a cusp anomalous dimension from it. Moreover, it is not clear if and how conjecture \eqref{conjecture} can be extended to the case with different ranks.
Still, the ABJ theory is expected to be integrable (it was proven to be so in a particular sector in the limit of \cite{Bianchi:2016rub}), therefore a derivation of its Bremsstrahlung function from integrability is also foreseeable. This, together with a deeper understanding of the ABJ supersymmetric cusp, would grant a firmer handle on the conjecture for the exact interpolating function of the ABJ model \cite{Cavaglia:2016ide}.
In conclusion, the lifting to the ABJ theory is challenging and requires more thinking, but is certainly an interesting direction to pursue.

In this paper we have found the exact expression of a cusped 1/2--BPS WL at three loops and at finite $N$, eq. (\ref{eq:logW}). A natural extension of this computation is the evaluation of the same 1/2--BPS cusp, but with open geometric angle ($\varphi \neq 0$), along the lines of \cite{Grozin:2015kna}.
This task, if performed using the HQET description, demands the evaluation of the 71 master integrals pointed out in \cite{Grozin:2015kna}, in three dimensions.
However,  this should be simpler than in four dimensions, as the leading divergence of the cusp integrals at loop $l$ is a pole of order $l$ in three dimensions, instead of $2l$ in four.
Consequently, the relevant integrals should be expanded up to terms of transcendentality 2 in the three loop case (compared to transcendental order 5).
We have preliminary evidence that some of the master integrals required for the computation do not evaluate to generalized polylogarithms, rather elliptic sectors appear.
Elliptic functions are likely to pop up at higher order in the $\epsilon$--expansion or they might cancel out when summing all the diagrams, yet their presence can hinder the evaluation of the master integrals. In particular, at a difference with respect to the four dimensional case, we expect that a canonical form \cite{Henn:2013pwa} does not exist for all of them.

\acknowledgments

S.P. thanks the Galileo Galilei Institute for Theoretical Physics (GGI) for the hospitality and INFN for partial support during the completion of this work, within the program ``New Developments in AdS3/CFT2 Holography''.  This work has been supported in part by Italian Ministero dell'Istruzione, Universit\`a e Ricerca (MIUR), Della Riccia Foundation and Istituto Nazionale di Fisica Nucleare (INFN) through the ``Gauge Theories, Strings, Supergravity" (GSS) and ``Gauge And String Theory" (GAST) research projects.

\vfill
\newpage

\appendix
	 
\section{Spinor  and  group conventions}\label{app:spinors}

We work in euclidean three dimensional space with coordinates $x^\mu = (x^0, x^1, x^2)$. The Dirac matrices satisfying the Clifford algebra $\{ \g^\mu , \g^\nu \} = 2 \d^{\mu\nu} \mathbb{I}$ are chosen to be
\beq
\label{diracmatrices}
(\g^\mu)_\a^{\; \, \b} = \{ -\s^3, \s^1, \s^2 \}
\eeq
with matrix product 
\beq
\label{prod}
(\g^\mu \g^\nu)_\a^{\; \, \b} \equiv (\g^\mu)_\a^{\; \, \g} (\g^\nu)_\g^{\; \, \b}
\eeq
The set of matrices \eqref{diracmatrices} satisfies satisfies the following set of identities
\begin{align}
& \g^\mu \g^\nu = \d^{\mu \nu} \mathds{1} - i \varepsilon^{\mu\nu\rho} \g^\rho
\non\\
& \g^\mu \g^\nu \g^\rho = \d^{\mu\nu} \g^\rho - \d^{\mu\rho} \g^\nu+  \d^{\nu\rho} \g^\mu  - i \varepsilon^{\mu\nu\rho} \mathds{1}
 \\
&
\g^\mu \g^\nu \g^\rho \g^\s -  \g^\s \g^\rho \g^\nu \g^\mu = -2i \left( \d^{\mu\nu} \varepsilon^{\rho\s \eta}  + \d^{\rho \s}  \varepsilon^{\mu\nu\eta} + \d^{\nu\eta} \varepsilon^{\rho \mu \s} +
\d^{\mu\eta} \varepsilon^{\nu\rho\s}  \right) \g^\eta\non
 \end{align}
which allows us to simplify the fermionic contributions to the WL. The relevant traces are instead given by
\begin{align}
\Tr (\g^\mu \g^\nu) = 2 \d^{\mu\nu}\  \ \ \ \ \ \ \ \ 
\textrm{and}\  \ \ \ \ \ \ \ \ 
\Tr (\g^\mu \g^\nu \g^\rho) = -2i \varepsilon^{\mu\nu\rho}
\end{align}
Spinor indices are raised and lowered by means of the $\epsilon-$tensor: 
\beq
\psi^\a = \varepsilon^{\a\b} \psi_\b \quad  \quad \psi_\a = \varepsilon_{\a\b}  \psi^\b    
\eeq
with $\varepsilon^{12} = - \varepsilon_{12} = 1$. In particular,  the antisymmetric combination of two spinors  can be reduced to scalar contractions:
\begin{align}
\psi_\a \chi_\b - \psi_\b \chi_\a = \varepsilon_{\a\b} \psi^\g \chi_\g\equiv \varepsilon_{\a\b} \psi\chi,
\ \ \ \ \ \ 
 \psi^\a \chi^\b - \psi^\b \chi^\a = -\varepsilon^{\a\b} \psi^\g \chi_\g\equiv-\varepsilon^{\a\b} \psi \chi
\end{align}
The gamma matrices with two lower indices, $(\g^\mu)_{\a  \b} \equiv   (\g^\mu)_\a^{\; \, \g} \varepsilon_{\b \g}$, are then given by
\beq
(\g^\mu)_{\a \b} = \{ -\s^1, -\s^3, i \mathds{1} \}
\eeq 
and they obey the  useful identity.
\begin{equation}
(\g^\mu)_{\a\b} (\g_\mu)_{\g\delta} = - \varepsilon_{\a\g} \varepsilon_{\b\delta} - \varepsilon_{\a\delta} \varepsilon_{\b\g} 
\end{equation} 
Under complex conjugation the gamma matrices transform as follows: $[(\g^\mu)_\a^{\; \, \b}]^* = (\g^\mu)^\b_{\; \, \a} \equiv \e^{\b\g} (\g^\mu)_\g^{\; \, \d}  \e_{\a\d}$. As a consequence, the hermitian conjugate of the vector bilinear  can be rewritten as follows
\beq
(\psi \g^\mu \chi)^\dagger =(\psi^\a (\g^\mu)_\a^{\; \, \b} \chi_\b)^\dagger = \bar{\chi}_\b (\g^\mu)^\b_{\; \, \a} \bar{\psi}^\a = \bar{\chi}^\b  (\g^\mu)_\b^{\; \, \a} \bar{\psi}_\a \equiv 
\bar{\chi} \g^\mu \bar{\psi}
\eeq 
where we have taken $ (\chi_\beta)^\dagger=\bar\chi_\beta$ and $ (\psi^\alpha)^\dagger=\bar\psi^\alpha$.

\noindent
The U$(N)$  generators are defined as $T^A = (T^0, T^a)$, where $T^0 =
\frac{1}{\sqrt{N}}\mathds{1}$ and $T^a$ ($a=1,\ldots, N^2-1$) are an orthonormal  set of traceless
$N\times N$ hermitian matrices.  The generators are normalized as
\be
\Tr( T^A T^B )= \delta^{AB}
\ee 
The structure constant are then defined by 
\be
[T^A,T^B]=i f^{AB}{}_C T^C
\ee
In the paper  we shall often use  the double notation and  the fields will carry two indices in the fundamental representation
of the gauge groups. An index in the fundamental representation of U$(N_1)$ will be generically by the  lowercase roman indices $i,j,k,\dots$, while for an index  in the fundamental representation of U$(N_2)$ we shall use the hatted lowercase roman indices $\hat i, \hat j, \hat k,\dots$

\section{Basic facts on ABJM action}\label{app:ABJM}

Here we collect some basic features on the action for general U$(N_1)_k \times$ U$(N_2)_{-k}$ ABJ(M) theories. The gauge sector contains two gauge fields 
$(A_\mu)_i{~}^j$  and $(\hat {A}_\mu)_{\hat i}{~}^{\hat{j}}$ belonging respectively to
the adjoint of U$(N_1)$ and U$(N_2)$. The matter sector instead consists of  the complex fields
$(C_I)_i{~}^{\hat j}$ and $(\bar {C}^I)_{\hat{i}}{~}^j$ as well as the fermions $(\psi_I)_i{~}^{\hat j}$ and $(\bar {\psi}^I)_{\hat{i}}{~}^j$ . The fields $(C_I, \bar\psi^I)$ transform in 
the $({\bf N_1},{\bf \bar N_2})$ of the gauge group while the couple $(\bar C^I,\psi_I)$ belongs to the
representation $({\bf \bar N_1},{\bf N_2})$.
The additional capitol index $I = 1,2,3,4$  belongs to the R--symmetry group $SU(4)$. In
order to quantize the theory at the perturbative level, we  introduce the usual Feynman gauge--fixing  for both gauge fields and the  two corresponding sets of ghosts $(\bar c,c)$ and $(\bar{\hat c},\hat c)$. Then the action contains four different contributions
\beq
\label{action}
S = S_{\mathrm{CS}} \big|_{\mathrm{g.f.}}+ S_{\mathrm{mat}} + S_{\mathrm{pot}}^{\mathrm{bos}} + S_{\mathrm{pot}}^{\mathrm{ferm}} 
\eeq 
where
\begin{subequations}
\begin{align}
\label{action1}
S_{\mathrm{CS}}\big|_{\mathrm{g.f.}} &=\frac{k}{4\pi}\int d^3x\,\varepsilon^{\mu\nu\rho} \Big\{ i \, \Tr \!\left(\hat{A}_\mu\partial_\nu 
\hat{A}_\rho+\frac{2}{3} i \hat{A}_\mu \hat{A}_\nu \hat{A}_\rho \right) \!-\! i \, \Tr \left( A_\mu\partial_\nu A_\rho+\frac{2}{3} i A_\mu A_\nu A_\rho\! \right)\!\non \\
&~   + \, \Tr \Big[ \frac{1}{\xi}  (\pa_\mu A^\mu)^2 -\frac{1}{\xi} ( \pa_\mu \hat{A}^\mu )^2 + \pa_\mu \bar{c} D^\mu c  
  - \pa_\mu \bar{\hat{c}} D^\mu \hat{c} \Big] \Big\}
\\
\label{action2}
S_{\mathrm{mat}} =& \int d^3x \, \Tr \Big[ D_\mu C_I D^\mu \bar{C}^I - i \bar{\Psi}^I \g^\mu D_\mu \Psi_I \Big] 
\\
 S_{\mathrm{pot}}^{\mathrm{bos}} =& -\frac{4\pi^2}{3 k^2} \int d^3x \, \Tr \Big[ C_I \bar{C}^I C_J \bar{C}^J C_K \bar{C}^K + \bar{C}^I C_I \bar{C}^J C_J \bar{C}^K C_K\non
 \\
&~ \qquad \qquad \qquad \qquad + 4 C_I \bar{C}^J C_K \bar{C}^I C_J \bar{C}^K - 6 C_I \bar{C}^J C_J \bar{C}^I C_K \bar{C}^K \Big] 
\\
 S_{\mathrm{pot}}^{\mathrm{ferm}} =&  -\frac{2\pi i}{k} \int d^3x \, \Tr \Big[ \bar{C}^I C_I \Psi_J \bar{\Psi}^J - C_I \bar{C}^I \bar{\Psi}^J \Psi_J
+2 C_I \bar{C}^J \bar{\Psi}^I \Psi_J 
\non \\
&~ \qquad \qquad  - 2 \bar{C}^I C_J \Psi_I \bar{\Psi}^J - \e_{IJKL} \bar{C}^I\bar{\Psi}^J \bar{C}^K \bar{\Psi}^L + \e^{IJKL} C_I \Psi_J C_K \Psi_L \Big]
\end{align}
\end{subequations}
 Here the invariant $SU(4)$ tensors $\epsilon_{IJKL}$ and $\epsilon^{IJKL}$ are defined by setting $\e_{1234}=\e^{1234} =1$.   The covariant derivatives used  in \eqref{action}   are instead given by
\bea
\label{covariant}
D_\mu C_I &=& \pa_\mu C_I + i A_\mu C_I - i C_I \hat{A}_\mu,
\quad \quad 
D_\mu \bar{C}^I = \pa_\mu \bar{C}^I - i \bar{C}^I A_\mu + i \hat{A}_\mu \bar{C}^I
\non \\
D_\mu \bar{\Psi}^I  &=& \pa_\mu \bar{\Psi}^I + i A_\mu \bar{\Psi}^I - i \bar{\Psi}^I \hat{A}_\mu,
\quad  \quad
D_\mu \Psi_I = \pa_\mu \Psi_I - i \Psi_I A_\mu + i \hat{A}_\mu \Psi_I  
\eea
 The action \eqref{action}, in absence of the  gauge-fixing terms and of the corresponding ghosts,  it is 
invariant under the following ${\cal N}=6$ SUSY transformations
 \bea
\label{susy2}
&& \d C_I = - 2 \Theta_{IJ} \bar{\Psi}^J
\non \\
&& \d \bar{C}^I =   2 \bar{\Theta}^{IJ} \Psi_J
\non \\
&& \d \Psi_I^\a = -2i \Theta_{IJ}^\b (\g^\mu)_\b^{\; \a}  D_\mu \bar{C}^J + \frac{4\pi i}{k} \TH_{IJ}^\a (\bar{C}^J C_K \bar{C}^K -\bar{C}^K C_K \bar{C}^J ) 
+  \frac{8\pi i}{k} \TH_{KL}^\a \bar{C}^K C_I \bar{C}^L  
\non \\
&& \d \bar{\Psi}^I_\a = 2i \bar{\Theta}^{IJ \, \b} (\g^\mu)_{\b \a}  D_\mu C_J - \frac{4\pi i}{k} \bar{\TH}^{IJ}_\a (C_K \bar{C}^K C_J  -C_J \bar{C}^K C_K ) 
-  \frac{8\pi i}{k} \bar{\TH}^{KL}_\a C_L \bar{C}^I C_K 
\non \\
&& \d A_\mu = \frac{4\pi i}{k} \bar{\TH}^{IJ} \g_\mu C_I \Psi_J - \frac{4\pi i}{k} \TH_{IJ} \g_\mu \bar{\Psi}^I \bar{C}^J
\non \\
&& \d \hat{A}_\mu = \frac{4\pi i}{k} \bar{\TH}^{IJ} \g_\mu \Psi_J  C_I  - \frac{4\pi i}{k} \TH_{IJ} \g_\mu \bar{C}^J \bar{\Psi}^I  
\eea

\section{Feynman rules}\label{app:FeynmanRules}

We use the Fourier transform definition
\begin{equation}\label{eq:Fourier}
\int \frac{d^{3-2\e}p}{(2 \p)^{3-2\e}} \frac{p^{\m}}{(p^2)^s} e^{i p \cdot (x-y)} =  \frac{\G(\frac{3}{2}-s-\e)}{4^s \pi^{3/2-\e}\Gamma(s)} \big(-i \partial^{\m}_x \big)\frac{1}{(x-y)^{2(3/2-s-\e)}}
\end{equation}
From  the action  \eqref{action}  we read the following Feynman rules\footnote{In euclidean space we define the functional generator as $Z \sim \int e^{-S}$.}
\begin{itemize}
\item Vector propagators in Landau gauge
\bea
\label{treevector}
 \langle (A_\mu)_i{}^j (x) (A_\nu)_{k}{}^\ell(y) \rangle^{(0)} &=&  \d^{\ell}_i \delta_k^j  \, \left( \frac{2\pi i}{k} \right) \frac{\G(\frac32-\e)}{2\pi^{\frac32 -\e}} \varepsilon_{\mu\nu\rho} \frac{(x-y)^\rho}{[(x-y)^2]^{\frac32 -\e} }
\non \\
&=&  \d^{\ell}_i \delta_k^j   \left( \frac{2\pi }{k} \right) \varepsilon_{\mu\nu\rho} \, \int \frac{d^np}{(2\pi)^n}   \frac{p^\rho}{p^2} e^{ip(x-y)}
\non \\
\non \\
 \langle (\hat{A}_\mu)_{\hat{i}}{}^{\hat{j}} (x) ( \hat{A}_\nu)_{\hat k}{}^{\hat\ell }(y) \rangle^{(0)} &=&  - \d^{\hat\ell}_{\hat i} \delta_{\hat k}^{\hat j}     \, \left( \frac{2\pi i}{k} \right) \frac{\G(\frac32-\e)}{2\pi^{\frac32 -\e}} \varepsilon_{\mu\nu\rho} \frac{(x-y)^\rho}{[(x-y)^2]^{\frac32 -\e} }
\non \\
&=& -\d^{\hat\ell}_{\hat i} \delta_{\hat k}^{\hat j} \left( \frac{2\pi }{k} \right) \varepsilon_{\mu\nu\rho} \, \int \frac{d^np}{(2\pi)^n}   \frac{p^\rho}{p^2} e^{ip(x-y)}
\eea
\item Scalar propagator
\bea
\label{scalar}
\langle (C_I)_i{}^{\hat{j}} (x) (\bar{C}^J)_{\hat{k}}{}^l(\; y) \rangle^{(0)}  &=& \d_I^J \d_i^l \d_{\hat{k}}^{\hat{j}} \, \frac{\G(\frac12 -\e)}{4\pi^{\frac32-\e}} 
\, \frac{1}{[(x-y)^2]^{\frac12 -\e}}
\non \\
&=& \d_I^J \d_i^l \d_{\hat{k}}^{\hat{j}} \, \int \frac{d^np}{(2\pi)^n}   \frac{e^{ip(x-y)} }{p^2}  
\eea
\item Fermion propagator
\bea
\label{treefermion}
\langle (\psi_I^\a)_{\hat{i}}{}^{ j}  (x) (\bar{\psi}^J_\b )_k{}^{ \hat{l}}(y) \rangle^{(0)} &=&  i \, \d_I^J \d_{\hat{i}}^{\hat{l}} \d_{k}^{j} \, 
\frac{\G(\frac32 - \e)}{2\pi^{\frac32 -\e}} \,  \frac{(\g^\mu)^\a_{\; \, \b} \,  (x-y)_\mu}{[(x-y)^2]^{\frac32 - \e}}
\non \\
&=& \d_I^J \d_{\hat{i}}^{\hat{l}} \d_{k}^{j} \, \int \frac{d^np}{(2\pi)^n}   \frac{(\g^\mu)^\a_{\; \, \b} \, p_\mu  }{p^2}  e^{ip(x-y)}
\eea
\item Gauge cubic vertex
\beq
\label{gaugecubic}
i \frac{k}{12\pi} \varepsilon^{\mu\nu\rho} \int d^3x \, f^{abc} A_\mu^a A_\nu^b A_\rho^c
\eeq
\item Gauge-fermion cubic vertex
\beq
\label{gaugefermion}
-\int d^3x \, \Tr \Big[ \bar{\Psi}^I \g^\mu \Psi_I A_\mu - \bar{\Psi}^I \g^\mu \hat{A}_\mu \Psi_I  \Big]
\eeq 
\end{itemize}

\noindent
At one--loop, if we work at finite $N_1$ and $N_2$,  beside corrections to the tree level vector propagators, a contribution to the mixed   $\langle A \hat{A}\rangle$  propagator is generated. They read
\begin{subequations}
\begin{align}
\label{1vector}
 \langle (A_\mu)_i{}^j (x) &(A_\nu)_{k}{}^\ell(y) \rangle^{(1)} =  \d^{\ell}_i \delta_k^j    \left( \frac{2\pi }{k} \right)^2 N_2 \, \frac{\G^2(\frac12-\e)}{4\pi^{3 -2\e}} 
\left[ \frac{\d_{\mu\nu}}{ [(x- y)^2]^{1-2\e}} - \pa_\mu \pa_\nu \frac{[(x-y)^2]^{2\e}}{4\e(1+2\e)} \right]  
\non  \\
  =&   \d^{\ell}_i \delta_k^j    \left( \frac{2\pi }{k} \right)^2 N_2 \, \frac{\G^2(\frac12-\e)\G(\frac12 +\e)}{\G(1-2\e) 2^{1-2\e} \pi^{\frac32 -\e}} 
\, \int \frac{d^np}{(2\pi)^n}   \frac{e^{ip(x-y)}}{(p^2)^{\frac12 +\e}}  \left( \d_{\mu \n} - \frac{p_\mu p_\nu}{p^2} \right) 
\\
 \langle (\hat{A}_\mu)_{\hat{i}}{}^{\hat{j}} (x) &( \hat{A}_\nu)_{\hat k}{}^{\hat\ell }(y) \rangle^{(1)} = \d^{\hat\ell}_{\hat i} \delta_{\hat k}^{\hat j}     \left( \frac{2\pi }{k} \right)^2 N_1 \, \frac{\G^2(\frac12-\e)}{4\pi^{3 -2\e}} 
\left[ \frac{\d_{\mu\nu}}{ [(x- y)^2]^{1-2\e}} - \pa_\mu \pa_\nu \frac{[(x-y)^2]^{2\e}}{4\e(1+2\e)} \right] \non \\
 =& \d^{\hat\ell}_{\hat i} \delta_{\hat k}^{\hat j}   \left( \frac{2\pi }{k} \right)^2 N_1 \, \frac{\G^2(\frac12-\e)\G(\frac12 +\e)}{\G(1-2\e) 2^{1-2\e} \pi^{\frac32 -\e}} 
\, \int \frac{d^np}{(2\pi)^n}   \frac{e^{ip(x-y)}}{(p^2)^{\frac12 +\e}} \left( \d_{\mu \n} - \frac{p_\mu p_\nu}{p^2} \right) 
\\\label{1vector3}
 \langle (A_\mu)_{i}{}^{j} (x) &( \hat{A}_\nu)_{\hat k}{}^{\hat\ell }(y) \rangle^{(1)} = - \d_{i}{}^{j} \delta_{\hat k}{}^{\hat\ell }     \left( \frac{2\pi }{k} \right)^2 \, \frac{\G^2(\frac12-\e)}{4\pi^{3 -2\e}} 
\left[ \frac{\d_{\mu\nu}}{ [(x- y)^2]^{1-2\e}} - \pa_\mu \pa_\nu \frac{[(x-y)^2]^{2\e}}{4\e(1+2\e)} \right] \non \\
 =& -  \d_{i}{}^{j} \delta_{\hat k}{}^{\hat\ell }    \left( \frac{2\pi }{k} \right)^2 \, \frac{\G^2(\frac12-\e)\G(\frac12 +\e)}{\G(1-2\e) 2^{1-2\e} \pi^{\frac32 -\e}} 
\, \int \frac{d^np}{(2\pi)^n}   \frac{e^{ip(x-y)}}{(p^2)^{\frac12 +\e}} \left( \d_{\mu \n} - \frac{p_\mu p_\nu}{p^2} \right) 
\end{align}
\end{subequations}
The one-loop fermion propagator reads 
\bea
\label{1fermion}
&& \langle (\psi_I^\a)_{\hat{i}}{}^{\; j}  (x) (\bar{\psi}^J_\b)_k{}^{\; \hat{l}}(y) \rangle^{(1)} =  -i \,\left( \frac{2\pi}{k} \right) \,  \d_I^J \d_{\hat{i}}^{\hat{l}} \d_{k}^{j} \,  \, \d^\a_{\; \, \b}
\, (N_1-N_2) \frac{\G^2(\frac12 - \e)}{16 \pi^{3-2\e}} \, \frac{1}{[(x-y)^2]^{1 - 2\e}}
\non \\
&~& \qquad= - \left( \frac{2\pi i}{k} \right) \, \d_I^J \d_{\hat{i}}^{\hat{l}} \d_{k}^{j} \,  \, \d^\a_{\; \, \b} \, (N_1-N_2) \frac{\G^2(\frac12 - \e) \G(\frac12 + \e)}{\G(1-2\e) 2^{3-2\e} \pi^{\frac32 -\e}} \, 
\int \frac{d^np}{(2\pi)^n}   \frac{e^{ip(x-y)} }{(p^2)^{\frac12 +\e}}
\eea 
and is proportional to the difference $(N_1-N_2)$ of the ranks of the gauge groups. Hence it vanishes in the ABJM limit and is not needed in our computation.

\section{Two--loop self--energy corrections to the fermions}\label{sec:2loopferm}

In this appendix we spell out the computation of the two--loop corrections to the fermion two--point functions. Throughout this section we work with different $N_1, N_2$ group ranks.

\begin{figure}[ht]
\centering 
\includegraphics[scale=0.38]{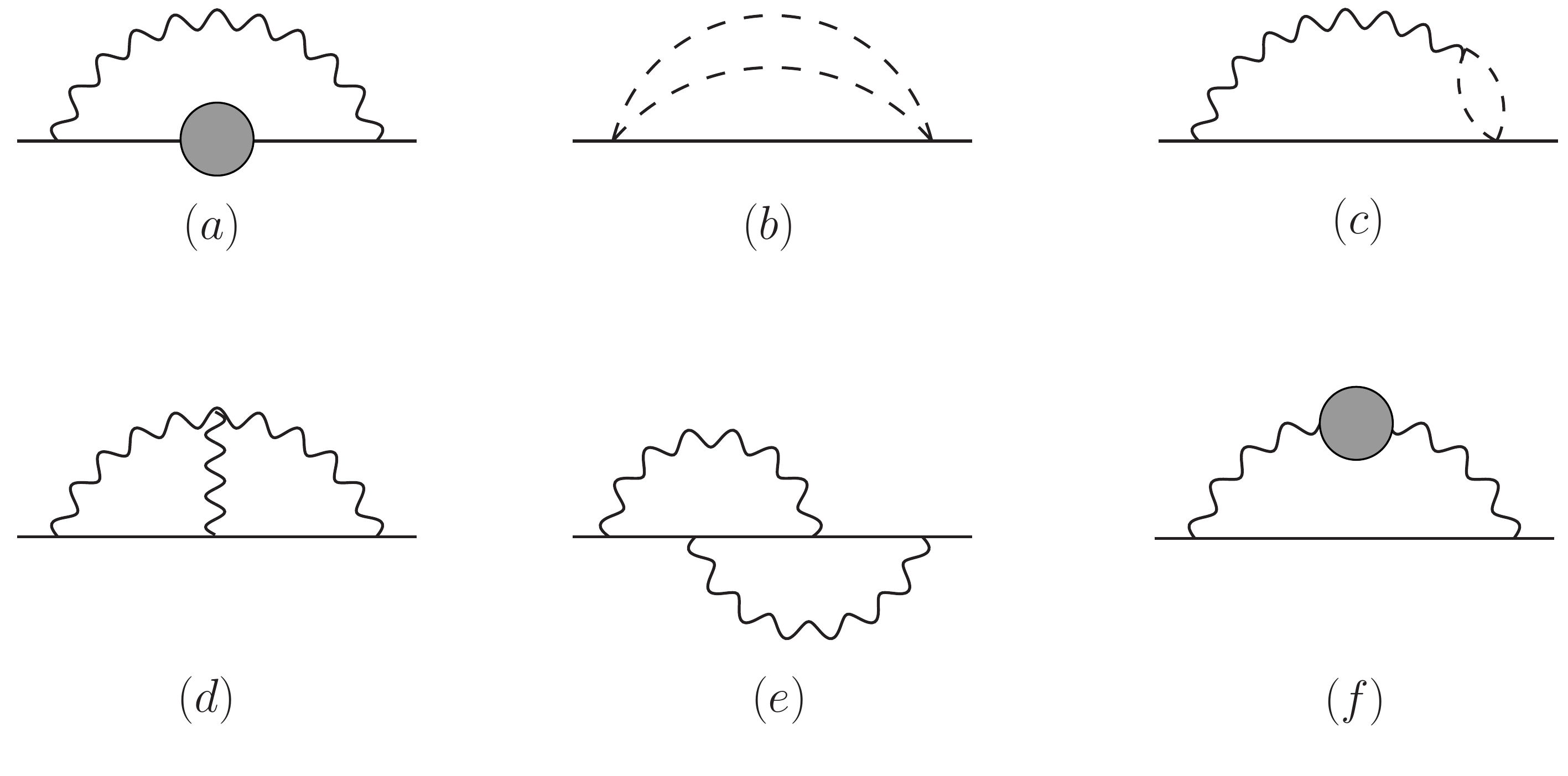}
\caption{Diagrams contributing to the fermion propagator at two loops.} \label{fig:2Lferm}
\end{figure}

The relevant Feynman diagrams are depicted in figure \ref{fig:2Lferm}.
Considering all the possible permutations, the contributions of the diagrams in figure \ref{fig:2Lferm}
are given by
\begin{equation} \label{f2loop}
\langle\psi^\a_I(x)\psi_\b^J(y)\rangle^{(2)}=  \left(\frac{2 \pi}{k}\right)^2\frac{1}{(4\pi)^{3-2\e}} \,\mathcal{C}_{(i)} \,\frac{\G(1/2-\e)^3 \G(2\e)}{\G(3/2-3\e)}\, \int \frac{d^d p}{(2\pi)^d} \, e^{i p\cdot(x-y)}\frac{p^{\a}_{\phantom{\a}\b} \, \d^{\phantom{I}J}_{I}}{(p^2)^{1+2\e}}  
\end{equation}
where 
\begin{align}
\mathcal{C}_{(a)} =& \,\, \frac{2(N_1-N_2)^2}{3} \\
\mathcal{C}_{(b)} = & \,\,  - \frac{56 N_1N_2}{3}   \\ 
\mathcal{C}_{(c)} = & \,\,\,\, 0 \\ 
\mathcal{C}_{(d)} =  & - \frac{2(N_1^2+N_2^2)}{3}\,\,\left(1-\pi \frac{\G(5/2-3\e)\G(1/2+\e)}{\G(2-2\e)\G(\e)\G(1-\e)} \,\right) \\ 
\mathcal{C}_{(e)} = & \frac{ 2N_1N_2}{3}\,\, \left(1-\pi \frac{\G(5/2-3\e)\G(1/2+\e)}{\G(2-2\e)\G(\e)\G(1-\e)} \,\right) \\ 
\mathcal{C}_{(f)}  = & \frac{16 N_1 N_2}{3}\,\frac{1-4\e}{1+2\e}
\end{align}
Summing all the contributions we get
\begin{align}
\S_{i} \, \mathcal{C}_{(i)} & =  - \frac{2N_1N_2(7+30\e)}{(1+2\e)} + \frac{2\pi}{3}(N_1^2+N_2^2-N_1N_2) \frac{\G(5/2-3\e)\G(1/2+\e)}{\G(2-2\e)\G(\e)\G(1-\e)} \non \\[0.1cm]
 & = -14 N_1N_2  + \left(-32 N_1N_2 +(N_1^2 - N_1 N_2 + N_2^2) \frac{\pi^2}{2}\right) \e + \mathcal{O}(\e^2)
\end{align}
\begin{figure}[ht]
\centering     
\includegraphics[scale=0.42]{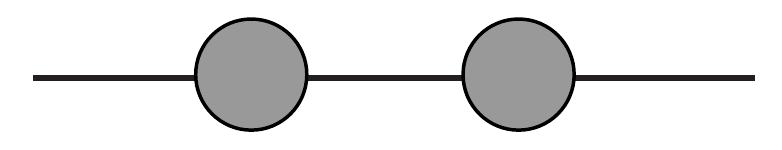}
\caption{Reducible contribution to the two-loop fermion propagator.}  \label{2Lfermred}
\end{figure}
The reducible two-loop correction of figure \ref{2Lfermred} gives
\begin{align} \label{f2loopred}
\langle\psi^\a_I(x)\psi_\b^J(y)\rangle^{(2)}_{red} = & \left(\frac{2 \pi}{k}\right)^2(N_1-N_2)^2 \frac{\G^2(1/2+\e) \G^4(1/2-\e)}{(4\pi)^{3-2\e}\G^2(1-2\e)}   \int \frac{d^d p}{(2\pi)^d} \, e^{i p\cdot(x-y)}\frac{p^{\a}_{\phantom{\a}\b} \, \d^{\phantom{I}J}_{I}}{(p^2)^{1+2\e}} 
\end{align}
Each correction to the fermion propagator can be Fourier transformed back to configuration space
\begin{equation}
\int \frac{d^d p}{(2\pi)^d} \, e^{i p\cdot(x-y)}\frac{p_{\nu} }{(p^2)^{1+2\e}} \rightarrow \frac{\G(1/2-3\e)}{\pi^{3/2-\e}4^{1+2\e}\G(1+2\e)}(-i\partial^x_{\nu})\frac{1}{[(x-y)^2]^{1/2-3\e}}
\end{equation}

\section{Useful identities on the cusp}
\label{App:cusp}

We parametrize a point on the line forming the cusp as 
\beq
x_i^\mu = (0 , \tau_i, 0) = \tau_i\, v^{\mu}
\eeq 
Simple identities that turn out to be useful along the calculation are  
\begin{subequations}
\label{Iden1}
\bea
\label{I1}
&& (x_i - x_j)^2 = \tau_{ij}^2
\\
\label{I2}
&& (x_i \cdot x_j) = \tau_{i} \tau_j
\\
\label{I3}
&&  (\dot{x}_i \cdot \dot{x}_j) = 1
\\
\label{I4}
&& (x_i \cdot \dot{x}_j) = \tau_{i}
\eea
\end{subequations}
The spinor couplings $\eta$, $\bar \eta$ for the fermions and the matrices $M$ for the scalars appearing in the superconnection \eqref{supermatrix} obey the identities (at $\varphi=0$)
\begin{subequations}
\label{Iden2}
\begin{align}
\label{id1}
& n_{1I}n_1^I = n_{2I}n_2^I = 1\\
& n_{1I}n_2^I = \cos \frac{\theta}{2}\label{eq:etagamma2}   \\
& \eta \bar \eta = 2 i\\
& \eta \gamma_{\mu} \bar \eta =  
2i\, \delta^{\mu}_1 = 2i\, v^{\mu} \label{eq:etagamma}\\
& (\eta \gamma_{\mu} \bar \eta)\, x_{ij}^{\mu} = 
2 i \, \tau_{ij}
 \label{eq:etagammax}
 \\
& \Tr({\cal M}_1) = \Tr({\cal M}_2) = \frac12 \Tr({\cal M}_1^2) = \frac12 \Tr({\cal M}_2^2) = \Tr({\cal M}_1^3) = \Tr({\cal M}_2^3)  \nonumber\\ & = \Tr({\cal M}_1^2{\cal M}_2) = \Tr({\cal M}_2^2{\cal M}_1) = 2 \label{eq:traces}\\
& \Tr({\cal M}_1 {\cal M}_2) = 4 \cos^2 \frac{\theta}{2}
\label{matrices} \\
& n_1 {\cal M}_1 \bar n_1 = n_2 {\cal M}_2 \bar n_2 = -1 \\
& n_1 {\cal M}_1 \bar n_2 = n_2 {\cal M}_1 \bar n_1 = n_1 {\cal M}_2 \bar n_2 = n_2 {\cal M}_2 \bar n_1 = -\cos \frac{\theta}{2} \\
& n_1 {\cal M}_2 \bar n_1 = n_2 {\cal M}_1 \bar n_2 = -\cos \theta \label{nMn}
\end{align}
\end{subequations}
Identity \eqref{eq:etagamma}, and the formulae above it, are widely used in the computation, since they apply whenever a diagram contains a fermion arch.
Identities \eqref{matrices} have been used in the diagrams with scalar insertions of figure \ref{fig:scalar}, while identities \eqref{nMn} are relevant for diagram $(3)_d$.

\section{Master integrals definitions and expansions}\label{app:masters}

We define the HQET planar integrals at one, two and three loops by the following products of propagators ($d=3-2\epsilon$)
\begin{align}\label{eq:masterintegrals}
\text{one loop:} & \,\,\, G_{a_1,a_2} \equiv \int \frac{d^dk_1}{(2\pi)^d}\, \frac{1}{(2k_1\cdot \tilde v+1)^{a_1} (k_1^2)^{a_2}} \nonumber\\
\text{two loops:} & \,\,\, G_{a_1,a_2} \equiv \int \frac{d^dk_1\, d^dk_2}{(2\pi)^{2d}}\, \frac{1}{P_1^{a_1}\, P_2^{a_2}\, P_4^{a_3} P_5^{a_4}\, P_7^{a_5}} \nonumber\\
\text{three loops:} &  \,\,\,G_{a_1,\dots,a_9} \equiv \int \frac{d^dk_1\, d^dk_2\, d^dk_3}{(2\pi)^{3d}}\, \prod_{i=1}^{9}\frac{1}{P_i^{a_i}}
\end{align}
where the explicit propagators read
\begin{align}
& P_1 = (2k_1\cdot \tilde v+1) ,\qquad P_2 = (2k_2\cdot \tilde v+1) ,\qquad P_3 = (2k_3\cdot \tilde v+1) \nonumber\\&
P_4 = k_1^2 ,\qquad P_5 = k_2^2 ,\qquad P_6 = k_3^2 \nonumber\\&
P_7 = (k_1-k_2)^2 ,\qquad P_8 = (k_2-k_3)^2 ,\qquad P_9 = (k_1-k_3)^2 
\end{align}
and $\tilde v^2=-1$.
At three loops we also need to consider the non--planar topology 
\begin{align}
   \,\,\,J_{a_1,\dots,a_9} \equiv \int \frac{d^dk_1\, d^dk_2\, d^dk_3}{(2\pi)^{3d}}\, \prod_{i=1}^{9}\frac{1}{Q_i^{a_i}}
\end{align}
with  propagators given by
\begin{align}
& Q_1 = (2k_1\cdot \tilde v+1) ,\qquad Q_2 = (2k_2\cdot \tilde v+1) ,\qquad Q_3 = (2k_3\cdot \tilde v+1) \nonumber\\&
Q_4 = k_1^2 ,\qquad Q_5 = k_2^2 ,\qquad Q_6 = (k_1+k_2-k_3)^2 \nonumber\\&
Q_7 = (k_1-k_2)^2 ,\qquad Q_8 = (k_2-k_3)^2 ,\qquad Q_9 = (k_1-k_3)^2 
\end{align}
At one loop there is only one master integral which can be chosen to be $G_{1,1}$.
The HQET bubble integral evaluates for generic indices \vspace{0.2cm}
\begin{equation}
\begin{minipage}{3cm}\vspace{-0.45cm}
\includegraphics[scale=0.35]{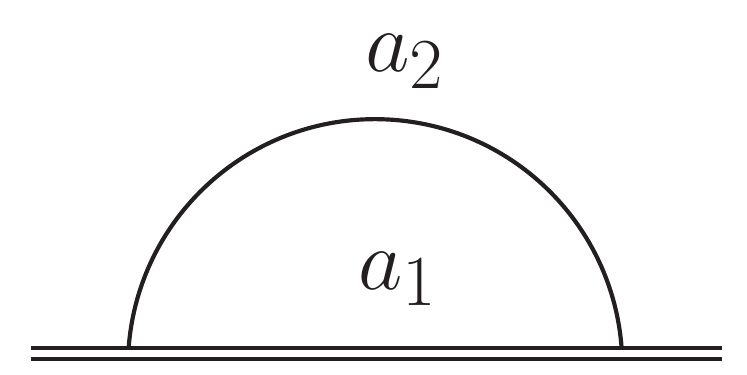}
\end{minipage} = G_{a_1,a_2}  = \frac{1}{(4 \pi)^{d/2}}\frac{\Gamma(a_1+ 2 a_2-d) \Gamma(d/2-a_2)}{\Gamma(a_1)\Gamma(a_2)}
\end{equation}
Fixing the index of the HQET and standard propagators to unity, this formula generalizes to multiloop bubble integrals by integrating bubble sub--topologies iteratively\vspace{0.2cm}
\begin{equation}
\begin{minipage}{3cm}\vspace{-0.45cm}
\includegraphics[scale=0.27]{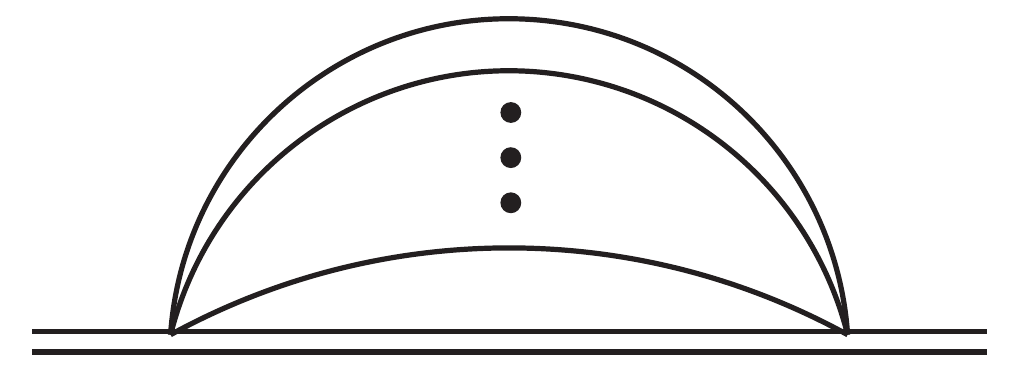}
\end{minipage} = I(n,d) =  \frac{1}{(4 \pi)^{d/2}}\frac{\Gamma(1-n(d-4))\Gamma^n(d/2-1)}{(1-n(d-2))_{2n}} 
\end{equation}
where in the denominator we used the Pochhammer symbol $(a)_n= \Gamma(a+n)/\Gamma(a) $. 
In view of the  three--loop calculation we need the explicit expansion 
\begin{equation}\label{1lmast}
G_{1,1} = \sqrt{\pi} \bigg[\frac{1}{2\e} + \log 2 + \e \left(\frac{7 \pi^2}{24} + \log^2 2 \right)  \bigg] + {\cal O}\left(\epsilon ^2\right)
\end{equation}
omitting a factor $e^{- \gamma_E \epsilon}/(4\pi)^{d/2}$ .

At two loops two master integrals appear which we choose to be $G_{1,0,1,0,1}$ and $G_{1,1,1,1,0}$.
They can be expressed exactly in terms of the function introduced above \vspace{0.3cm}
\begin{align}
\begin{minipage}{3cm}\vspace{-0.45cm}
\includegraphics[scale=0.27]{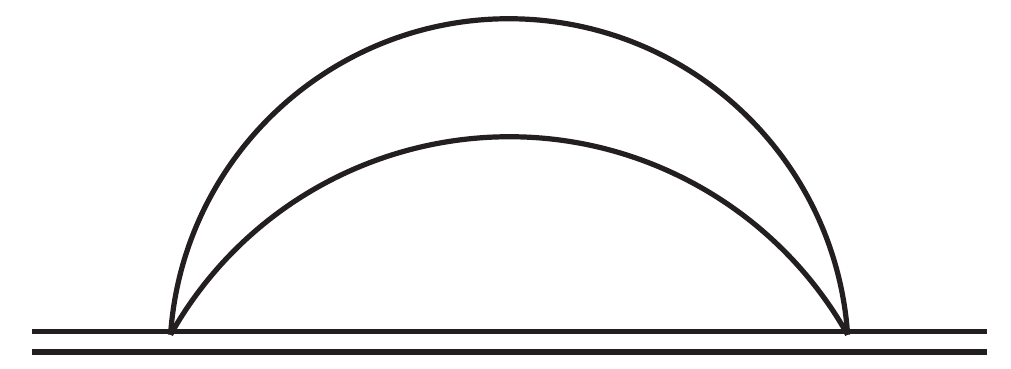}
\end{minipage} = G_{1,0,1,0,1} & = I(2,d)\\[0.3cm]
\begin{minipage}{3cm}\vspace{-0.45cm}
\includegraphics[scale=0.27]{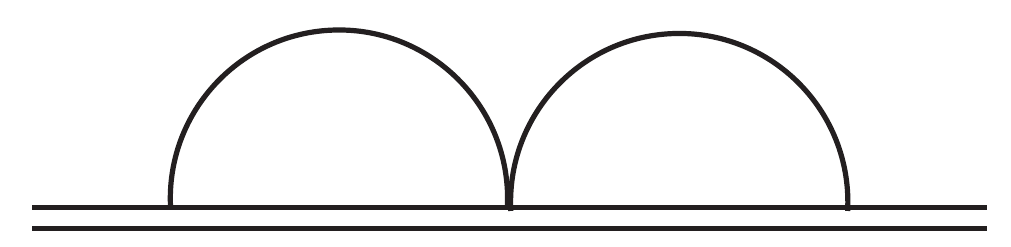}
\end{minipage} = G_{1,1,1,1,0}  & = I(1,d)^2 
\end{align}
These functions can be straightforwardly expanded in power series of $\epsilon$
\begin{align}
G_{1,0,1,0,1} & =\pi \bigg[- \frac{1}{4\e}-\frac{1}{2}(2\log2+2) + \e \left(4  + \frac{11\pi^2}{24} + 4 \log2 + 2 \log^2 2\right) \bigg] + {\cal O}\left(\epsilon^2 \right)  \\ 
 G_{1,1,1,1,0}  & = \pi \bigg[ \frac{1}{4\e^2}+\frac{\log2}{\e} + \left(\frac{7 \pi^2}{24} + 2 \log^2 2 \right)    \bigg] + {\cal O}\left(\epsilon \right)
\end{align}
omitting  $e^{- 2 \gamma_E \epsilon}/(4\pi)^{d}$ .

\begin{figure}[ht]
\begin{center}
\includegraphics[scale=0.42]{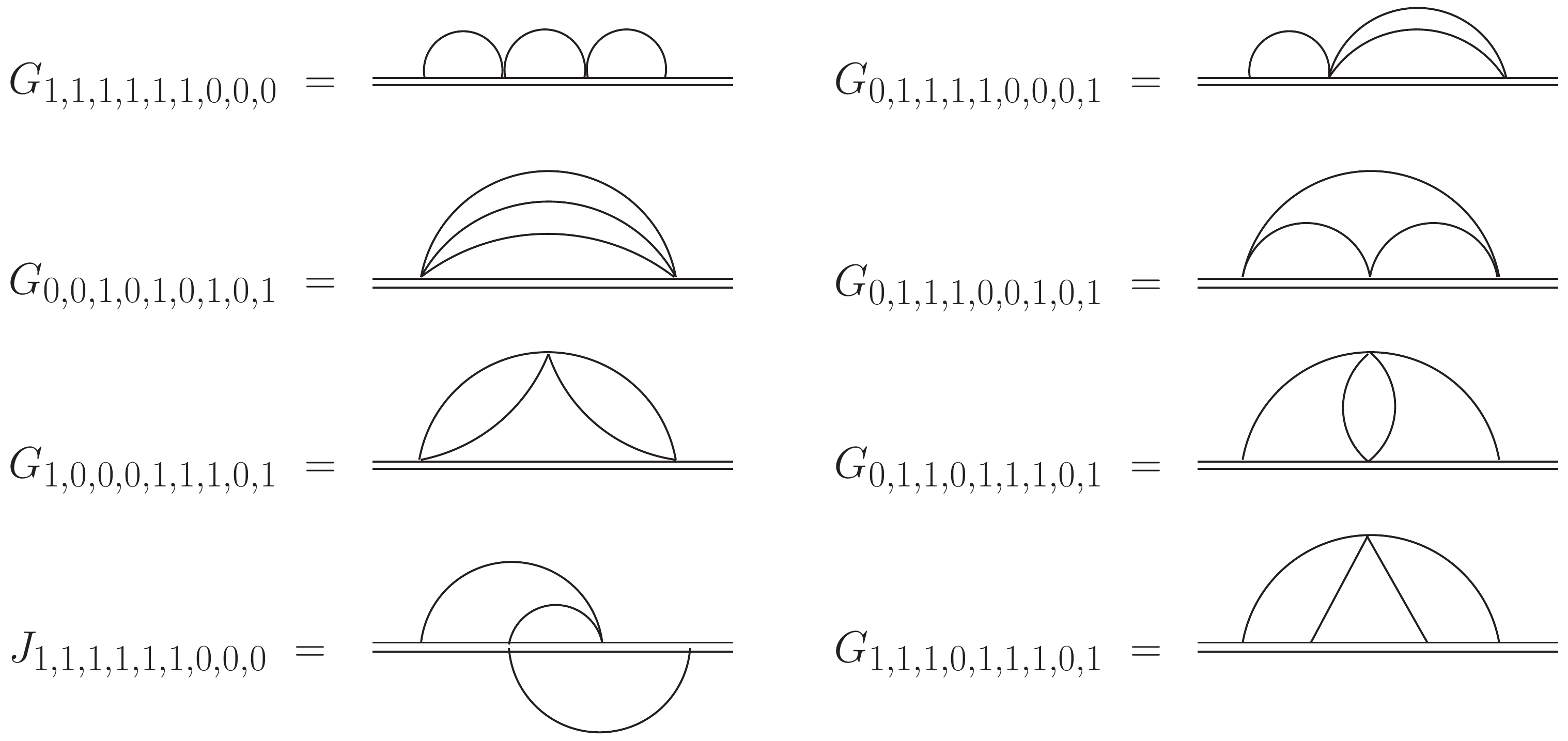}
\caption{Three loop master integrals}
\label{fig:3lmasters}
\end{center}
\end{figure}

At three loops there are eight master integrals that, following \cite{Grozin:2000jv}, we choose to be the ones in figure \ref{fig:3lmasters}. 
The master integrals can be evaluated in $d=3-2\e$ dimensions and expanded up to the needed order
\begin{align}
G_{1,1,1,1,1,1,0,0,0} & = \frac{1}{4}\bigg[\frac{1}{2\e^3}+\frac{3 \log 2}{\e^2}  +\frac{7\pi^2 +72 \log^2 2}{8 \e} + \frac{1}{4} (21 \pi^2 \log 2+ 72 \log^3 2 -2 \zeta (3) )  \bigg] \non \\&
 + {\cal O}\left(\epsilon \right) \non \\
G_{0,1,1,1,1,0,0,0,1} & = - \frac{1}{4}\bigg[\frac{1}{2 \e^2}+\frac{4 + 6 \log 2}{2\e}   + \frac{1}{24} (192+ 29 \pi^2 + 216 \log^2 2 +288 \log2 )   \bigg]+ {\cal O}\left(\epsilon \right)\non\\
G_{0,0,1,0,1,0,1,0,1} & =\frac{1}{4}\bigg[\frac{1}{3\e}+ 3 + 2\log2   + \frac{\e}{4} (5 \pi^2 +6(14+2\log2 (6+2\log2)) )  \bigg] + {\cal O}\left(\epsilon^2 \right)\non  \\
G_{0,1,1,1,0,0,1,0,1} & =  -\frac{1}{6}\bigg[\frac{1}{\e^2}+\frac{6 + 6 \log 2}{\e}   + \frac{1}{12} (37 \pi^2 +216(2+\log^2 2 +2\log2) )  \bigg] + {\cal O}\left(\epsilon \right) \non \\
G_{1,0,0,0,1,1,1,0,1} & = \frac{1}{6}\bigg[\frac{\pi^2}{\e} + 10 \pi^2 \log2  \bigg] + {\cal O}\left(\epsilon \right) \non  \\
G_{0,1,1,0,1,1,1,0,1} & = 14  \zeta(3) + {\cal O}\left(\epsilon \right)\non \\
G_{1,1,1,0,1,1,1,0,1} & =\frac{2\pi^{2}}{3} + {\cal O}\left(\epsilon \right) \non  \\
J_{1,1,1,1,1,1,0,0,0} & = \non \frac{1}{4}\bigg[\frac{1}{6\e^3}+\frac{\log 2}{\e^2}+\frac{7 \pi^2 + 72 \log^2 2 }{24\e}\bigg] + {\cal O}\left(\epsilon^0 \right)
\end{align}
where an overall factor $\pi^{3/2}\, e^{-3 \gamma_E \epsilon}/(4\pi)^{3d/2}$ is omitted.

\section{Results for two--loop diagrams}\label{app:diagrams2loops}

Here we provide the list of  results for the two--loop diagrams of figure \ref{2ldiag}, considering only the  1PI configurations on the cusp. 
\begin{align}
 (2)_a& =  \frac{ N^2 C_{\theta}^2}{4 \e}+{\cal O}\left(\epsilon \right)  \\
 (2)_b& = \frac{N^2}{2}\bigg[-\frac{1}{\e}+ 3  \bigg] +{\cal O}\left(\epsilon \right)  \\
 (2)_c& =  0 \\
(2)_d& =  \frac{N^2}{4}\bigg[\frac{C_{\theta}}{\e^2}+ \frac{-4C_{\theta}+1}{\e}+ 4(C_{\theta}-1) + \frac{\pi^2}{6}(9C_{\theta}-2)  \bigg] +{\cal O}\left(\epsilon \right)  \\
(2)_{e}  &= \frac{N^2}{4}\bigg[\frac{C_{\theta}( C_{\theta}-2)}{2\e^2}+\frac{C_{\theta}(4-3C_{\theta})}{\e} + 8C_{\theta}(C_{\theta}-1) \non \\
& \hspace{1cm}  +\frac{11 \pi^2 C_{\theta}}{12}(-2+C_{\theta})\bigg] +{\cal O}\left(\epsilon \right)
\end{align}
A common factor $\left(\frac{(e^{-\gamma_E}16\pi)^{\epsilon}}{k}\right)^2$ is omitted.

\section{Results for three--loop diagrams}\label{app:diagrams}

Here we list the results for the HQET-1PI part of the diagrams. The results include the nonplanar corrections and are already given by the sum over different configurations belonging to the same topology, following the classification of section \ref{sec:diag3l}.
A common factor $\left(\frac{(e^{-\gamma_E}16\pi)^{\epsilon}}{k}\right)^3$ is understood.

\begin{align}
(3)_a & = -\frac{N \left(N^2+1\right)}{24 \epsilon ^2}
+{\cal O}\left(\epsilon ^0\right) \\[0.2cm]
(3)_b &= \frac{N^3+N}{24 \epsilon ^2}+\frac{N \left(N^2+1\right) \pi ^2}{96 \epsilon }
+{\cal O}\left(\epsilon ^0\right)
\\[0.2cm]
(3)_c &= \frac{N C_{\theta } \left(4 C_{\theta }+N^2 \left(C_{\theta } \left(3 C_{\theta }-4\right)-2\right)+2\right)}{48 \epsilon ^2}
+\nonumber \\&
+\frac{N C_{\theta } \left(C_{\theta } \left(4 C_{\theta }-7 N^2 C_{\theta }+4 N^2-4\right)+2 \left(N^2-1\right)\right)}{24 \epsilon }
+{\cal O}\left(\epsilon ^0\right)
\\[0.2cm]
(3)_d & = \frac{N \left(N^2-1\right) C_{\theta } \left(C_{\theta }+1\right)}{6 \epsilon ^2}-\frac{4 \left(N \left(N^2-1\right) C_{\theta }\right)}{6 \epsilon }
+{\cal O}\left(\epsilon ^0\right)
\\[0.2cm]
 (3)_e & = \frac{N^3 C_{\theta }}{12 \epsilon ^3}+\frac{8 N^3+\left(N-25 N^3\right) C_{\theta }}{48 \epsilon ^2}
+ \\&
+\frac{N \left(\left(\left(318+67 \pi ^2\right) N^2-\pi ^2-30\right) C_{\theta }-\left(324+23 \pi ^2\right) N^2-3 \pi ^2+36\right)}{288 \epsilon }
+{\cal O}\left(\epsilon ^0\right)\nonumber
\\[0.2cm]
 (3)_f & = \frac{ N \left(1-N^2\right)}{3 \epsilon ^2}-\frac{N \left(N^2-1\right) \left(2 \left(-8+\pi ^2\right) C_{\theta }-\pi ^2-16\right)}{12 \epsilon }
+{\cal O}\left(\epsilon ^0\right)
\\[0.2cm]
 (3)_g & = \frac{N \left(N^2-1\right) \left(\left(-84+9 \pi ^2\right) C_{\theta }-\pi ^2+24\right)}{576 \epsilon }
+{\cal O}\left(\epsilon ^0\right)
\\[0.2cm]
(3)_h & = \frac{N \left(N^2-1\right) \left(\left(-84+9 \pi ^2\right) C_{\theta }+\pi ^2+24\right)}{288 \epsilon }
+{\cal O}\left(\epsilon ^0\right)
\\[0.2cm]
 (3)_i &= -\frac{N \left(\left(N^2+2\right) C_{\theta }-4 N^2+4\right)}{12 \epsilon ^2}+\frac{N \left(\left(13 N^2+2\right) C_{\theta }-20 \left(N^2-1\right)\right)}{12 \epsilon }
+{\cal O}\left(\epsilon ^0\right)
\\[0.2cm]
 (3)_j & = -\frac{7 N \left(N^2-1\right) C_{\theta }}{48 \epsilon ^2}+\frac{N \left(N^2-1\right) \left(-64+\pi ^2\right) C_{\theta }}{192 \epsilon }
+{\cal O}\left(\epsilon ^0\right)
\\[0.2cm]
 (3)_k &= \frac{N \left(N^2-1\right) C_{\theta }}{24 \epsilon ^3}-\frac{N \left(N^2-1\right) \left(7 C_{\theta }-2\right)}{24 \epsilon ^2}
\\&
+\frac{N \left(N^2-1\right)}{288 \epsilon } \Big[8 \left(-6+\pi ^2\right) C_{\theta }^2+\left(264+29 \pi ^2\right) C_{\theta }-16 \left(9+\pi ^2\right)\Big]
+{\cal O}\left(\epsilon ^0\right)\nonumber
\\[0.2cm]
 (3)_l& = \frac{N C_{\theta } \left(3 C_{\theta } N^2-7 N^2+1\right)}{24 \epsilon ^3}+\frac{N \left(-26 C_{\theta }^2 N^2+\left(42 N^2-3\right) C_{\theta }-6 N^2+2\right)}{24 \epsilon ^2}
\nonumber\\&
+\frac{N}{288 \epsilon } \Big[\left(\left(1344+107 \pi ^2\right) N^2+16 \left(-3+\pi ^2\right)\right) C_{\theta }^2
\\&
-3 \left(\left(624+85 \pi ^2\right) N^2+\pi ^2-72\right) C_{\theta }+24 \left(N^2 \left(22+\pi ^2\right)-10\right)\Big]
+{\cal O}\left(\epsilon ^0\right)\nonumber
\\[0.2cm]
 (3)_m & = \frac{N^3 \left(C_{\theta }-4\right) \left(C_{\theta }-2\right) C_{\theta }}{96 \epsilon ^3}-\frac{N C_{\theta } \left(N^2 \left(6 C_{\theta }^2-26 C_{\theta }+22\right)+2\right)}{48 \epsilon ^2}
\nonumber\\&
+\frac{N C_{\theta }}{1152 \epsilon } \bigg[48 \left(C_{\theta } \left(\left(19 N^2-4\right) C_{\theta }-64 N^2+4\right)+46 N^2+2\right)
 \\&
+\pi ^2 \left(16 \left(\left(C_{\theta }-2\right) C_{\theta }+2\right)+N^2 \left(5 C_{\theta }-12\right) \left(9 C_{\theta }-26\right)\right)\bigg]
+{\cal O}\left(\epsilon ^0\right) \nonumber
\end{align}

\bibliographystyle{JHEP}

\bibliography{biblio}

\providecommand{\href}[2]{#2}\begingroup\raggedright\begin{thebibliography}{10}

\bibitem{Bianchi:2014laa}
M.~S. Bianchi, L.~Griguolo, M.~Leoni, S.~Penati and D.~Seminara, \emph{{BPS
  Wilson loops and Bremsstrahlung function in ABJ(M): a two loop analysis}},
  \href{http://dx.doi.org/10.1007/JHEP06(2014)123}{\emph{JHEP} {\bfseries 06}
  (2014) 123}, [\href{https://arxiv.org/abs/1402.4128}{{\ttfamily 1402.4128}}].

\bibitem{Maldacena:1997re}
J.~M. Maldacena, \emph{{The Large N limit of superconformal field theories and
  supergravity}}, {\emph{Adv.Theor.Math.Phys.} {\bfseries 2} (1998) 231--252},
  [\href{https://arxiv.org/abs/hep-th/9711200}{{\ttfamily hep-th/9711200}}].

\bibitem{Witten:1998qj}
E.~Witten, \emph{Anti-de {S}itter space and holography}, {\emph{Adv. Theor.
  Math. Phys.} {\bfseries 2} (1998) 253--291},
  [\href{https://arxiv.org/abs/hep-th/9802150}{{\ttfamily hep-th/9802150}}].

\bibitem{Gubser:1998bc}
S.~S. Gubser, I.~R. Klebanov and A.~M. Polyakov, \emph{Gauge theory correlators
  from non-critical string theory},
  \href{http://dx.doi.org/10.1016/S0370-2693(98)00377-3}{\emph{Phys. Lett.}
  {\bfseries B428} (1998) 105--114},
  [\href{https://arxiv.org/abs/hep-th/9802109}{{\ttfamily hep-th/9802109}}].

\bibitem{Maldacena:1998im}
J.~M. Maldacena, \emph{{W}ilson loops in large {N} field theories},
  \href{http://dx.doi.org/10.1103/PhysRevLett.80.4859}{\emph{Phys.Rev.Lett.}
  {\bfseries 80} (1998) 4859--4862},
  [\href{https://arxiv.org/abs/hep-th/9803002}{{\ttfamily hep-th/9803002}}].

\bibitem{Rey:1998ik}
S.-J. Rey and J.-T. Yee, \emph{{Macroscopic strings as heavy quarks in large N
  gauge theory and anti-de Sitter supergravity}},
  \href{http://dx.doi.org/10.1007/s100520100799}{\emph{Eur.Phys.J.} {\bfseries
  C22} (2001) 379--394},
  [\href{https://arxiv.org/abs/hep-th/9803001}{{\ttfamily hep-th/9803001}}].

\bibitem{Pestun:2007rz}
V.~Pestun, \emph{{Localization of gauge theory on a four-sphere and
  supersymmetric Wilson loops}},
  \href{http://dx.doi.org/10.1007/s00220-012-1485-0}{\emph{Commun. Math. Phys.}
  {\bfseries 313} (2012) 71--129},
  [\href{https://arxiv.org/abs/0712.2824}{{\ttfamily 0712.2824}}].

\bibitem{Kapustin:2009kz}
A.~Kapustin, B.~Willett and I.~Yaakov, \emph{{Exact Results for Wilson Loops in
  Superconformal Chern-Simons Theories with Matter}},
  \href{http://dx.doi.org/10.1007/JHEP03(2010)089}{\emph{JHEP} {\bfseries 1003}
  (2010) 089}, [\href{https://arxiv.org/abs/0909.4559}{{\ttfamily 0909.4559}}].

\bibitem{Erickson:2000af}
J.~Erickson, G.~Semenoff and K.~Zarembo, \emph{{W}ilson loops in {$\mathcal{N}
  = 4$} supersymmetric {Y}ang-{M}ills theory},
  \href{http://dx.doi.org/10.1016/S0550-3213(00)00300-X}{\emph{Nucl.Phys.}
  {\bfseries B582} (2000) 155--175},
  [\href{https://arxiv.org/abs/hep-th/0003055}{{\ttfamily hep-th/0003055}}].

\bibitem{Drukker:2000rr}
N.~Drukker and D.~J. Gross, \emph{{An Exact prediction of N=4 SUSYM theory for
  string theory}}, \href{http://dx.doi.org/10.1063/1.1372177}{\emph{J. Math.
  Phys.} {\bfseries 42} (2001) 2896--2914},
  [\href{https://arxiv.org/abs/hep-th/0010274}{{\ttfamily hep-th/0010274}}].

\bibitem{Correa:2012at}
D.~Correa, J.~Henn, J.~Maldacena and A.~Sever, \emph{{An exact formula for the
  radiation of a moving quark in N=4 super Yang Mills}},
  \href{http://dx.doi.org/10.1007/JHEP06(2012)048}{\emph{JHEP} {\bfseries 06}
  (2012) 048}, [\href{https://arxiv.org/abs/1202.4455}{{\ttfamily 1202.4455}}].

\bibitem{Drukker:1999zq}
N.~Drukker, D.~J. Gross and H.~Ooguri, \emph{{W}ilson loops and minimal
  surfaces},
  \href{http://dx.doi.org/10.1103/PhysRevD.60.125006}{\emph{Phys.Rev.}
  {\bfseries D60} (1999) 125006},
  [\href{https://arxiv.org/abs/hep-th/9904191}{{\ttfamily hep-th/9904191}}].

\bibitem{Drukker:2011za}
N.~Drukker and V.~Forini, \emph{{Generalized quark-antiquark potential at weak
  and strong coupling}},
  \href{http://dx.doi.org/10.1007/JHEP06(2011)131}{\emph{JHEP} {\bfseries 06}
  (2011) 131}, [\href{https://arxiv.org/abs/1105.5144}{{\ttfamily 1105.5144}}].

\bibitem{Bombardelli:2009ns}
D.~Bombardelli, D.~Fioravanti and R.~Tateo, \emph{{Thermodynamic Bethe Ansatz
  for planar AdS/CFT: A Proposal}},
  \href{http://dx.doi.org/10.1088/1751-8113/42/37/375401}{\emph{J. Phys.}
  {\bfseries A42} (2009) 375401},
  [\href{https://arxiv.org/abs/0902.3930}{{\ttfamily 0902.3930}}].

\bibitem{Gromov:2009bc}
N.~Gromov, V.~Kazakov, A.~Kozak and P.~Vieira, \emph{{Exact Spectrum of
  Anomalous Dimensions of Planar N = 4 Supersymmetric Yang-Mills Theory: TBA
  and excited states}},
  \href{http://dx.doi.org/10.1007/s11005-010-0374-8}{\emph{Lett. Math. Phys.}
  {\bfseries 91} (2010) 265--287},
  [\href{https://arxiv.org/abs/0902.4458}{{\ttfamily 0902.4458}}].

\bibitem{Arutyunov:2009ur}
G.~Arutyunov and S.~Frolov, \emph{{Thermodynamic Bethe Ansatz for the AdS(5)
  $\times $ S(5) Mirror Model}},
  \href{http://dx.doi.org/10.1088/1126-6708/2009/05/068}{\emph{JHEP} {\bfseries
  05} (2009) 068}, [\href{https://arxiv.org/abs/0903.0141}{{\ttfamily
  0903.0141}}].

\bibitem{Correa:2012hh}
D.~Correa, J.~Maldacena and A.~Sever, \emph{{The quark anti-quark potential and
  the cusp anomalous dimension from a TBA equation}},
  \href{http://dx.doi.org/10.1007/JHEP08(2012)134}{\emph{JHEP} {\bfseries 08}
  (2012) 134}, [\href{https://arxiv.org/abs/1203.1913}{{\ttfamily 1203.1913}}].

\bibitem{Drukker:2012de}
N.~Drukker, \emph{{Integrable Wilson loops}},
  \href{http://dx.doi.org/10.1007/JHEP10(2013)135}{\emph{JHEP} {\bfseries 10}
  (2013) 135}, [\href{https://arxiv.org/abs/1203.1617}{{\ttfamily 1203.1617}}].

\bibitem{Gromov:2012eu}
N.~Gromov and A.~Sever, \emph{{Analytic Solution of Bremsstrahlung TBA}},
  \href{http://dx.doi.org/10.1007/JHEP11(2012)075}{\emph{JHEP} {\bfseries 11}
  (2012) 075}, [\href{https://arxiv.org/abs/1207.5489}{{\ttfamily 1207.5489}}].

\bibitem{Gromov:2013qga}
N.~Gromov, F.~Levkovich-Maslyuk and G.~Sizov, \emph{{Analytic Solution of
  Bremsstrahlung TBA II: Turning on the Sphere Angle}},
  \href{http://dx.doi.org/10.1007/JHEP10(2013)036}{\emph{JHEP} {\bfseries 10}
  (2013) 036}, [\href{https://arxiv.org/abs/1305.1944}{{\ttfamily 1305.1944}}].

\bibitem{Bonini:2015fng}
M.~Bonini, L.~Griguolo, M.~Preti and D.~Seminara, \emph{{Bremsstrahlung
  function, leading Lüscher correction at weak coupling and localization}},
  \href{http://dx.doi.org/10.1007/JHEP02(2016)172}{\emph{JHEP} {\bfseries 02}
  (2016) 172}, [\href{https://arxiv.org/abs/1511.05016}{{\ttfamily
  1511.05016}}].

\bibitem{Gromov:2013pga}
N.~Gromov, V.~Kazakov, S.~Leurent and D.~Volin, \emph{{Quantum Spectral Curve
  for Planar $\mathcal{N} =$ Super-Yang-Mills Theory}},
  \href{http://dx.doi.org/10.1103/PhysRevLett.112.011602}{\emph{Phys. Rev.
  Lett.} {\bfseries 112} (2014) 011602},
  [\href{https://arxiv.org/abs/1305.1939}{{\ttfamily 1305.1939}}].

\bibitem{Gromov:2014caa}
N.~Gromov, V.~Kazakov, S.~Leurent and D.~Volin, \emph{{Quantum spectral curve
  for arbitrary state/operator in AdS$_{5}$/CFT$_{4}$}},
  \href{http://dx.doi.org/10.1007/JHEP09(2015)187}{\emph{JHEP} {\bfseries 09}
  (2015) 187}, [\href{https://arxiv.org/abs/1405.4857}{{\ttfamily 1405.4857}}].

\bibitem{Gromov:2015dfa}
N.~Gromov and F.~Levkovich-Maslyuk, \emph{{Quantum Spectral Curve for a cusped
  Wilson line in $ \mathcal{N}=4 $ SYM}},
  \href{http://dx.doi.org/10.1007/JHEP04(2016)134}{\emph{JHEP} {\bfseries 04}
  (2016) 134}, [\href{https://arxiv.org/abs/1510.02098}{{\ttfamily
  1510.02098}}].

\bibitem{Fiol:2015spa}
B.~Fiol, E.~Gerchkovitz and Z.~Komargodski, \emph{{Exact Bremsstrahlung
  Function in $N=2$ Superconformal Field Theories}},
  \href{http://dx.doi.org/10.1103/PhysRevLett.116.081601}{\emph{Phys. Rev.
  Lett.} {\bfseries 116} (2016) 081601},
  [\href{https://arxiv.org/abs/1510.01332}{{\ttfamily 1510.01332}}].

\bibitem{Aharony:2008ug}
O.~Aharony, O.~Bergman, D.~L. Jafferis and J.~Maldacena, \emph{{$\mathcal{N} =
  6$} superconformal {C}hern-{S}imons-matter theories, {M2}-branes and their
  gravity duals},
  \href{http://dx.doi.org/10.1088/1126-6708/2008/10/091}{\emph{JHEP} {\bfseries
  0810} (2008) 091}, [\href{https://arxiv.org/abs/0806.1218}{{\ttfamily
  0806.1218}}].

\bibitem{Aharony:2008gk}
O.~Aharony, O.~Bergman and D.~L. Jafferis, \emph{{Fractional M2-branes}},
  \href{http://dx.doi.org/10.1088/1126-6708/2008/11/043}{\emph{JHEP} {\bfseries
  0811} (2008) 043}, [\href{https://arxiv.org/abs/0807.4924}{{\ttfamily
  0807.4924}}].

\bibitem{Berenstein:2008dc}
D.~Berenstein and D.~Trancanelli, \emph{Three-dimensional {$\mathcal{N} = 6$}
  {SCFT}'s and their membrane dynamics},
  \href{http://dx.doi.org/10.1103/PhysRevD.78.106009}{\emph{Phys. Rev.}
  {\bfseries D78} (2008) 106009},
  [\href{https://arxiv.org/abs/0808.2503}{{\ttfamily 0808.2503}}].

\bibitem{Drukker:2008zx}
N.~Drukker, J.~Plefka and D.~Young, \emph{{Wilson loops in 3-dimensional N=6
  supersymmetric Chern-Simons Theory and their string theory duals}},
  \href{http://dx.doi.org/10.1088/1126-6708/2008/11/019}{\emph{JHEP} {\bfseries
  11} (2008) 019}, [\href{https://arxiv.org/abs/0809.2787}{{\ttfamily
  0809.2787}}].

\bibitem{Chen:2008bp}
B.~Chen and J.-B. Wu, \emph{{Supersymmetric Wilson Loops in N=6 Super
  Chern-Simons-matter theory}},
  \href{http://dx.doi.org/10.1016/j.nuclphysb.2009.09.015}{\emph{Nucl. Phys.}
  {\bfseries B825} (2010) 38--51},
  [\href{https://arxiv.org/abs/0809.2863}{{\ttfamily 0809.2863}}].

\bibitem{Rey:2008bh}
S.-J. Rey, T.~Suyama and S.~Yamaguchi, \emph{{Wilson Loops in Superconformal
  Chern-Simons Theory and Fundamental Strings in Anti-de Sitter Supergravity
  Dual}}, \href{http://dx.doi.org/10.1088/1126-6708/2009/03/127}{\emph{JHEP}
  {\bfseries 0903} (2009) 127},
  [\href{https://arxiv.org/abs/0809.3786}{{\ttfamily 0809.3786}}].

\bibitem{Drukker:2009hy}
N.~Drukker and D.~Trancanelli, \emph{{A Supermatrix model for N=6 super
  Chern-Simons-matter theory}},
  \href{http://dx.doi.org/10.1007/JHEP02(2010)058}{\emph{JHEP} {\bfseries 02}
  (2010) 058}, [\href{https://arxiv.org/abs/0912.3006}{{\ttfamily 0912.3006}}].

\bibitem{Griguolo:2012iq}
L.~Griguolo, D.~Marmiroli, G.~Martelloni and D.~Seminara, \emph{{The
  generalized cusp in ABJ(M) N = 6 Super Chern-Simons theories}},
  \href{http://dx.doi.org/10.1007/JHEP05(2013)113}{\emph{JHEP} {\bfseries 05}
  (2013) 113}, [\href{https://arxiv.org/abs/1208.5766}{{\ttfamily 1208.5766}}].

\bibitem{Lewkowycz:2013laa}
A.~Lewkowycz and J.~Maldacena, \emph{{Exact results for the entanglement
  entropy and the energy radiated by a quark}},
  \href{http://dx.doi.org/10.1007/JHEP05(2014)025}{\emph{JHEP} {\bfseries 05}
  (2014) 025}, [\href{https://arxiv.org/abs/1312.5682}{{\ttfamily 1312.5682}}].

\bibitem{Marino:2009jd}
M.~Marino and P.~Putrov, \emph{{Exact Results in ABJM Theory from Topological
  Strings}}, \href{http://dx.doi.org/10.1007/JHEP06(2010)011}{\emph{JHEP}
  {\bfseries 1006} (2010) 011},
  [\href{https://arxiv.org/abs/0912.3074}{{\ttfamily 0912.3074}}].

\bibitem{Drukker:2010nc}
N.~Drukker, M.~Marino and P.~Putrov, \emph{{From weak to strong coupling in
  ABJM theory}},
  \href{http://dx.doi.org/10.1007/s00220-011-1253-6}{\emph{Commun.Math.Phys.}
  {\bfseries 306} (2011) 511--563},
  [\href{https://arxiv.org/abs/1007.3837}{{\ttfamily 1007.3837}}].

\bibitem{Forini:2012bb}
V.~Forini, V.~G.~M. Puletti and O.~Ohlsson~Sax, \emph{{The generalized cusp in
  $AdS_4 \times CP^3$ and more one-loop results from semiclassical strings}},
  \href{http://dx.doi.org/10.1088/1751-8113/46/11/115402}{\emph{J. Phys.}
  {\bfseries A46} (2013) 115402},
  [\href{https://arxiv.org/abs/1204.3302}{{\ttfamily 1204.3302}}].

\bibitem{Aguilera-Damia:2014bqa}
J.~Aguilera-Damia, D.~H. Correa and G.~A. Silva, \emph{{Semiclassical partition
  function for strings dual to Wilson loops with small cusps in ABJM}},
  \href{http://dx.doi.org/10.1007/JHEP03(2015)002}{\emph{JHEP} {\bfseries 03}
  (2015) 002}, [\href{https://arxiv.org/abs/1412.4084}{{\ttfamily 1412.4084}}].

\bibitem{Correa:2014aga}
D.~H. Correa, J.~Aguilera-Damia and G.~A. Silva, \emph{{Strings in $AdS_4
  \times \mathbb{CP}^{3}$ Wilson loops in $\mathcal N=$6 super
  Chern-Simons-matter and bremsstrahlung functions}},
  \href{http://dx.doi.org/10.1007/JHEP06(2014)139}{\emph{JHEP} {\bfseries 06}
  (2014) 139}, [\href{https://arxiv.org/abs/1405.1396}{{\ttfamily 1405.1396}}].

\bibitem{Korchemsky:1987wg}
G.~P. Korchemsky and A.~V. Radyushkin, \emph{{Renormalization of the Wilson
  Loops Beyond the Leading Order}},
  \href{http://dx.doi.org/10.1016/0550-3213(87)90277-X}{\emph{Nucl. Phys.}
  {\bfseries B283} (1987) 342--364}.

\bibitem{Gervais:1979fv}
J.-L. Gervais and A.~Neveu, \emph{{The Slope of the Leading Regge Trajectory in
  Quantum Chromodynamics}},
  \href{http://dx.doi.org/10.1016/0550-3213(80)90397-1}{\emph{Nucl. Phys.}
  {\bfseries B163} (1980) 189--216}.

\bibitem{Dotsenko:1979wb}
V.~S. Dotsenko and S.~N. Vergeles, \emph{{Renormalizability of Phase Factors in
  the Nonabelian Gauge Theory}},
  \href{http://dx.doi.org/10.1016/0550-3213(80)90103-0}{\emph{Nucl. Phys.}
  {\bfseries B169} (1980) 527--546}.

\bibitem{Arefeva:1980zd}
I.~{\relax Ya}. Arefeva, \emph{{Quantum Contour Field Equations}},
  \href{http://dx.doi.org/10.1016/0370-2693(80)90529-8}{\emph{Phys. Lett.}
  {\bfseries B93} (1980) 347--353}.

\bibitem{Gatheral:1983cz}
J.~G.~M. Gatheral, \emph{{Exponentiation of Eikonal Cross-sections in
  Nonabelian Gauge Theories}},
  \href{http://dx.doi.org/10.1016/0370-2693(83)90112-0}{\emph{Phys. Lett.}
  {\bfseries B133} (1983) 90--94}.

\bibitem{Klemm:2012ii}
A.~Klemm, M.~Marino, M.~Schiereck and M.~Soroush,
  \emph{{Aharony-Bergman-Jafferis-Maldacena Wilson loops in the Fermi gas
  approach}}, \href{http://dx.doi.org/10.5560/ZNA.2012-0118}{\emph{Z.
  Naturforsch.} {\bfseries A68} (2013) 178--209},
  [\href{https://arxiv.org/abs/1207.0611}{{\ttfamily 1207.0611}}].

\bibitem{Bianchi:2016yzj}
M.~S. Bianchi, L.~Griguolo, M.~Leoni, A.~Mauri, S.~Penati and D.~Seminara,
  \emph{{Framing and localization in Chern-Simons theories with matter}},
  \href{http://dx.doi.org/10.1007/JHEP06(2016)133}{\emph{JHEP} {\bfseries 06}
  (2016) 133}, [\href{https://arxiv.org/abs/1604.00383}{{\ttfamily
  1604.00383}}].

\bibitem{Bianchi:2016vvm}
M.~S. Bianchi, L.~Griguolo, M.~Leoni, A.~Mauri, S.~Penati and D.~Seminara,
  \emph{{The quantum 1/2 BPS Wilson loop in ${\cal N}=4$ Chern-Simons-matter
  theories}}, \href{http://dx.doi.org/10.1007/JHEP09(2016)009}{\emph{JHEP}
  {\bfseries 09} (2016) 009},
  [\href{https://arxiv.org/abs/1606.07058}{{\ttfamily 1606.07058}}].

\bibitem{Chetyrkin:1980pr}
K.~G. Chetyrkin, A.~L. Kataev and F.~V. Tkachov, \emph{{New Approach to
  Evaluation of Multiloop Feynman Integrals: The Gegenbauer Polynomial $\times$
  Space Technique}},
  \href{http://dx.doi.org/10.1016/0550-3213(80)90289-8}{\emph{Nucl. Phys.}
  {\bfseries B174} (1980) 345--377}.

\bibitem{Mauri:2013vd}
A.~Mauri, A.~Santambrogio and S.~Scoleri, \emph{{The Leading Order Dressing
  Phase in ABJM Theory}},
  \href{http://dx.doi.org/10.1007/JHEP04(2013)146}{\emph{JHEP} {\bfseries 04}
  (2013) 146}, [\href{https://arxiv.org/abs/1301.7732}{{\ttfamily 1301.7732}}].

\bibitem{Grozin:2014hna}
A.~Grozin, J.~M. Henn, G.~P. Korchemsky and P.~Marquard, \emph{{Three Loop Cusp
  Anomalous Dimension in QCD}},
  \href{http://dx.doi.org/10.1103/PhysRevLett.114.062006}{\emph{Phys. Rev.
  Lett.} {\bfseries 114} (2015) 062006},
  [\href{https://arxiv.org/abs/1409.0023}{{\ttfamily 1409.0023}}].

\bibitem{Grozin:2015kna}
A.~Grozin, J.~M. Henn, G.~P. Korchemsky and P.~Marquard, \emph{{The three-loop
  cusp anomalous dimension in QCD and its supersymmetric extensions}},
  \href{http://dx.doi.org/10.1007/JHEP01(2016)140}{\emph{JHEP} {\bfseries 01}
  (2016) 140}, [\href{https://arxiv.org/abs/1510.07803}{{\ttfamily
  1510.07803}}].

\bibitem{Siegel:1979wq}
W.~Siegel, \emph{{Supersymmetric Dimensional Regularization via Dimensional
  Reduction}},
  \href{http://dx.doi.org/10.1016/0370-2693(79)90282-X}{\emph{Phys. Lett.}
  {\bfseries B84} (1979) 193--196}.

\bibitem{Bianchi:2013rma}
M.~Bianchi, G.~Giribet, M.~Leoni and S.~Penati, \emph{{The 1/2 BPS Wilson loop
  in ABJ(M) at two loops: The details}},
  \href{http://dx.doi.org/10.1007/JHEP10(2013)085}{\emph{JHEP} {\bfseries 1310}
  (2013) 085}, [\href{https://arxiv.org/abs/1307.0786}{{\ttfamily 1307.0786}}].

\bibitem{Griguolo:2013sma}
L.~Griguolo, G.~Martelloni, M.~Poggi and D.~Seminara, \emph{{Perturbative
  evaluation of circular 1/2 BPS Wilson loops in $\mathcal{N} =$ 6 Super
  Chern-Simons theories}},
  \href{http://dx.doi.org/10.1007/JHEP09(2013)157}{\emph{JHEP} {\bfseries 1309}
  (2013) 157}, [\href{https://arxiv.org/abs/1307.0787}{{\ttfamily 1307.0787}}].

\bibitem{Bianchi:2013pva}
M.~S. Bianchi, G.~Giribet, M.~Leoni and S.~Penati, \emph{{Light-like Wilson
  loops in ABJM and maximal transcendentality}},
  \href{http://dx.doi.org/10.1007/JHEP08(2013)111}{\emph{JHEP} {\bfseries 08}
  (2013) 111}, [\href{https://arxiv.org/abs/1304.6085}{{\ttfamily 1304.6085}}].

\bibitem{Grozin:2000jv}
A.~G. Grozin, \emph{{Calculating three loop diagrams in heavy quark effective
  theory with integration by parts recurrence relations}},
  \href{http://dx.doi.org/10.1088/1126-6708/2000/03/013}{\emph{JHEP} {\bfseries
  03} (2000) 013}, [\href{https://arxiv.org/abs/hep-ph/0002266}{{\ttfamily
  hep-ph/0002266}}].

\bibitem{Chetyrkin:2003vi}
K.~G. Chetyrkin and A.~G. Grozin, \emph{{Three loop anomalous dimension of the
  heavy light quark current in HQET}},
  \href{http://dx.doi.org/10.1016/S0550-3213(03)00490-5}{\emph{Nucl. Phys.}
  {\bfseries B666} (2003) 289--302},
  [\href{https://arxiv.org/abs/hep-ph/0303113}{{\ttfamily hep-ph/0303113}}].

\bibitem{Tkachov:1981wb}
F.~V. Tkachov, \emph{{A Theorem on Analytical Calculability of Four Loop
  Renormalization Group Functions}},
  \href{http://dx.doi.org/10.1016/0370-2693(81)90288-4}{\emph{Phys. Lett.}
  {\bfseries B100} (1981) 65--68}.

\bibitem{Chetyrkin:1981qh}
K.~G. Chetyrkin and F.~V. Tkachov, \emph{{Integration by Parts: The Algorithm
  to Calculate beta Functions in 4 Loops}},
  \href{http://dx.doi.org/10.1016/0550-3213(81)90199-1}{\emph{Nucl. Phys.}
  {\bfseries B192} (1981) 159--204}.

\bibitem{Laporta:1996mq}
S.~Laporta and E.~Remiddi, \emph{{The Analytical value of the electron (g-2) at
  order alpha**3 in QED}},
  \href{http://dx.doi.org/10.1016/0370-2693(96)00439-X}{\emph{Phys. Lett.}
  {\bfseries B379} (1996) 283--291},
  [\href{https://arxiv.org/abs/hep-ph/9602417}{{\ttfamily hep-ph/9602417}}].

\bibitem{Laporta:2001dd}
S.~Laporta, \emph{{High precision calculation of multiloop Feynman integrals by
  difference equations}},
  \href{http://dx.doi.org/10.1016/S0217-751X(00)00215-7,
  10.1142/S0217751X00002157}{\emph{Int. J. Mod. Phys.} {\bfseries A15} (2000)
  5087--5159}, [\href{https://arxiv.org/abs/hep-ph/0102033}{{\ttfamily
  hep-ph/0102033}}].

\bibitem{Smirnov:2008iw}
A.~V. Smirnov, \emph{{Algorithm FIRE -- Feynman Integral REduction}},
  \href{http://dx.doi.org/10.1088/1126-6708/2008/10/107}{\emph{JHEP} {\bfseries
  10} (2008) 107}, [\href{https://arxiv.org/abs/0807.3243}{{\ttfamily
  0807.3243}}].

\bibitem{Smirnov:2013dia}
A.~V. Smirnov and V.~A. Smirnov, \emph{{FIRE4, LiteRed and accompanying tools
  to solve integration by parts relations}},
  \href{http://dx.doi.org/10.1016/j.cpc.2013.06.016}{\emph{Comput. Phys.
  Commun.} {\bfseries 184} (2013) 2820--2827},
  [\href{https://arxiv.org/abs/1302.5885}{{\ttfamily 1302.5885}}].

\bibitem{Smirnov:2014hma}
A.~V. Smirnov, \emph{{FIRE5: a C++ implementation of Feynman Integral
  REduction}}, \href{http://dx.doi.org/10.1016/j.cpc.2014.11.024}{\emph{Comput.
  Phys. Commun.} {\bfseries 189} (2015) 182--191},
  [\href{https://arxiv.org/abs/1408.2372}{{\ttfamily 1408.2372}}].

\bibitem{Lee:2012cn}
R.~N. Lee, \emph{{Presenting LiteRed: a tool for the Loop InTEgrals
  REDuction}},  \href{https://arxiv.org/abs/1212.2685}{{\ttfamily 1212.2685}}.

\bibitem{Lee:2013mka}
R.~N. Lee, \emph{{LiteRed 1.4: a powerful tool for reduction of multiloop
  integrals}}, \href{http://dx.doi.org/10.1088/1742-6596/523/1/012059}{\emph{J.
  Phys. Conf. Ser.} {\bfseries 523} (2014) 012059},
  [\href{https://arxiv.org/abs/1310.1145}{{\ttfamily 1310.1145}}].

\bibitem{Lee:2010hk}
K.-M. Lee and S.~Lee, \emph{{1/2-BPS Wilson Loops and Vortices in ABJM Model}},
  \href{http://dx.doi.org/10.1007/JHEP09(2010)004}{\emph{JHEP} {\bfseries 09}
  (2010) 004}, [\href{https://arxiv.org/abs/1006.5589}{{\ttfamily 1006.5589}}].

\bibitem{Lietti:2017gtc}
M.~Lietti, A.~Mauri, S.~Penati and J.-j. Zhang, \emph{{String theory duals of
  Wilson loops from Higgsing}},
  \href{https://arxiv.org/abs/1705.02322}{{\ttfamily 1705.02322}}.

\bibitem{Dorn:1986dt}
H.~Dorn, \emph{{Renormalization of Path Ordered Phase Factors and Related
  Hadron Operators in Gauge Field Theories}},
  \href{http://dx.doi.org/10.1002/prop.19860340104}{\emph{Fortsch. Phys.}
  {\bfseries 34} (1986) 11--56}.

\bibitem{Craigie:1980qs}
N.~S. Craigie and H.~Dorn, \emph{{On the Renormalization and Short Distance
  Properties of Hadronic Operators in {QCD}}},
  \href{http://dx.doi.org/10.1016/0550-3213(81)90372-2}{\emph{Nucl. Phys.}
  {\bfseries B185} (1981) 204--220}.

\bibitem{Aoyama:1981ev}
S.~Aoyama, \emph{{The Renormalization of the String Operator in {QCD}}},
  \href{http://dx.doi.org/10.1016/0550-3213(82)90023-2}{\emph{Nucl. Phys.}
  {\bfseries B194} (1982) 513--534}.

\bibitem{Knauss:1984rx}
D.~Knauss and K.~Scharnhorst, \emph{{Two Loop Renormalization of Nonsmooth
  String Operators in {Yang-Mills} Theory}},
  \href{http://dx.doi.org/10.1002/andp.19844960413}{\emph{Annalen Phys.}
  {\bfseries 41} (1984) 331--344}.

\bibitem{Preti:2017}
M.~Preti, \emph{{WiLE: a Mathematic package for weak coupling expansion of
  Wilson loops in ABJ(M) theory}}, {\emph{to appear} (2017) }.

\bibitem{Gromov:2016rrp}
N.~Gromov and F.~Levkovich-Maslyuk, \emph{{Quark-anti-quark potential in $
  \mathcal{N} =$ 4 SYM}},
  \href{http://dx.doi.org/10.1007/JHEP12(2016)122}{\emph{JHEP} {\bfseries 12}
  (2016) 122}, [\href{https://arxiv.org/abs/1601.05679}{{\ttfamily
  1601.05679}}].

\bibitem{Cavaglia:2014exa}
A.~Cavagli{\`a}, D.~Fioravanti, N.~Gromov and R.~Tateo, \emph{{Quantum Spectral
  Curve of the $\mathcal N=$ 6 Supersymmetric Chern-Simons Theory}},
  \href{http://dx.doi.org/10.1103/PhysRevLett.113.021601}{\emph{Phys. Rev.
  Lett.} {\bfseries 113} (2014) 021601},
  [\href{https://arxiv.org/abs/1403.1859}{{\ttfamily 1403.1859}}].

\bibitem{Bombardelli:2017vhk}
D.~Bombardelli, A.~Cavagli{\`a}, D.~Fioravanti, N.~Gromov and R.~Tateo,
  \emph{{The full Quantum Spectral Curve for $AdS_4/CFT_3$}},
  \href{https://arxiv.org/abs/1701.00473}{{\ttfamily 1701.00473}}.

\bibitem{Nishioka:2008gz}
T.~Nishioka and T.~Takayanagi, \emph{{On Type IIA Penrose Limit and N=6
  Chern-Simons Theories}},
  \href{http://dx.doi.org/10.1088/1126-6708/2008/08/001}{\emph{JHEP} {\bfseries
  08} (2008) 001}, [\href{https://arxiv.org/abs/0806.3391}{{\ttfamily
  0806.3391}}].

\bibitem{Grignani:2008is}
G.~Grignani, T.~Harmark and M.~Orselli, \emph{{The SU(2) $\times$ SU(2) sector
  in the string dual of N=6 superconformal Chern-Simons theory}},
  \href{http://dx.doi.org/10.1016/j.nuclphysb.2008.10.019}{\emph{Nucl. Phys.}
  {\bfseries B810} (2009) 115--134},
  [\href{https://arxiv.org/abs/0806.4959}{{\ttfamily 0806.4959}}].

\bibitem{Gaiotto:2008cg}
D.~Gaiotto, S.~Giombi and X.~Yin, \emph{{Spin Chains in N=6 Superconformal
  Chern-Simons-Matter Theory}},
  \href{http://dx.doi.org/10.1088/1126-6708/2009/04/066}{\emph{JHEP} {\bfseries
  04} (2009) 066}, [\href{https://arxiv.org/abs/0806.4589}{{\ttfamily
  0806.4589}}].

\bibitem{Gromov:2014eha}
N.~Gromov and G.~Sizov, \emph{{Exact Slope and Interpolating Functions in N=6
  Supersymmetric Chern-Simons Theory}},
  \href{http://dx.doi.org/10.1103/PhysRevLett.113.121601}{\emph{Phys. Rev.
  Lett.} {\bfseries 113} (2014) 121601},
  [\href{https://arxiv.org/abs/1403.1894}{{\ttfamily 1403.1894}}].

\bibitem{Bonini:2016fnc}
M.~Bonini, L.~Griguolo, M.~Preti and D.~Seminara, \emph{{Surprises from the
  resummation of ladders in the ABJ(M) cusp anomalous dimension}},
  \href{http://dx.doi.org/10.1007/JHEP05(2016)180}{\emph{JHEP} {\bfseries 05}
  (2016) 180}, [\href{https://arxiv.org/abs/1603.00541}{{\ttfamily
  1603.00541}}].

\bibitem{Bianchi:2016rub}
M.~S. Bianchi and M.~Leoni, \emph{{An exact limit of the
  Aharony-Bergman-Jafferis-Maldacena theory}},
  \href{http://dx.doi.org/10.1103/PhysRevD.94.045011}{\emph{Phys. Rev.}
  {\bfseries D94} (2016) 045011},
  [\href{https://arxiv.org/abs/1605.02745}{{\ttfamily 1605.02745}}].

\bibitem{Cavaglia:2016ide}
A.~Cavagli{\`a}, N.~Gromov and F.~Levkovich-Maslyuk, \emph{{On the Exact
  Interpolating Function in ABJ Theory}},
  \href{http://dx.doi.org/10.1007/JHEP12(2016)086}{\emph{JHEP} {\bfseries 12}
  (2016) 086}, [\href{https://arxiv.org/abs/1605.04888}{{\ttfamily
  1605.04888}}].

\bibitem{Henn:2013pwa}
J.~M. Henn, \emph{{Multiloop integrals in dimensional regularization made
  simple}},
  \href{http://dx.doi.org/10.1103/PhysRevLett.110.251601}{\emph{Phys.Rev.Lett.}
  {\bfseries 110} (2013) 251601},
  [\href{https://arxiv.org/abs/1304.1806}{{\ttfamily 1304.1806}}].

\end{thebibliography}\endgroup

\end{document}